\documentclass[12pt]{revtex4-1}
\pdfoutput=1
\usepackage{hyperref, graphicx, multirow, chngpage}
\usepackage{amsmath,amssymb,amsfonts,physics}
\usepackage{caption}
\usepackage{tikz-feynman}
\usepackage{float}
\usepackage[makeroom]{cancel} 

\usepackage{xcolor}
\usepackage{ulem}

\begin{document}

\renewcommand\tablename{{\small Table}}

\renewcommand{\vec}[1]{{\boldsymbol{#1}}} 
\newcommand\Wtilde{\widetilde{W}}
\newcommand\bfd{\overset\leftrightarrow{D}}
\newcommand\htilde{\widetilde{H}}


\title{Primary Observables for\\ Electroweak Gauge Boson Collider Signals}

\author{Cristien Arzate, Spencer Chang, Gabriel Jacobo }

\address{Department of Physics and Institute for Fundamental Science\\ 
University of Oregon, Eugene, Oregon 97403}

\begin{abstract}
In this paper, we determine a basis for the on-shell 4-point amplitudes $VVVV$ for Standard Model gauge bosons  $V=W^\pm, Z, \gamma, g$.
Following previous work, this completes the analysis of 3- and 4-point amplitudes for the Standard Model and could be used for model independent 
searches for beyond the Standard Model physics at colliders.  Our results include a Lagrangian parametrization for the ``primary'' amplitudes, where including additional derivatives leads to the Mandelstam ``descendant'' amplitudes, and upper bounds on the coupling constants from imposing unitarity.  We also perform an estimate for the sensitivity for new $Z$ decays at the HL-LHC, finding that $Z\to \gamma \bar{\ell}\ell$ could be searched for, but that other decay modes, like $Z\to (\gamma\gamma\gamma, \gamma gg)$, are too small to be discovered after imposing unitarity constraints.

\end{abstract}
\maketitle

\noindent
\section{Introduction}
\label{sec:intro}
Recently there has been enormous progress in determining the general structure of on-shell amplitudes for the Standard Model \cite{Shadmi:2018xan, Durieux:2019eor, Durieux:2020gip,Dong:2022jru,Liu:2023jbq, Chang:2022crb, Bradshaw_2023}.  In addition to being of interest theoretically, these results could enable broad model independent searches for beyond the Standard Model physics, without relying on the standard effective field theory (EFT) parameterizations (i.e.~SMEFT \cite{Buchmuller:1985jz,Grzadkowski:2010es} and HEFT \cite{Feruglio:1992wf}).  Indeed, amplitudes may be a better way to connect experiment and theory, given the direct connection to experimental analyses and since amplitudes do not require the EFT assumptions of power counting and do not suffer from ambiguous basis issues of Lagrangian operators.  

In Refs.~\cite{Chang:2022crb, Bradshaw_2023}, the full structure of on-shell 3- and 4-point amplitudes involving the Higgs and top quark were determined.  This leaves 4-point gauge boson amplitudes as the remaining ones to be analyzed.  In this paper, we complete this study for both massive ($W^\pm,Z$) and massless ($\gamma, g$) gauge boson interactions.  At the HL-LHC and future colliders (e.g.~$e^+ e^-$ colliders at the $Z$ pole), we stand to increase our sample of $W/Z$ particles by orders of magnitude.  Our study of the amplitudes, then allows us to consider if there are interesting amplitudes for three body decays like $Z\to \gamma\gamma\gamma$ beyond those considered in the past.   
  
The rest of this paper is organized as follows:  Section~\ref{sec:indptamps} describes what amplitudes we will explore and how to determine independent amplitudes.  Section~\ref{sec:HilbertSeries} discusses the Hilbert series results for our gauge boson operators.  In Section~\ref{sec:pheno}, we discuss some relevant phenomenological issues, such as unitarity bounds on coupling strengths and also rough estimates for $Z$ decays at the HL-LHC.  Section~\ref{sec:amplitudes} is the main body of results, where we list the operators for the primary amplitudes.  In Section~\ref{sec:Zdecays}, we estimate which $Z$ decay amplitudes are interesting for exploration at the HL-LHC.  
Finally in Section~\ref{sec:conclusions}, we conclude.

\section{Finding Independent Amplitudes/Couplings for Electroweak Gauge Bosons}
\label{sec:indptamps}
To find the most general on-shell amplitudes for gauge bosons, we impose invariance under $\mathrm{SU}(3)_c\times \mathrm{U}(1)_{em}$ and Lorentz symmetry.  For 3- and 4-point interactions, this gives the following list:  
\begin{align}
\text{3pt}:  \bar{f}fV, hhV, hVV, VVV
\quad \text{4pt}:  hhhV, \bar{f}f hV, \bar{f}fVV, hhVV, hVVV, VVVV
\end{align}
where $f$ is a fermion, $h$ is a Higgs boson, and $V$ is any gauge boson.  To fully characterize these 4-point interactions, we also need additional 3-point interactions for exchange diagrams, which add 
\begin{align}
\text{3pt additional}:  \text{$hhh$, $\bar{ f } f h$}
\end{align}
Of these couplings,  the 3-point  and 4-point couplings, except for $VVVV$, have been determined (e.g.~\cite{Chang_2023, Bradshaw_2023}), so in this paper this leaves the following 4-point couplings to analyze:
	\begin{equation}
	\begin{split}
		VVVV: & \quad WWWW, WWZZ, ZZZZ, WWZ\gamma, ZZZ\gamma, WW\gamma \gamma, WWgg, \\
		& \quad ZZ\gamma \gamma, ZZgg, Z\gamma gg, Z\gamma \gamma \gamma, Zggg, \gamma \gamma \gamma \gamma, \gamma \gamma gg, \gamma ggg, gggg.
	\end{split}
	\end{equation}

In \cite{Chang_2023, Bradshaw_2023}, an approach for determining a basis for independent operators for 3- and 4-point on-shell amplitudes was developed and explained in detail. Here, we will briefly summarize the 3-step process of $1)$ enumerating an over-complete basis of amplitudes, $2)$ determining the independent primary amplitudes, and $3)$ checking the result against a Hilbert series calculation.  For those interested in the details, please refer to the discussion in \cite{Chang_2023, Bradshaw_2023}.

For step one, we use the fact that local on-shell 4-point gauge boson amplitudes are Lorentz invariants involving gauge boson polarization and momenta contracted with the metric or the Levi-Civita tensor.
For processes with massless gauge bosons, we use the associated field strength tensors to maintain gauge invariance and satisfy the Ward identity. 
We distinguish amplitudes that have no factors of Mandelstam variables from those with such factors.  We refer to the former as primary amplitudes and the latter as descendant amplitudes, following the terminology of  \cite{Chang_2023, Bradshaw_2023}.  By allowing for arbitrary Mandelstam factors in the descendants\footnote{For processes with identical particles, these Mandelstam factors must be invariant under the relevant crossing symmetries, as we will discuss later.}, we parameterize the most general on-shell 4-point amplitudes.
However, for on-shell 3-point amplitudes, all Mandelstam invariants  can be written in terms of the particle masses, so there are only primary amplitudes.
A process with amplitude $\mathcal{ M }$ can then be written as a linear sum $\mathcal{ M } = \sum_a C_a \mathcal{ M }_a$ of these parametrized amplitudes $\mathcal{ M }_a$. Each $\mathcal{ M }_a$ will have an associated Lagrangian operator $\mathcal{ O }_a$ with mass dimension $d_{ \mathcal{ O } }$ and a dimensionless coupling $c_a$. The associated Lagrangian terms can then be written as 
	\begin{equation}
		\mathcal{ L }_{ \textrm{amp} } = \sum_a \frac{ c_a }{ v^{ d_{ \mathcal{ O } - 4 } } } \mathcal{ O }_a.
	\end{equation}
The factors of Higgs vev in the denominator are chosen such that the coupling remains dimensionless at any given mass dimension. 

For the 4-point amplitudes we consider here, there are commonly two or more identical bosons. 
In such cases, the amplitudes need to be symmetric under crossing exchange of the identical bosons.  Assume we have the process $p_1(\epsilon_1) + p_2 (\epsilon_2) \to p_3(\epsilon_3) + p_4 (\epsilon_4)$, where the incoming particles are identical (note that for the moment, we are not assuming any gauge charges for them), with the amplitude ${ \mathcal{ M } }(12;34)$ where $i=1\cdots 4$ is shorthand for $p_i, \epsilon_i$.  
We can then form symmetric and anti-symmetric combinations under $1, 2$ exchange,
	\begin{equation}
		{ \mathcal{ M } }_{ \pm }(12;34) = \frac{1}{2}\left[ { \mathcal{ M } }(12;34) \pm { \mathcal{ M } }(21;34) \right],
	\end{equation}
we can use these to construct general amplitudes with Mandelstam invariants. Since under $1 \leftrightarrow 2$, $s$ is invariant and $t \leftrightarrow u$, the most general $1 \leftrightarrow 2$ symmetric amplitude is, 
	\begin{equation}
		\mathcal{ M }_{12}(12;34)= \mathcal{ M }_{12}(21;34) = F\left( s, (t-u)^2 \right) { \mathcal{ M } }_+(12;34) + (t-u)G\left( s, (t-u)^2 \right) { \mathcal{ M } }_-(12;34),
	\end{equation}
where the polynomial functions $F$ and $G$ are $1 \leftrightarrow 2$ exchange symmetric.

For the case of three identical bosons, the amplitudes should first be symmetrized for the first two bosons and then that result should be symmetrized with respect to exchanges with the third particle. The result of this yields
	\begin{align}
	\begin{split}
		\mathcal{ M }_{123}(12;34) & = H\left( s, (t-u)^2 \right) \mathcal{ M }_{12}(12;34) + H\left( t, (s-u)^2 \right) \mathcal{ M }_{12}(13;24) \\
		& + H\left( u, (t-s)^2 \right) \mathcal{ M }_{12}(32;14)
	\end{split}
	\end{align}
where exchange of incoming and outgoing particles has a minus sign and complex conjugation, e.g.~$1\leftrightarrow 3$ takes $p_1\to -p_3, \epsilon_1 \to \epsilon^*_3$ and $p_3\to -p_1, \epsilon_3 \to \epsilon^*_1.$	
The arguments of $H$ have been chosen such that under $i \leftrightarrow j$ exchange, with $i, j = 1,2,3$, the sum of the three terms is invariant, i.e.~$\mathcal{ M }_{123}(12;34)$ is invariant under the permutations of $1,2,3$ in the argument of the function.     This follows since under $1 \leftrightarrow 3$ exchange, $s \leftrightarrow u$ and under $2 \leftrightarrow 3$ exchange, $s \leftrightarrow t$.  Finally, for the case of four identical bosons, the previous result needs to be symmetrized with respect to the fourth particle.  This requires replacing  $\mathcal{ M }_{12}(12;34)$ with $\mathcal{ M }_{12,34}(12;34)=\frac{1}{2}[\mathcal{ M }_{12}(12;34)+\mathcal{ M }_{12}(12;43)]$.  

For step two, we take this over-complete basis of amplitudes and find the independent ones.  To do this, we will work in increasing mass dimension for $\mathcal{O}_a$.  
In the center of mass frame for the $1 + 2 \to 3+4$, the weighted amplitude $E^n_\textrm{cm} \mathcal{M}$ is a polynomial of the kinematic variables $E_{\textrm{cm}}, \cos{\theta_\textrm{cm}}, \sin{\theta_\textrm{cm}},|\vec{p}_\textrm{initial}|, |\vec{p}_\textrm{final}|$.\footnote{In some cases, there can be inverse factors of $E_\textrm{cm}$, which we can get rid of by taking the power $n$ to be large enough.}
To proceed, we simplify the amplitude by replacing even powers of $\sin{\theta_\textrm{cm}}, |\vec{p}_\textrm{initial}|, |\vec{p}_\textrm{final}|$ with their solution in terms of $\cos{\theta_\textrm{cm}}$ and $E_\textrm{cm}$ leading to the general amplitude
\begin{align}
 E^n_\textrm{cm} \mathcal{ M } =&   P + Q \sin{\theta_\textrm{cm}} + R |\vec{p}_\textrm{initial}| + S  |\vec{p}_\textrm{final}|+ T \sin{\theta_\textrm{cm}} |\vec{p}_\textrm{initial}|+ U \sin{\theta_\textrm{cm}}  |\vec{p}_\textrm{final}| \nonumber \\  & + V  |\vec{p}_\textrm{initial}|  |\vec{p}_\textrm{final}| + W \sin{\theta_\textrm{cm}} |\vec{p}_\textrm{initial}|  |\vec{p}_\textrm{final}|
\end{align}  
where $P, Q, R, S, T, U, V, W$ are polynomials in $\cos{\theta_\textrm{cm}}, E_\textrm{cm}$.  
As argued in \cite{Chang_2023}, for a redundancy to occur, i.e.~$\mathcal{ M }=0$, one needs each of the $P, \cdots, W$ polynomials to independently vanish.  Since the coefficients in the polynomials depend on the couplings, $c_a$, one can use a numerical singular value decomposition to find how many redundancies there are and by process of elimination, find an independent set of couplings $c_a$ with corresponding amplitudes $\mathcal{M}_a$.

The third step provides a complementary constraint on the number of independent amplitudes.  To do so, we determine the number of independent Lagrangian operators using the Hilbert series, which counts the number of operators at each mass dimension \cite{Lehman_2015, Henning_2015, Lehman_2016, Henning:2015alf, Henning_2017, Gr_f_2021, Gr_f_2023}.   Since there is one-to-one correspondence between non-redundant operators $\mathcal{O}_a$ and amplitudes $\mathcal{M}_a$ (e.g.~\cite{Shadmi:2018xan}), this counting can be used to check the numerical analysis.  
More will be said about the specifics of the Hilbert series in the next section.

\section{Hilbert Series}
\label{sec:HilbertSeries}

The Hilbert series is a tool that provides the number of gauge invariant independent operators in a given EFT \cite{Lehman_2015, Henning:2015daa, Lehman_2016, Henning:2015alf, Henning_2017, Gr_f_2021, Gr_f_2023}. The Hilbert series counts the number of independent operators while taking into account symmetry constraints, equations of motion, and redundancies due to integration by parts. In Eq. \ref{eqn:Hilbert4pt}, we list the Hilbert series for 4-point interactions involving only electroweak gauge bosons. The Hilbert series for the other 3- and 4-point operators aforementioned can be found in \cite{Chang_2023, Bradshaw_2023}.
\begin{align} \footnotesize
\begin{split} 
& H_{WWWW} = \frac{2q^4 + 16 q^6 + 22q^8 + 7 q^{10} -2 q^{12}}{(1-q^2)(1-q^4)},\quad  H_{WWZZ} = \frac{2q^4 + 27 q^6 + 40q^8 + 14 q^{10} -2 q^{12}}{(1-q^2)(1-q^4)}, \\
& H_{ZZZZ} = \frac{q^4 + 4 q^6 + 8q^8  +11 q^{10} +5 q^{12}-2q^{14}}{(1-q^4)(1-q^6)}, \quad H_{WWZ\gamma} = \frac{22 q^6 + 34 q^8 + (2-4)q^{10}}{(1-q^2)^2}, \\
& H_{ZZZ\gamma} = \frac{4 q^6 + 14 q^8 + 22q^{10}+12 q^{12} + (4-2)q^{14}}{(1-q^4)(1-q^6)},\quad H_{WW\gamma\gamma}= H_{WWgg} = \frac{3 q^6 + 19 q^8 + 14q^{10} + (2 - 2)q^{ 12 }}{(1-q^2)(1-q^4)}, \\
& H_{ZZ\gamma\gamma}= H_{ZZgg} = \frac{3 q^6 + 13 q^8 + 7q^{10}-2q^{12}}{(1-q^2)(1-q^4)}, \quad H_{Z\gamma gg}=  \frac{12 q^8 + 12q^{10}+(2-2)q^{12}}{(1-q^2)(1-q^4)}, \\
& H_{Z\gamma\gamma\gamma} = \frac{4 q^8 + 10q^{10}+8q^{12}+(4-2)q^{14}}{(1-q^4)(1-q^6)}, \quad H_{Zggg} = \frac{6 q^8 + 18q^{10}+16q^{12}+(8-2)q^{14}+2q^{16}}{(1-q^4)(1-q^6)},\\
& H_{\gamma\gamma\gamma\gamma} = \frac{3 q^8 + 5q^{10}+q^{12}-2q^{14}}{(1-q^4)(1-q^6)}, \quad H_{\gamma\gamma gg} = \frac{7 q^8 + 5q^{10}-2q^{12}}{(1-q^2)(1-q^4)},\\
& H_{\gamma ggg} = \frac{4 q^8 + 12q^{10}+8q^{12}+(6-2)q^{14}+4q^{16}}{(1-q^4)(1-q^6)}, \quad H_{gggg} = \frac{9 q^8 + 14q^{10}+16q^{12}+(9-2)q^{14}+(2-4)q^{16}}{(1-q^4)(1-q^6)}.\\
\end{split}
\label{eqn:Hilbert4pt}
\end{align}
The correct way to interpret these Hilbert series is to take the exponent of each $q$ to be the mass dimension and the corresponding coefficients to be the number of independent operators minus the number of redundancies at that mass dimension.  When evaluating the Hilbert series, one cannot tell if there is such a cancellation.  
Only by looking at the independent amplitudes can we resolve this ambiguity, so using those results we have written out terms with a cancellation explicitly, e.g. $(6-2)q^{14}$ in the $H_{\gamma ggg}$ shows that there are 6 new primaries and 2 redundancies appearing at dimension 14.  
The appearance of these negative terms in the coefficients means that descendants of primary operators at a lower mass dimension become redundant to operators at the corresponding mass dimension of the negative term and that the higher dimensional operators and their descendants should be discarded from the set of independent operators.

To illustrate this in a specific example, take the Hilbert series for $WWWW$. The numerator looks like
	\begin{equation}
		2q^4 + 16 q^6 + 22q^8 + 7 q^{10} -2 q^{12}.
	\end{equation}
So in terms of primary operators at dimension 4 there are two operators, at dimension 6 there are 16 operators, at dimension 8 there are 22 operators, at dimension 10, there are seven operators, and at dimension 12 there are two redundancies. Now, to see the descendant structure of each operator, the denominator should be Taylor expanded. For example, the negative term turns out to be 
	\begin{equation}
		H_{WWWW} \supset -\frac{2 q^{12}}{(1-q^2)(1-q^4)} = -2 q^{12} \left( 1 + q^2 + q^4 + \cdots \right) \left( 1 + q^4 + q^8 + \cdots \right).
	\end{equation}
This says that if we consider the channel $W^+ W^+ \rightarrow W^- W^-$, then there are two redundant operators at dimension 12 which have descendant structures of the form $s^n (t-u)^{ 2 m }\mathcal{ O }$. The reason that they follow that specific structure is because of the exchange symmetries that the operators have to obey, namely a symmetry under $1 \leftrightarrow 2$ and $3 \leftrightarrow 4$. 
We'll later find out that the redundant operators are descendants of two dimension 8 operators, allowing us to rewrite the negative term, along with the positive dimension 8 term, as 
	\begin{align}
	\begin{split}
		H_{WWWW} & \supset \frac{22 q^8 - 2 q^{12}}{(1-q^2)(1-q^4)} = \frac{20 q^8 + 2 q^8 \left( 1 - q^{4} \right)}{(1-q^2)(1-q^4)} = \frac{20 q^8 }{ (1-q^2)(1-q^4) } + \frac{ 2 q^8 }{ (1-q^2) }\\
		& = 20 q^{8} \left( 1 + q^2 + q^4 + \cdots \right) \left( 1 + q^4 + q^8 + \cdots \right) + 2 q^{8} \left( 1 + q^2 + q^4 + \cdots \right).
	\end{split}
	\end{align}
From this we see that the correct interpretation of the independent operators is that two dimension 8 primary operators have a descendant structure of $s^n \mathcal{ O }$, while the other 20 dimension 8 primaries have a descendant structure of $s^n (t-u)^{ 2 m } \mathcal{ O }$. This means that for the former two primary operators, we can throw out their descendants of the form $s^n (t-u)^{ 2 m } \mathcal{ O }$ for $n \geq 0$ and $m \geq 1$ and still have an independent set of operators. Again, this Hilbert series interpretation must be checked with the amplitudes to confirm this explanation.  As another example for what the denominators mean, consider the denominator for the Hilbert series for $\gamma \gamma \gamma \gamma$.
	\begin{equation}
		\frac{1}{\left( 1 - q^4 \right) \left( 1 - q^6 \right)} = \left( 1 + q^4 + q^8 + \cdots \right) \left( 1 + q^6 + q^{ 12 } + \cdots \right)
	\end{equation}
The first set of parentheses says that there are powers of a 4-dimensional function of Mandelstam variables and the second set of parentheses says that are powers of a 6-dimensional function of Mandelstam variables. Because the $\gamma \gamma \gamma \gamma$ interaction should have exchange symmetries between all pairs of particles, the descendant structure should have the form $(s^2 + t^2 + u^2)^n (stu)^m$, in agreement with the dimensional analysis.

As mentioned earlier, some of the coefficients in the Hilbert series are written as a positive integer minus a negative integer. For example, this occurs in the Hilbert series for the $ZZZ\gamma$ interaction at dimension 14 as shown by the term $(4 - 2)q^{14}$.  When evaluating the Hilbert series this coefficient would be $2$, but in this case, by studying the amplitudes, we find that there are four new primary operators and two redundancies at  mass dimension 14, so we write the coefficient in this way to make this explicit. This also means that, for a given interaction, at mass dimensions higher than what we have explored, there could be terms with coefficients of zero not because there are no primary operators present, but because there are the same number of redundancies as primaries at that dimension. An example of this happening occurs for the $Z \gamma g g$ interaction at mass dimension 12, which we've written the term as $(2-2)q^{12}$. Therefore, it is not guaranteed that we have enumerated all possible primary operators, since there can be cancellations with the redundancies. However, because those would appear at very high mass dimension, they are phenomenologically unimportant and so don't warrant much concern.  This possibility is the reason why we have analyzed operators up to at least the first mass dimension that has a zero coefficient in the numerator and up to dimension 16 for operators of $Z ggg$, $\gamma ggg$, and $gggg$ interactions.

\section{Phenomenology}
\label{sec:pheno}
\subsection{Unitarity}
\label{subsec:unitarity}

As in \cite{Chang_2023, Bradshaw_2023}, we use unitarity constraints to place upper bounds on the couplings of the operators we have enumerated. We know that the SM does not violate tree-level unitarity at high energies (e.g.~\cite{PhysRevD.10.1145, Liu:2022alx}), therefore a deviation from the SM will violate it at some high energy scale $E_{ \textrm{max} }$. Our constraints will depend upon this scale and we roughly expect $E_{ \textrm{max} }$ above a TeV to be consistent with current LHC analyses, but values lower than a TeV to possibly be in tension. To compute the bounds, we follow the same techniques developed by \cite{Chang_2020, Abu_Ajamieh_2021, Abu_Ajamieh_2022} (see also \cite{Falkowski_2019}).

For each operator we create a schematic SMEFT realization of it in order to compute our unitarity bounds. To illustrate this, we turn to the $WWWW$ interaction where there is a dimension 6 primary operator $i W^{ + \mu } \widetilde{ W }^+_{ \nu \rho } D_{ \mu } W^{ - \nu } W^{ - \rho } + \textrm{h.c.}$. To realize this operator in SMEFT, one needs at least four Higgs doublets.  The non-field strength $W$'s come from covariant derivatives acting on the doublets, leading to four covariant derivatives, which by integration by parts can act on just three of the doublets. 
The dual field strength tensor $\widetilde{ W }^a_{ \mu \nu }$ needs to be contracted with the $\textrm{SU}(2)$ generators $T^a$.
This leads to a SMEFT operator, which we simplify into a schematic form: 
	\begin{equation}
		i H^{ \dag } T^a D^{ \mu } H \widetilde{ W }^a_{ \nu \rho } D_{ \mu }^{ \enspace \nu } H^{ \dag } D^{ \rho } H \rightarrow D^4 H^4 \widetilde{ W }^a_{ \mu \nu }.
	\end{equation}

Primary operators that have either zero or one field strength tensor have a SMEFT operator with at least four Higgs doublets by the following argument where we try to use only two Higgs doublets. If there are no field strength tensors, the SMEFT operator has the schematic form $D^n ( H^\dag H$) and, by integration by parts, can always be reduced to factors of $D^\mu D_\mu$ acting on one of the Higgs doublets.  Using the equations of motion, each $D^\mu D_\mu$ can be removed and replaced with expressions that involve no gauge bosons.  By iterating, we see that this doesn't realize the four gauge boson amplitude we wanted, thus we need operators with four Higgs doublets, like $D^n (H^\dag H H^\dag H)$.  Since there are four Higgs doublets, it is not possible to write all Lorentz invariants in terms of $D^\mu D_\mu$'s, so the above argument cannot be applied.   A similar argument works for operators with one field strength, $D^n ( H^\dag W_{\mu\nu} H)$, since derivative pairs on the three fields can also be moved on to individual fields.  From Lorentz invariance, these are either of the form 1) $W^{\mu\nu}(D_\mu D_\nu H) \propto W^{\mu\nu} W_{\mu\nu} H$, and thus actually have two field strengths, or 2) $(D^\mu D_\mu H)$ or $(D^\mu D_\mu W_{\alpha\beta})$ which can be reduced by equations of motion to expressions with fewer covariant derivatives or more than one field strength, respectively.  For operators with two or three field strength tensors, these arguments no longer work and we   only need two Higgs doublets instead of four for the SMEFT operator.  Finally, if we have four field strength tensors, then we do not need any Higgs doublets for the SMEFT operator.   

To calculate the coupling constraints, we need to estimate two quantities: the scattering amplitude and phase space factor for the highest and lowest particle multiplicity of an interaction process. By using these two quantities we can put upper bounds on the coupling strengths of operators. Note that because we are calculating approximate bounds we only care about factors of $v$ and neglect $O(1)$ factors like $\sqrt{ 2 }$, $g$, $g'$, $\sin{ \theta_W }$, and $\cos{ \theta_W }$. At high energy our amplitudes will then be of the form
	\begin{equation}
		\mathcal{ M }\left( \phi_1 \cdots \phi_k \rightarrow \phi_{ k + 1 } \cdots \phi_n \right) \sim c_a \frac{ E^m }{ v^{m+n-4} },
	\end{equation}
where $E$ is the total energy, $v$ is the Higgs vev, $k$ is the total number of incoming particles, and $n$ is the total number of particles.  The allowed values of $n$ are determined by the SMEFT operator and can be varied by setting Higgs doublets to their vev. To obtain $m$, we count the energy scaling of the various quantities in the schematic SMEFT operator. The scaling behavior of various quantities are:
	\begin{equation}
		D \sim E, \quad B_{ \mu \nu } \sim E, \quad W^a_{ \mu \nu } \sim E, \quad  G^b_{ \mu \nu } \sim E.
	\end{equation}
The value of $m$ is just the total energy scaling of the operator and the power of $v$ is given by dimensional analysis.  
Next, the unitarity bound on the amplitude depends on the the phase space factor of the initial and final states:
	\begin{equation}
		\mathcal{ M }\left( \phi_1 \cdots \phi_k \rightarrow \phi_{ k + 1 } \cdots \phi_n \right) \lesssim \frac{ 1 }{ \sqrt{ \Phi_k(E) \Phi_{ n - k }(E) } }
	\end{equation}
where $\Phi_k(E)$ is the total phase space for $k$ particles with center of mass energy $E$.  We follow \cite{Chang_2023} and work in the massless limit where approximately,
	\begin{equation}
		\Phi_k(E) \sim \frac{ 1 }{ 8 \pi } \left( \frac{ E }{ 4 \pi } \right)^{ 2 k - 4 }.
	\end{equation}
 Since unitarity is violated at some energy $E_{ \textrm{max} }$, we get that the unitarity bound on the couplings is
	\begin{equation} \label{eq:uppercbound}
		c_a \lesssim 2(4\pi)^{n-3} \left(\frac{v}{E_\textrm{max}}\right)^{m+n-4}.  
	\end{equation}
Now, recall that the interaction of the operator $i W^{ + \mu } \widetilde{ W }^+_{ \nu \rho } D_{ \mu } W^{ - \nu } W^{ - \rho } + \textrm{h.c.}$ has a schematic SMEFT form of $D^4 H^4 \widetilde{ W }^a_{ \mu \nu }$. We now evaluate the amplitudes at high energy by using the equivalence theorem, using the Nambu-Goldstone bosons in $H$ for longitudinal $W$'s and $Z$'s.  We find that the best bounds come from using the derivative part of the covariant derivatives and not the transverse gauge bosons.  Thus, for the schematic operator $D^4 H^4 \widetilde{ W }^a_{ \mu \nu }$, we can have the 5-point process $\phi \phi \rightarrow \phi \phi W_T$ or the 4-point process $\phi \phi \rightarrow \phi W_T$, where $\phi$ is a Nambu-Goldstone boson.  Under our approximation, these have amplitudes $\mathcal{ M }\left( \phi \phi \rightarrow \phi \phi W_T \right) \sim c_a \frac{ E^5 }{ v^6 }$ and $\mathcal{ M }\left( \phi \phi \rightarrow \phi   W_T \right) \sim c_a \frac{ E^5 }{ v^5 }$.
For the first case $n = 5$, $m = 5$,  and for the second case $n = 4$, $m = 5$.  Respectively, these lead to the bounds $c_a \lesssim \frac{0.07}{E_\text{TeV}^6}$ and $c_a \lesssim \frac{0.02}{E_\text{TeV}^5}$, where $E_\text{TeV} \equiv E_\text{max}/\text{TeV}.$  Thus the higher multiplicity amplitude is more stringent for higher  $E_\text{TeV}$ and the lower multiplicity is more stringent for lower  $E_\text{TeV}$, where they cross at  $E_\text{max} = 4\pi v$ for the bound $c_a \lesssim 2(4\pi)^{1-m}$. In this way we proceed to calculate unitarity bounds on coupling strengths for all enumerated operators.

\subsection{Electroweak Gauge Boson Decays}
In this subsection, we give estimates for modifications to electroweak gauge boson decays.  Two body decays of the $Z$ are quite well covered, as $Z\to (\gamma\gamma,gg)$ are forbidden by the Landau-Yang theorem and $Z\to \bar{f}f$ were studied at LEP1 for vector and axial couplings \cite{ALEPH:2005ab}.  This leaves only the dipole couplings to fermions, which  interfere with the Standard Model with a rate proportional to the fermion mass \cite{Gupta:2014rxa}. 
We will now discuss the on-shell 3-body decay modes of $Z$ bosons which are allowed by the SM, which are
	\begin{equation}
		Z \to (\gamma\gamma\gamma, \gamma g g, g g g, \bar{f}f\gamma, \bar{f}f g).
	\end{equation}
These decay modes occur in the Standard Model at higher order, so there can be interference with the new amplitudes.   

We ignore the masses of the fermions, so that the mass of the $Z$ is the only relevant energy scale. Then we can  approximate the new decay amplitudes as
	\begin{equation}
		\mathcal{ M }_{ \mathcal{ O } }(Z \rightarrow 3) \simeq \frac{ c_{ \mathcal{ O } } }{ v^{ d_{ \mathcal{ O } } - 4 } } m_Z^{ d_{ \mathcal{ O } } - 4 }, 
	\end{equation}
where $v$ is the Higgs vev, $c_{ \mathcal{ O } }$ are couplings and $m_Z$ is the mass of the $Z$ boson.
If the SM amplitude is larger than the BSM one, then interference between the SM and BSM amplitudes forms the most significant contribution to the total decay amplitude. Making the same approximations as in \cite{Bradshaw_2023}, we estimate that the branching ratios including interference are
	\begin{align}
	\begin{split}
		& \textrm{BR}( Z \rightarrow 3 )_{ \textrm{BSM} } \approx \frac{ m_Z }{ 512 \pi^3 \Gamma_Z } \left| \mathcal{ M }(Z \rightarrow 3)_{ \textrm{SM} } + \mathcal{ M }(Z \rightarrow 3)_{ \textrm{BSM} } \right|^2
	\end{split}
	\end{align}
where we have approximated both the SM and BSM amplitudes as constants.

\section{Independent Amplitudes for Electroweak Gauge Bosons}
\label{sec:amplitudes}

In this section, we discuss the independent primary operators for $VVVV$ interaction amplitudes. We will also check the number of operators and redundancies with the Hilbert series for each interaction. In the second column of Tables \ref{tab:wwww1}-\ref{tab:ggggg2}, we list the operators for the primary amplitudes.  In addition, we give the amplitude's CP transformation, the dimension of the operator, the schematic form of a SMEFT operator realization, and the unitarity bounds on the coupling strength. An example of how these unitarity bounds are calculated can be found in subsection \ref{subsec:unitarity}.  Primary operators and their descendants are $\textrm{SU}(3)_c \times \textrm{U}(1)_{ em }$ invariant so the covariant derivatives only involve the gluon and photon, whereas the covariant derivatives for the SMEFT operators are $\textrm{SU}(3)_c \times \textrm{SU}(2)_L \times \textrm{U}(1)_{Y}$ invariant.  Finally, for operators that have nontrivial $\textrm{SU}(3)$ contractions, we will add a column to specify this.  In the following, we discuss each interaction's Table(s) in detail and describe how the amplitudes and their redundancies agree with the Hilbert Series in Eqn.~\ref{eqn:Hilbert4pt}. 

Tables \ref{tab:wwww1} and \ref{tab:wwww2} list the primary operators for $W^+W^+W^-W^-$ interactions up to dimension 10. To analyze the amplitude, we assume the process $W_1^+W_2^+\rightarrow W_3^+ W_4^+$.  In terms of primary operators, the Hilbert series predicts that there should be two dimension 4 operators, 16 dimension 6 operators, 22 dimension 8 operators, seven dimension 10 operators, and two redundancies appearing at dimension 12.  From our amplitude enumeration procedure, we find agreement with the Hilbert series and find that the dimension 12 descendant operators $s^2 \mathcal{ O }^{ WWWW }_{ 25 }$ and $s^2 \mathcal{ O }^{ WWWW }_{ 35 }$ are redundant, with $s = (p_{ W^+_1 } + p_{ W^+_2 })^2$. Because descendants of redundant operators continue to be redundant, the operators $s^n (t-u)^{ 2m } \mathcal{ O }^{ WWWW }_{ 25 }$ and $s^n (t-u)^{ 2m } \mathcal{ O }^{ WWWW }_{ 35 }$, for $n \geq 2$, $m \geq 0$ should be removed in order to form a set of independent operators.  Remember that since $t=(p_{ W^+_1 } - p_{ W^+_3 })^2$ and $u=(p_{ W^+_1 } - p_{ W^+_4})^2$, $s$ and $(t-u)^2$ are the Mandelstam invariants which respect the interchanges  $W_1^+ \leftrightarrow W_2^+$ and $W_3^+ \leftrightarrow W_4^+$. 

In Tables \ref{tab:zzww1}-\ref{tab:zzww3} we list the primary operators for the $ZZW^+W^-$ interactions up to dimension 10 after considering the process $ZZ\to W^+W^-$. The Hilbert series states that for these there should be two dimension 4 operators, 27 dimension 6 operators, 40 dimension 8 operators, 14 dimension 10 operators, and two redundancies which appear at dimension 12. Our findings are in agreement with the Hilbert series. The redundancies and their descendants are given by the operators $s^n (t-u)^{ 2 m } \mathcal{ O }^{ ZZWW }_{ 39}$ and $s^n (t-u)^{ 2 m } \mathcal{ O }^{ ZZWW }_{ 54}$, for $n \geq 2$ and $m \geq 0$. To form a set independent operators, they should all be omitted.

We list the primary operators for $ZZZZ$ interactions up to dimension 12 in Table \ref{tab:zzzz}. We achieve full agreement with the Hilbert series, from which we expect that there should be one dimension 4 operator, four dimension 6 operators, eight dimension 8 operators, 11 dimension 10 operators, five dimension 12 operators, and two redundancies appearing at dimension 14. Operators $x^n y^m \mathcal{ O }^{ ZZZZ }_{ 22}$ and $x^n y^m \mathcal{ O }^{ ZZZZ }_{ 23}$ for $n\geq1, m\geq 0$ are redundant, where $x = s^2 + t^2 + u^2$ and $y = s t u$, and should not be included in a set of independent operators.

In Tables \ref{tab:wwzgam1}-\ref{tab:wwzgam2}, by using the process $W^+W^- \rightarrow Z\gamma $, we enumerate the primary operators for the $W^+ W^- Z\gamma$ interaction up to dimension 10. We agree with the Hilbert series that there are 22 dimension 6 operators, 34 dimension 8 operators, two dimension 10 operators, and four redundancies that appear at dimension 10. These redundancies and their descendants correspond to operators $s^n t^m \mathcal{ O }^{ WW\gamma Z }_{ 44}$, $s^n t^m \mathcal{ O }^{ WW\gamma Z }_{ 49}$, $s^n t^m \mathcal{ O }^{ WW\gamma Z }_{ 54}$, and $s^n t^m \mathcal{ O }^{ WW\gamma Z }_{ 55}$, with $n \geq 1$, $m \geq 0$. To form a complete set of independent operators they should be removed.

In Tables \ref{tab:zzzgam1}-\ref{tab:zzzgam2} we list the primary operators for the $ZZZ\gamma$ interaction up to dimension 14. We obtain full agreement with the Hilbert series. There are four dimension 6 operators, 14 dimension 8 operators, 22 dimension 10 operators, 12 dimension 12 operators, four dimension 14 operators, and two redundancies that appear at dimension 14. To form a set of independent operators, $x^n y^m \mathcal{ O }^{ ZZZ \gamma }_{ 28}$ and $x^n y^m \mathcal{ O }^{ ZZZ \gamma }_{ 29}$, with $x = s^2 + t^2 + u^2$, $y = s t u$, $n \geq 1$ and $m \geq 0$, should be omitted.

The primary operators for the $W^+ W^- \gamma \gamma$ interaction up to dimension 12 are listed in Table \ref{tab:wwgamgam} and they agree with the expectations from the Hilbert series. There are three dimension 6 operators, 19 dimension 8 operators, 14 dimension 10 operators, two dimension 12 operators, and two redundancies that show up at dimension 12. Note that in this case, the coefficient for $q^{12}$ exactly cancels between the two operators and two redundancies. The following operators and their descendants should be removed to maintain an independent set of operators: $s^n (t-u)^{ 2 m } \mathcal{ O }^{ WW \gamma \gamma }_{ 9}$ and $s^n (t-u)^{ 2 m } \mathcal{ O }^{ WW \gamma \gamma }_{ 17}$, with $n \geq 0$, and $m \geq 1$. The primary operators for the $W^+ W^- gg$ interaction up to dimension 12 can be obtained by replacing $F_{ \mu \nu } \rightarrow G^A_{ \mu \nu }$ and contracting the $\textrm{SU}(3)$ indices with $\delta_{AB}$. The redundancies of the $W^+ W^- \gamma \gamma$ interaction apply to the corresponding operators of the $W^+ W^- gg$ interaction.

Primary operators for the $ZZ \gamma \gamma$ interaction up to dimension 10 are tabulated in Table \ref{tab:zzgamgam}. Our results agree with the Hilbert series, from which we expect there to be three dimension 6 operators, 13 dimension 8 operators, 7 dimension 10 operators, and two redundancies appearing at dimension 12. The operators and descendants $s^n (t-u)^{ 2 m } \mathcal{ O }^{ ZZ \gamma \gamma }_{ 9}$ and $s^n (t-u)^{ 2 m } \mathcal{ O }^{ ZZ \gamma \gamma }_{ 12}$, with $n \geq 0$ and $m \geq 1$, should be omitted in order to form a set of independent operators. The primary operators for the $ZZ gg$ interaction up to dimension 12 can be obtained by making the replacement $F_{ \mu \nu } \rightarrow G^A_{ \mu \nu }$ and contracting the $\textrm{SU}(3)$ indices with  $\delta_{AB}$. The redundancies of the $ZZ \gamma \gamma$ interaction apply to the corresponding operators of the $ZZ gg$ interaction.

Table \ref{tab:zgamgg} lists the primary operators for the $Z \gamma g g$ interaction up to dimension 12. Agreeing with the Hilbert series, we  find 12 dimension 8 operators, 12 dimension 10 operators, two dimension 12 operators, and two redundancies at dimension 12. To form a set of independent operators, the operators and descendants $s^n (t-u)^{ 2 m } \mathcal{ O }^{ Z \gamma g g }_{ 4}$, $s^n (t-u)^{ 2 m } \mathcal{ O }^{ Z \gamma g g }_{ 9 }$, with $n \geq 0$, $m \geq 1$, should be removed. 

The primary operators for the $Z \gamma \gamma \gamma$ interaction up to dimension 14 are enumerated in Table \ref{tab:zgamgamgam}. The Hilbert series predicts that there are four dimension 8 operators, ten dimension 10 operators, eight dimension 12 operators, four dimension 14 operators, and two redundancies that appear at dimension 14. Our results are in agreement with this prediction. In order to form a set of independent operators, the operators and descendants $x^n y^m \mathcal{ O }^{ Z \gamma \gamma \gamma }_{ 9}$ and $x^n y^m \mathcal{ O }^{ Z \gamma \gamma \gamma }_{ 13 }$, with $x = s^2 + t^2 + u^2$, $y = s t u$, $n \geq 1$, and $m \geq 0$, should be omitted. Symmetric (in any $g \leftrightarrow g$ particle exchange of their kinematic variables) primary operators for the $Zggg$ interaction up to dimension 14 can be obtained by making the replacement $F_{ \mu \nu } \rightarrow G^A_{ \mu \nu }$ and contracting $\textrm{SU}(3)$ indices with the fully symmetric structure constant tensor $d_{ ABC }$. The redundancies of the $Z \gamma \gamma \gamma$ interaction apply to the corresponding operators of the $Zggg$ interaction.

In Table \ref{tab:zggg}, we list the antisymmetric (in any $g \leftrightarrow g$ particle exchange of kinematics) primary operators for the $Z g g g$ interaction up to dimension 16. Note that  the $\textrm{SU}(3)$ indices of the gluon field strengths $G^A_{ \mu \nu }$ are suppressed in our notation and taken to be contracted with the fully anti-symmetric structure constant tensor $f_{ABC}$, so that under the combined color and kinematic exchange, the gluons obey Bose-Einstein statistics. When taken with the symmetric $Z g g g$ operators obtained from modifications to operators in Table \ref{tab:zgamgamgam}, which was discussed in the last paragraph, our results agree with the Hilbert series. We find that there are six dimension 8 operators, 18 dimension 10 operators, 16 dimension 12 operators, eight dimension 14 operators, two dimension 16 operators, and two redundancies that appear at dimension 14. These redundancies come from operators involving the fully symmetric structure constant tensor $d_{ ABC }$ and can be obtained from Table \ref{tab:zgamgamgam}. They, along with their descendants, should be removed to form a set of independent operators. 

In Table \ref{tab:gamgamgamgam}, we have a list of the primary operators for the $\gamma \gamma \gamma \gamma$ interaction up to dimension 12. The Hilbert series predicts three dimension 8 operators, five dimension 10 operators, one dimension 12 operator, and two redundancies at dimension 14, which are the results that we find. We can form a set of independent operators by removing the descendants  $x^n y^{m} \mathcal{ O }^{ \gamma \gamma \gamma \gamma }_{ 6  }$ and $x^{n} y^{m} \mathcal{ O }^{ \gamma \gamma \gamma \gamma }_{ 7 }$, with $n \geq 1$, and $m \geq 0$, $x = s^2 + t^2 + u^2$, $y = s t u$ which are redundant

In Table \ref{tab:gamgamgg}, we can find the primary operators for the $\gamma \gamma g g$ interaction up to dimension 10. There are seven dimension 8 operators, and five dimension 10 operators. This result is consistent with the corresponding Hilbert series. At dimension 12, there are two redundancies such that we should omit the descendants  $s^n (t-u)^{ 2 m } \mathcal{ O }^{ \gamma \gamma g g }_{ 4  }$ and $s^{m+1} (t-u)^{ 2 n } \mathcal{ O }^{ \gamma \gamma g g }_{ 6  }$, with $n \geq 0$, and $m \geq 1$, in order to have a list of independent operators.

In Tables \ref{tab:ggggamma1}-\ref{tab:ggggamma2}, we enumerate a set of primary operators for the $\gamma ggg$ interaction up to dimension 16. We find four dimension 8 operators, twelve dimension 10 operators, eight dimension 12 operators, six dimension 14 operators, four dimension 16 operators, and two redundancies at dimension 14, agreeing with the Hilbert series. We should remove the descendant operators  $x^n y^m \mathcal{ O }^{ \gamma ggg}_{ 8 }$, $x^n y^m \mathcal{ O }^{ \gamma ggg}_{ 16  }$, with $x = s^2 + t^2 + u^2$, $y = s t u$, $n \geq 1$ and $m \geq 0$ to exclude redundancies. In this case, we are including in the tables the fully symmetric and anti-symmetric structure constants $d_{ABC}$ and $f_{ABC}$, respectively, which implicitly contract the SU(3) indices of the $G^{A}_{\mu \nu}$'s.

 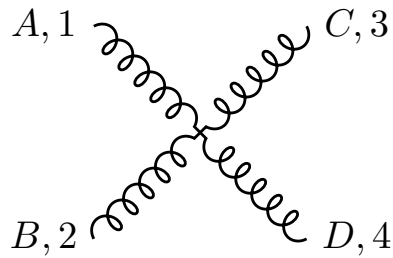
\begin{figure}[H]
 \centering
 \tikzfeynmanset{
mygluon/.style={
decoration={coil,amplitude=1.2mm,segment length=2.25mm, aspect=0.75},decorate}
}

\resizebox{0.35\textwidth}{!}{
 \begin{tikzpicture}
 
   \begin{feynman}
     \vertex (a);
     \vertex [below right=of a] (e);
     \vertex [above right=of e] (c);
     \vertex [below left=of e] (b);
     \vertex [below right=of e] (d);
    
     \diagram* {
       (a) -- [mygluon, thick] (e),
       (b) -- [mygluon, thick] (e),
       (c) -- [mygluon, thick] (e),
       (d) -- [mygluon, thick] (e),
     };
    
     \node at ([xshift=-5mm] a) {$A,1$};
     \node at ([xshift=-5mm] b) {$B,2$};
     \node at ([xshift=5mm] c) {$C,3$};
     \node at ([xshift=5mm] d) {$D,4$};
   \end{feynman}
 \end{tikzpicture}
  }
 \centering
 \caption{4-gluon interaction. Labels correspond to color indices, momenta and polarization and we consider the process $(A,1) + (B,2) \to (C,3) + (D,4)$.}
 \label{fig:gluons}
 \end{figure}

Four gluon scattering proceeds as in Figure~\ref{fig:gluons}.
An example Lagrangian operator for four gluon interactions is
$f(T^A, T^B, T^C, T^C) G^{A}_{\mu \nu} G^{B \nu \rho} G^{C}_{\rho \sigma} G^{D \sigma \mu}$, where we have chosen a structure for the contraction of the Lorentz indices and $f$ represents a configuration of the trace of the generators. In this case, we have two ways to express $f$ (up to trivial permutations) to keep the amplitude invariant: $Tr(T^A T^B) Tr(T^C T^D)$ and $Tr(T^A T^B T^C T^D)$. Considering that the gluons are identical, this gives two different possible amplitudes:\\
\begin{flalign}
\begin{split}\label{type1}
    M_{Tr^2}(1A \ 2B; 3C \ 4D)= & Tr(T^A T^B) Tr(T^C T^D) M_{Tr^2}(12;34)  \\
    &+ Tr(T^A T^C) Tr(T^B T^D) M_{Tr^2}(13;24) \\
    &+ Tr(T^A T^D) Tr(T^B T^C) M_{Tr^2}(14;23),
\end{split}
\end{flalign}
\begin{flalign}
\begin{split}\label{type2}
    M_{Tr}(1A \ 2B; 3C \ 4D)=&Tr(T^A T^B T^C T^D) M_{Tr}(12;34)+Tr(T^A T^B T^D T^C) M_{Tr}(12;43) \\
     & +Tr(T^A T^C T^B T^D) M_{Tr}(13;24) +Tr(T^A T^C T^D T^B) M_{Tr}(13;42) \\
    & +Tr(T^A T^D T^B T^C) M_{Tr}(14;23) +Tr(T^A T^D T^C T^B) M_{Tr}(14;32) 
\end{split}
\end{flalign}
where the right hand side factorizes the amplitude into color factors and sub-amplitudes which only  depend on kinematics and polarizations.  Given the structure of the operators, the sub-amplitudes have the following identities under exchange of kinematics, $M_{Tr^2}(12;34) = M_{Tr^2}(21;34) = M_{Tr^2}(12;43) = M_{Tr^2}(21;43)= M_{Tr^2}(34;12)=M_{Tr^2}(43;12)=M_{Tr^2}(34;21)=M_{Tr^2}(43;21)$ and $M_{Tr}(12;34) = M_{Tr}(23;41) = M_{Tr}(34;12) = M(_{Tr}41;23)$.  By forming candidate sub-amplitudes with the correct symmetries under the exchange of kinematics, we can find the independent $M_{Tr^2}(12;34)$ and $M_{Tr}(12;34)$.  These will respectively lead to an independent set of $M_{Tr^2}(1A \ 2B; 3C \ 4D)$ and $M_{Tr}(1A \ 2B; 3C \ 4D)$. However, it is possible that there will still be redundancies between the two types of amplitudes $M_{Tr^2}(1A \ 2B; 3C \ 4D)$ and $M_{Tr}(1A \ 2B; 3C \ 4D)$.

In the following, we show that if there is a redundancy between the sub-amplitudes of the two types, then there will be a redundancy amongst the full amplitudes after putting in the color factors.  To see this, assume that there is a redundancy between two sub-amplitudes $\hat{M}(12;34)=M_{Tr^2}(12;34)=M_{Tr}(12;34)$. Now $\hat{M}$ is invariant under both of the permutation symmetries of the two sub-amplitudes and one can show that this means $\hat{M}$ is invariant under arbitrary permutations of the 4 particles.   Then we can show that  
\begin{flalign}
    M_{Tr^2}(1A \ 2B; 3C \ 4D)=&(Tr(T^A T^B) Tr(T^C T^D) +Tr(T^A T^C) Tr(T^B T^D)  \nonumber \\
    &+ Tr(T^A T^D) Tr(T^B T^C) )\hat{M}(12;34)_{\text{}}  \nonumber \\
    =& (Tr(T^A T^B T^C T^D) +Tr(T^A T^B T^D T^C) \nonumber \\
     & +Tr(T^A T^C T^B T^D) +Tr(T^A T^C T^D T^B)  \nonumber \\
    & +Tr(T^A T^D T^B T^C) +Tr(T^A T^D T^C T^B) ) \hat{M}(12;34)_{\text{}} \nonumber \\
    =&  M_{Tr}(1A \ 2B; 3C \ 4D)
\end{flalign}
which shows that the full amplitudes are also redundant.  
In the second equality, we have used the group theory identity for $\mathrm{SU}(3)$
\begin{equation}
    \sum_{\text{six distinct perms}}Tr(T^A T^B T^C T^D)=\sum_{\text{three distinct perms}}Tr(T^A T^B)Tr( T^C T^D)
\end{equation}
with the distinct permutations given in (\ref{type1}-\ref{type2}).  This implies that it is enough to look for the independent sub-amplitudes of both types to characterize the independent full amplitudes.  

For simplicity, we adopt the notation $Tr(T^2)Tr(T^2)$ and $Tr(T^4)$ to represent \\ $Tr(T^A T^B)Tr(T^C T^D)$ and $Tr(T^A T^B T^C T^D)$ respectively.  In Tables \ref{tab:ggggg}-\ref{tab:ggggg2}, we show a set of primary operators for $gggg$ interactions, up to dimension 16. We find nine dimension 8 operators, 14 dimension 10 operators, 16 dimension 12 operators, nine dimension 14 operators, two dimension 16 operators, two redundancies at dimension 14, and four redundancies at dimension 16, which agrees with the corresponding Hilbert series. The redundancies are the operators and descendants given by  $x^n y^m \mathcal{ O }^{gggg}_{ 4 }$, $x^n y^m \mathcal{ O }^{gggg}_{ 9   }$,  $x^n y^m \mathcal{ O }^{gggg}_{ 12 }$, $x^n y^m \mathcal{ O }^{gggg}_{ 14 }$, $x^n y^m \mathcal{ O }^{gggg}_{ 19 }$,  $x^n y^m \mathcal{ O }^{gggg}_{ 20 }$,   with $x = s^2 + t^2 + u^2$, $y = s t u$, $n \geq 0$ and $m \geq 1$, and should be removed if we want a set of independent operators.

\begin{table}[p]
\begin{adjustwidth}{-.5in}{-.5in}  
\begin{center}
\footnotesize
\centering
\renewcommand{\arraystretch}{0.9}
\setlength{\tabcolsep}{6pt}
\begin{tabular}{|c|c|c|c|c|c|}

\hline
\multirow{2}{*}{$i$} & \multirow{2}{*}{$\mathcal{O}_i^{WWWW}$}   & \multirow{2}{*}{CP} & \multirow{2}{*}{$d_{\mathcal{O}_i}$}& SMEFT & $c$ Unitarity  \\
 & & & & Operator Form & Bound \\

\hline 
1 & \scriptsize{$ W^{ + \mu } W^{ + }_{ \mu } W^{ - \nu } W^-_{ \nu } $} & $+$ & \multirow{2}{*}{4} & \multirow{2}{*}{ \scriptsize{$ D^4 H^4 $} } & \multirow{2}{*}{ $\frac{ 0.09 }{ E_\text{TeV}^4}$ }  \\

2 & \scriptsize{$ W^{ + \mu } W^{ + \nu } W^-_{ \mu } W^-_{ \nu } $} & $+$ &  &  &  \\

\hline
3 & \scriptsize{$ D^{ \rho } W^{ + \mu } W^{ + \nu } \left( W^-_{ \mu } \overleftrightarrow{ D }_{ \rho } W^-_{ \nu } \right) + \textrm{h.c.} $} & $+$ & \multirow{12}{*}{6} & \multirow{6}{*}{ \scriptsize{$ D^6 H^4 $} } & \multirow{12}{*}{  $ \frac{0.006}{E_\text{TeV}^6} $ }  \\

4 & \scriptsize{$ D_{ \nu } W^{ + \mu } D_{ \mu } W^{ + \nu } W^{ - \rho } W^-_{ \rho } + \textrm{h.c.} $} &$+$&  &  &  \\

5 & \scriptsize{$ W^{ + \mu } W^{ + \nu } W^{ - \rho } D_{ \mu \nu } W^-_{ \rho } + \textrm{h.c.} $} &$+$&  &  &  \\

6 & \scriptsize{$ D_{ \nu } W^{ + \mu } D_{ \mu } W^{ + \rho } W^{ - \nu } W^-_{ \rho } + \textrm{h.c.} $} &$+$&  &  &  \\

7 & \scriptsize{$ W^{ + \mu } W^{ + \nu } D_{ \mu \rho } W^{ - }_{ \nu } W^{ - \rho } + \textrm{h.c.} $} &$+$&  &  &  \\

8 & \scriptsize{$ D_{ \rho } W^{ + \mu } W^{ + \nu } D_{ \mu } W^{ - \rho } W^-_{ \nu } $} &$+$&  &  &  \\

\cline{5-5}

9 & \scriptsize{$ i \varepsilon^{ \mu \nu \rho \sigma } D_{ \mu } W^{ + \alpha } W^{ + }_{ \nu } D_{ \alpha } W^{ - }_{ \rho } W^-_{ \sigma } + \textrm{h.c.} $} &$+$&  & \scriptsize{$ \varepsilon D^6 H^4 $} &  \\

\cline{5-5}

10 & \scriptsize{$ i D_{ \nu } W^{ + \mu } D_{ \mu } W^{ + \nu } W^{ - \rho } W^-_{ \rho } + \textrm{h.c.} $} &$-$&  & \multirow{3}{*}{ \scriptsize{$ D^6 H^4 $} } &  \\

11 & \scriptsize{$ i W^{ + \mu } W^{ + \nu } W^{ - \rho } D_{ \mu \nu } W^-_{ \rho } + \textrm{h.c.} $} &$-$&  &  &  \\

12 & \scriptsize{$ i D_{ \nu } W^{ + \mu } D_{ \mu } W^{ + \rho } W^{ - \nu } W^-_{ \rho } + \textrm{h.c.} $} &$-$&  &  &  \\

\cline{5-5}

13 & \scriptsize{$ \varepsilon^{ \mu \nu \rho \sigma } D_{ \alpha } W^{ + }_{ \mu } W^{ + }_{ \nu } \left( W^{ - }_{ \rho } \overleftrightarrow{ D }^{ \alpha } W^-_{ \sigma } \right) + \textrm{h.c.} $} &$-$&  & \multirow{2}{*}{ \scriptsize{$ \varepsilon D^6 H^4 $} } &  \\

14 & \scriptsize{$ \varepsilon^{ \mu \nu \rho \sigma } D_{ \mu } W^{ + \alpha } W^{ + }_{ \nu } D_{ \alpha } W^{ - }_{ \rho } W^-_{ \sigma } + \textrm{h.c.} $} &$-$&  &  &  \\

\hline

15 & \scriptsize{$ i W^{ + \mu } \widetilde{ W }^{ + }_{ \nu \rho } D_{ \mu } W^{ - \nu } W^{ - \rho } + \textrm{h.c.} $} &$+$& \multirow{4}{*}{6} & \multirow{4}{*}{ \scriptsize{$ D^4 H^4 \widetilde{ W }^{ a }_{ \mu \nu } $} } & \multirow{4}{*}{ $ \frac{0.02}{E_\text{TeV}^5} $, $ \frac{0.07}{E_\text{TeV}^6} $ } \\

16 & \scriptsize{$ i D^{ \mu } W^{ + \nu } W^{ + }_{ \nu } \widetilde{ W }^{ - }_{ \mu \rho } W^{ - \rho } + \textrm{h.c.} $} &$+$&  &  &  \\

17 & \scriptsize{$ W^{ + \mu } \widetilde{ W }^{ + }_{ \nu \rho } D_{ \mu } W^{ - \nu } W^{ - \rho } + \textrm{h.c.} $} &$-$&  &  &  \\

18 & \scriptsize{$ D^{ \mu } W^{ + \nu } W^{ + }_{ \nu } \widetilde{ W }^{ - }_{ \mu \rho } W^{ - \rho } + \textrm{h.c.} $} &$-$&  &  &  \\

\hline

\end{tabular}
\begin{minipage}{5.7in}
\medskip
\caption{\label{tab:wwww1} \footnotesize Primary dimension 4 and 6 operators for $W^+ W^+ W^- W^-$ interactions. }
\end{minipage}
\end{center}
\end{adjustwidth}
\end{table}

\begin{table}[p]
\begin{adjustwidth}{-.5in}{-.5in}  
\begin{center}
\footnotesize
\centering
\renewcommand{\arraystretch}{0.9}
\setlength{\tabcolsep}{6pt}
\begin{tabular}{|c|c|c|c|c|c|}

\hline
\multirow{2}{*}{$i$} & \multirow{2}{*}{$\mathcal{O}_i^{WWWW}$}   & \multirow{2}{*}{CP} & \multirow{2}{*}{$d_{\mathcal{O}_i}$}& SMEFT & $c$ Unitarity  \\
 & & & & Operator Form & Bound \\

\hline 
19 & \scriptsize{$ D_{ \sigma } W^{ + \mu } D_{ \mu } W^{ + \nu } \left( D_{ \nu } W^{ - \rho } \overleftrightarrow{ D }^{ \sigma } W^-_{ \rho } \right) + \textrm{h.c.} $} & $+$ & \multirow{17}{*}{8} & \multirow{7}{*}{ \scriptsize{$ D^8 H^4 $} } & \multirow{17}{*}{ $ \frac{3 \times 10^{-4}}{E_\text{TeV}^8} $ }  \\

20 & \scriptsize{$ D_{ \rho \sigma } W^{ + \mu } D_{ \mu } W^{ + \nu } \left( W^{ - \rho } \overleftrightarrow{ D }^{ \sigma } W^-_{ \nu } \right) + \textrm{h.c.} $} &$+$&  &  &  \\

21 & \scriptsize{$ D_{ \rho \sigma } W^{ + }_{ \mu } W^{ + }_{ \nu } \left( D^{ \mu } W^{ - \rho } \overleftrightarrow{ D }^{ \sigma } W^{ - \nu } \right) + \textrm{h.c.} $} &$+$&  &  &  \\

22 & \scriptsize{$ D_{ \nu \rho \sigma } W^{ + \mu } D_{ \mu } W^{ + \nu } W^{ - \rho } W^{ - \sigma } + \textrm{h.c.} $} &$+$&  &  &  \\

23 & \scriptsize{$ D_{ \nu \rho } W^{ + \mu } D_{ \mu \sigma } W^{ + \nu } W^{ - \rho } W^{ - \sigma } + \textrm{h.c.} $} &$+$&  &  &  \\

24 & \scriptsize{$ D_{ \rho } W^{ + \mu } D_{ \mu \sigma } W^{ + \nu } D_{ \nu } W^{ - \rho } W^{ - \sigma } + \textrm{h.c.} $} &$+$&  &  &  \\

25 & \scriptsize{$ D_{ \rho \sigma } W^{ + \mu } W^{ + \nu } D_{ \mu \nu } W^{ - \rho } W^{ - \sigma } + \textrm{h.c.} $} &$+$&  &  &  \\

\cline{5-5}

26 & \scriptsize{$ \varepsilon^{ \mu \nu \rho \sigma } D_{ \mu \alpha } W^{ + \beta } D_{ \beta } W^{ + }_{ \nu } \left( i W^{ - }_{ \rho } \overleftrightarrow{ D }^{ \alpha } W^-_{ \sigma } \right) + \textrm{h.c.} $} &$+$&  & \multirow{2}{*}{ \scriptsize{$ \varepsilon D^8 H^4 $} } &  \\

27 & \scriptsize{$ \varepsilon^{ \mu \nu \rho \sigma } D_{ \mu \alpha } W^{ + \beta } D_{ \nu } W^{ + }_{ \beta } \left( i W^{ - }_{ \rho } \overleftrightarrow{ D }^{ \alpha } W^-_{ \sigma } \right) + \textrm{h.c.} $} &$+$&  &  &  \\

\cline{5-5}

28 & \scriptsize{$ D_{ \rho \sigma } W^{ + \mu } D_{ \mu } W^{ + \nu } \left( i W^{ - \rho } \overleftrightarrow{ D }^{ \sigma } W^-_{ \nu } \right) + \textrm{h.c.} $} &$-$&  & \multirow{3}{*}{ \scriptsize{$ D^8 H^4 $} } &  \\

29 & \scriptsize{$ D_{ \sigma } W^{ + \mu } D_{ \mu } W^{ + \nu } \left( i D_{ \nu } W^{ - \rho } \overleftrightarrow{ D }^{ \sigma } W^{ - }_{ \rho } \right) + \textrm{h.c.} $} &$-$&  &  &  \\

30 & \scriptsize{$ i D_{ \nu \rho \sigma } W^{ + \mu } D_{ \mu } W^{ + \nu } W^{ - \rho } W^{ - \sigma } + \textrm{h.c.} $} &$-$&  &  &  \\

\cline{5-5}

31 & \scriptsize{$ \varepsilon^{ \mu \nu \rho \sigma } D_{ \mu \alpha } W^{ + \beta } D_{ \beta } W^{ + }_{ \nu } \left( W^{ - }_{ \rho } \overleftrightarrow{ D }^{ \alpha } W^-_{ \sigma } \right) + \textrm{h.c.} $} &$-$&  & \multirow{5}{*}{ \scriptsize{$ \varepsilon D^8 H^4 $} } &  \\

32 & \scriptsize{$ \varepsilon^{ \mu \nu \rho \sigma } D_{ \mu \alpha } W^{ + \beta } D_{ \nu } W^{ + }_{ \beta } \left( W^{ - }_{ \rho } \overleftrightarrow{ D }^{ \alpha } W^-_{ \sigma } \right) + \textrm{h.c.} $} &$-$&  &  &  \\

33 & \scriptsize{$ \varepsilon^{ \mu \nu \rho \sigma } D_{ \mu \alpha } W^{ + \beta } W^{ + }_{ \nu } \left( D_{ \rho } W^{ - }_{ \beta } \overleftrightarrow{ D }^{ \alpha } W^-_{ \sigma } \right) + \textrm{h.c.} $} &$-$&  &  &  \\

34 & \scriptsize{$ \varepsilon^{ \mu \nu \rho \sigma } D_{ \mu } W^{ + \alpha } D_{ \nu \alpha } W^{ + \beta } D_{ \beta } W^{ - }_{ \rho } W^{ - }_{ \sigma } + \textrm{h.c.} $} &$-$&  &  &  \\

35 & \scriptsize{$ \varepsilon^{ \mu \nu \rho \sigma } D_{ \mu \beta } W^{ + \alpha } W^{ + }_{ \nu } D_{ \rho \alpha } W^{ - \beta } W^{ - }_{ \sigma } $} &$-$&  &  &  \\

\hline 

36 & \scriptsize{$ D_{ \sigma } W^{ + \mu } D_{ \mu } W^{ + }_{ \nu \rho } \left( W^{ - \rho } \overleftrightarrow{ D }^{ \sigma } W^{ - \nu } \right) + \textrm{h.c.} $} &$+$& \multirow{5}{*}{8} & \multirow{1}{*}{ \scriptsize{$ D^6 H^4 W^{ a }_{ \mu \nu } $} } & \multirow{5}{*}{ $ \frac{0.001}{E_\text{TeV}^7}, \frac{0.004}{E_\text{TeV}^8} $ } \\

\cline{5-5}

37 & \scriptsize{$ D_{ \sigma } W^{ + \mu } D_{ \mu } \widetilde{ W }^{ + }_{ \nu \rho } \left( i W^{ - \nu } \overleftrightarrow{ D }^{ \sigma } W^{ - \rho } \right) + \textrm{h.c.} $} &$+$&  & \multirow{4}{*}{ \scriptsize{$ D^6 H^4 \widetilde{ W }^{ a }_{ \mu \nu } $} } &  \\

38 & \scriptsize{$ i D^{ \nu \rho } W^{ + \mu } D_{ \mu } W^{ + }_{ \nu } \widetilde{ W }^{ - }_{ \rho \sigma } W^{ - \sigma } + \textrm{h.c.} $} &$+$&  &  &  \\

39 & \scriptsize{$ i D^{ \rho } W^{ + \mu } W^{ + \nu } D_{ \mu \nu } \widetilde{ W }^{ - }_{ \rho \sigma } W^{ - \sigma } + \textrm{h.c.} $} &$+$&  &  &  \\

40 & \scriptsize{$ D^{ \rho } W^{ + \mu } W^{ + \nu } D_{ \mu \nu } \widetilde{ W }^{ - }_{ \rho \sigma } W^{ - \sigma } + \textrm{h.c.} $} &$-$&  &  &  \\

\hline 

41 & \scriptsize{$ D_{ \nu \rho \alpha } W^{ + \mu } D_{ \mu \sigma } W^{ + \nu } \left( W^{ - \rho } \overleftrightarrow{ D }^{ \alpha } W^{ - \sigma } \right) + \textrm{h.c.} $} & $+$ & \multirow{7}{*}{10} & \multirow{2}{*}{ \scriptsize{$ D^{ 10 } H^4 $} } & \multirow{7}{*}{ $ \frac{2 \times 10^{ -5 }}{E_\text{TeV}^{10}} $ }  \\

42 & \scriptsize{$ D_{ \rho \sigma \alpha } W^{ + \mu } D_{ \mu } W^{ + \nu } \left( D_{ \nu } W^{ - \rho } \overleftrightarrow{ D }^{ \alpha } W^{ - \sigma } \right) + \textrm{h.c.} $} &$+$&  &  &  \\

\cline{5-5}

43 & \scriptsize{$ \varepsilon^{ \mu \nu \rho \sigma } D_{ \mu \beta \tau } W^{ + \alpha } D_{ \nu \alpha } W^{ + \beta } \left( i W^{ - }_{ \rho } \overleftrightarrow{ D }^{ \tau } W^{ - }_{ \sigma } \right) + \textrm{h.c.} $} &$+$&  & \multirow{2}{*}{ \scriptsize{$ \varepsilon D^{ 10 } H^4 $} } &  \\

44 & \scriptsize{$ \varepsilon^{ \mu \nu \rho \sigma } D_{ \mu \tau } W^{ + \alpha } D_{ \nu } W^{ + \beta } \left( i D_{ \alpha \beta } W^{ - }_{ \rho } \overleftrightarrow{ D }^{ \tau } W^{ - }_{ \sigma } \right) + \textrm{h.c.} $} &$+$&  &  &  \\ 

\cline{5-5}

45 & \scriptsize{$ D_{ \nu \rho \alpha } W^{ + \mu } D_{ \mu \sigma } W^{ + \nu } \left( i W^{ - \rho } \overleftrightarrow{ D }^{ \alpha } W^{ - \sigma } \right) + \textrm{h.c.} $} &$-$&  & \multirow{2}{*}{ \scriptsize{$ D^{ 10 } H^4 $} } &  \\

46 & \scriptsize{$ D_{ \rho \sigma \alpha } W^{ + \mu } D_{ \mu } W^{ + \nu } \left( i D_{ \nu } W^{ - \rho } \overleftrightarrow{ D }^{ \alpha } W^{ - \sigma } \right) + \textrm{h.c.} $} &$-$&  &  &  \\

\cline{5-5}

47 & \scriptsize{$ \varepsilon^{ \mu \nu \rho \sigma } D_{ \mu \beta \tau } W^{ + \alpha } D_{ \nu \alpha } W^{ + \beta } \left( W^{ - }_{ \rho } \overleftrightarrow{ D }^{ \tau } W^{ - }_{ \sigma } \right) + \textrm{h.c.} $} &$-$&  & \multirow{1}{*}{ \scriptsize{$ \varepsilon D^{ 10 } H^4 $} } &  \\

\hline

\end{tabular}
\begin{minipage}{5.7in}
\medskip
\caption{\label{tab:wwww2} \footnotesize Primary dimension 8 and 10 operators for the $W^+ W^+ W^- W^-$ interaction. There are two redundancies which appear at dimension 12 such that in order to form a set of independent operators,  $s^n (t-u)^{ 2 m } \mathcal{ O }^{ WWWW }_{ 25 }$ and $s^n (t-u)^{ 2 m } \mathcal{ O }^{ WWWW }_{ 35 }$, with $n \geq 2$, $m \geq 0$, should be omitted. }
\end{minipage}
\end{center}
\end{adjustwidth}
\end{table}

\begin{table}[p]
\begin{adjustwidth}{-.5in}{-.5in}  
\begin{center}
\footnotesize
\centering
\renewcommand{\arraystretch}{0.9}
\setlength{\tabcolsep}{6pt}
\begin{tabular}{|c|c|c|c|c|c|}

\hline
\multirow{2}{*}{$i$} & \multirow{2}{*}{$\mathcal{O}_i^{ZZWW}$}   & \multirow{2}{*}{CP} & \multirow{2}{*}{$d_{\mathcal{O}_i}$}& SMEFT & $c$ Unitarity  \\
 & & & & Operator Form & Bound \\

\hline 
1 & \scriptsize{$ Z^{ \mu } Z_{ \mu } W^{ + \nu } W^-_{ \nu } $} & $+$ & \multirow{2}{*}{4} & \multirow{2}{*}{ \scriptsize{$ D^4 H^4 $} } & \multirow{2}{*}{ $\frac{ 0.09 }{ E_\text{TeV}^4}$ }  \\

2 & \scriptsize{$ Z^{ \mu } Z^{ \nu } W^+_{ \mu } W^-_{ \nu } $} & $+$ &  &  &  \\

\hline
3 & \scriptsize{$ \partial^{ \rho } Z^{ \mu } Z^{ \nu } \left( W^+_{ \mu } \overleftrightarrow{ D }_{ \rho } W^-_{ \nu } + \textrm{h.c.} \right) $} & $+$ & \multirow{18}{*}{6} & \multirow{9}{*}{ \scriptsize{$ D^6 H^4 $} } & \multirow{18}{*}{  $ \frac{0.006}{E_\text{TeV}^6} $ }  \\

4 & \scriptsize{$ \partial_{ \nu } Z^{ \mu } \partial_{ \mu } Z^{ \nu } W^{ + \rho } W^-_{ \rho } $} &$+$&  &  &  \\

5 & \scriptsize{$ Z^{ \mu } Z^{ \nu } \left( W^{ + \rho } D_{ \mu \nu } W^-_{ \rho } + \textrm{h.c.} \right) $} &$+$&  &  &  \\

6 & \scriptsize{$ \partial_{ \rho } Z^{ \mu } \partial_{ \mu } Z^{ \nu } \left( W^{ + \rho } W^-_{ \nu } + \textrm{h.c.} \right) $} &$+$&  &  &  \\

7 & \scriptsize{$ Z^{ \mu } \partial_{ \mu \rho } Z^{ \nu } \left( W^{ + \rho } W^-_{ \nu } + \textrm{h.c.} \right) $} &$+$&  &  &  \\

8 & \scriptsize{$ \partial^{ \rho } Z^{ \mu } Z^{ \nu } \left( W^{ + }_{ \nu } D_{ \mu } W^-_{ \rho } + \textrm{h.c.} \right) $} &$+$&  &  &  \\

9 & \scriptsize{$ Z^{ \mu } \partial^{ \rho } Z^{ \nu } \left( W^{ + }_{ \nu } D_{ \mu } W^-_{ \rho } + \textrm{h.c.} \right) $} &$+$&  &  &  \\

10 & \scriptsize{$ \partial_{ \nu \rho } Z^{ \mu } Z_{ \mu } \left( W^{ + \nu } W^{ - \rho } \right) $} &$+$&  &  &  \\

11 & \scriptsize{$ \partial_{ \nu } Z^{ \mu } \partial_{ \rho } Z_{ \mu } \left( W^{ + \nu } W^{ - \rho } + \textrm{h.c.} \right) $} &$+$&  &  &  \\

\cline{5-5}

12 & \scriptsize{$ i \varepsilon^{ \mu \nu \rho \sigma } \partial_{ \mu } Z^{ \alpha } \partial_{ \alpha } Z_{ \nu } W^+_{ \rho } W^-_{ \sigma } $} &$+$&  & \multirow{3}{*}{ \scriptsize{$ \varepsilon D^6 H^4 $} } &  \\

13 & \scriptsize{$ \varepsilon^{ \mu \nu \rho \sigma } \partial^{ \alpha } Z_{ \mu } Z_{ \nu } \left( i W^+_{ \rho } D_{ \sigma } W^-_{ \alpha } + \textrm{h.c.} \right) $} &$+$&  &  &  \\

14 & \scriptsize{$ \varepsilon^{ \mu \nu \rho \sigma } \partial_{ \mu } Z^{ \alpha } Z_{ \nu } \left( i W^+_{ \rho } D_{ \sigma } W^-_{ \alpha } + \textrm{h.c.} \right) $} &$+$&  &  &  \\

\cline{5-5}

15 & \scriptsize{$ Z^{ \mu } \partial_{ \mu } Z^{ \nu } \left( i W^{ + \rho } D_{ \nu } W^-_{ \rho } + \textrm{h.c.} \right) $} &$-$&  & \multirow{4}{*}{ \scriptsize{$ D^6 H^4 $} } &  \\

16 & \scriptsize{$ \partial^{ \mu } Z^{ \nu } \partial_{ \nu } Z^{ \rho } \left( i W^+_{ \mu } W^-_{ \rho } + \textrm{h.c.} \right) $} &$-$&  &  &  \\

17 & \scriptsize{$ \partial^{ \mu } Z^{ \nu } Z^{ \rho } \left( i W^+_{ \rho } D_{ \nu } W^-_{ \mu } + \textrm{h.c.} \right) $} &$-$&  &  &  \\

18 & \scriptsize{$ Z^{ \mu } \partial^{ \rho } Z^{ \nu } \left( i W^+_{ \nu } D_{ \mu } W^-_{ \rho } + \textrm{h.c.} \right) $} &$-$&  &  &  \\

\cline{5-5}

19 & \scriptsize{$ \varepsilon^{ \mu \nu \rho \sigma } \partial_{ \alpha } Z_{ \mu } Z_{ \nu } \left( W^+_{ \rho } \overleftrightarrow{ D }^{ \alpha } W^-_{ \sigma } + \textrm{h.c.} \right) $} &$-$&  & \multirow{2}{*}{ \scriptsize{$ \varepsilon D^6 H^4 $} } &  \\

20 & \scriptsize{$ \varepsilon^{ \mu \nu \rho \sigma } \partial_{ \mu } Z^{ \alpha } Z_{ \nu } \left( W^+_{ \rho } D_{ \alpha } W^-_{ \sigma } + \textrm{h.c.} \right) $} &$-$&  &  &  \\

\hline

21 & \scriptsize{$ i Z^{ \alpha } \partial_{ \alpha } \widetilde{ Z }^{ \mu \nu } W^+_{ \mu } W^-_{ \nu } $} &$+$& \multirow{9}{*}{6} & \multirow{3}{*}{ \scriptsize{$ D^4 H^4 \widetilde{ W }^a_{ \mu \nu } $} } & \multirow{9}{*}{ $ \frac{0.02}{E_\text{TeV}^5} $, $ \frac{0.07}{E_\text{TeV}^6} $ } \\

22 & \scriptsize{$ Z^{ \mu } \widetilde{ Z }^{ \nu \rho } \left( i W^+_{ \nu } D_{ \rho } W^-_{ \mu } + \textrm{h.c.} \right) $} &$+$& &  &  \\

23 & \scriptsize{$ Z^{ \mu } Z^{ \nu } \left( i W^{ + \rho } D_{ \mu } \widetilde{ W }^-_{ \nu \rho } + \textrm{h.c.} \right) $} &$+$&  &  &  \\

\cline{5-5}

24 & \scriptsize{$ Z^{ \mu } \partial_{ \mu } Z^{ \nu \rho } \left( i W^{ + }_{ \nu } W^-_{ \rho } + \textrm{h.c.} \right) $} &$-$& & \scriptsize{$ D^4 H^4 W^{ a }_{ \mu \nu } $} &  \\

\cline{5-5}

25 & \scriptsize{$ \partial^{ \rho } \widetilde{ Z }^{ \mu \nu } Z_{ \mu } \left( W^{ + }_{ \rho } W^-_{ \nu } + \textrm{h.c.} \right) $} &$-$&  & \multirow{5}{*}{ \scriptsize{$ D^4 H^4 \widetilde{ W }^a_{ \mu \nu } $} } &  \\

26 & \scriptsize{$ Z^{ \mu } \widetilde{ Z }^{ \nu \rho } \left( W^+_{ \nu } D_{ \mu } W^-_{ \rho } + \textrm{h.c.} \right) $} &$-$&  &  &  \\

27 & \scriptsize{$ \partial_{ \nu } Z^{ \mu } \widetilde{ Z }^{ \nu \rho } \left( W^+_{ \mu } W^-_{ \rho } + \textrm{h.c.} \right) $} &$-$&  &  &  \\

28 & \scriptsize{$ Z^{ \mu } \partial_{ \mu } Z_{ \nu } \left( W^+_{ \rho } \widetilde{ W }^{ - \nu \rho } + \textrm{h.c.} \right) $} &$-$&  &  &  \\

29 & \scriptsize{$ \partial_{ \mu } Z^{ \rho } Z_{ \rho } \left( W^+_{ \nu } \widetilde{ W }^{ - \mu \nu } + \textrm{h.c.} \right) $} &$-$&  &  &  \\

\hline

\end{tabular}
\begin{minipage}{5.7in}
\medskip
\caption{\label{tab:zzww1} \footnotesize Primary dimension 4 and 6 operators for the $Z Z W^+ W^-$ interaction. }
\end{minipage}
\end{center}
\end{adjustwidth}
\end{table}

\begin{table}[p]
\begin{adjustwidth}{-.5in}{-.5in}  
\begin{center}
\scriptsize
\centering
\renewcommand{\arraystretch}{0.9}
\setlength{\tabcolsep}{6pt}
\begin{tabular}{|c|c|c|c|c|c|}

\hline
\multirow{2}{*}{$i$} & \multirow{2}{*}{$\mathcal{O}_i^{ZZWW}$}   & \multirow{2}{*}{CP} & \multirow{2}{*}{$d_{\mathcal{O}_i}$}& SMEFT & $c$ Unitarity  \\
 & & & & Operator Form & Bound \\

\hline 
30 & \scriptsize{$ \partial^{ \sigma } Z^{ \mu } \partial_{ \mu } Z^{ \nu } \left( W^{ + \rho } \overleftrightarrow{ D }_{ \sigma } D_{ \nu } W^-_{ \rho } + \textrm{h.c.} \right) $} & $+$ & \multirow{25}{*}{8} & \multirow{10}{*}{ \scriptsize{$ D^8 H^4 $} } & \multirow{25}{*}{ $ \frac{3 \times 10^{-4}}{E_\text{TeV}^8} $ }  \\

31 & \scriptsize{$ \partial_{ \rho \sigma } Z^{ \mu } \partial_{ \mu } Z^{ \nu } \left( W^{ + \rho } \overleftrightarrow{ D }^{ \sigma } W^-_{ \nu } + \textrm{h.c.} \right) $} &$+$&  &  &  \\

32 & \scriptsize{$ \partial^{ \sigma } Z^{ \mu } \partial^{ \rho } Z^{ \nu } \left( W^{ + }_{ \mu } \overleftrightarrow{ D }_{ \sigma } D_{ \nu } W^-_{ \rho } + \textrm{h.c.} \right) $} &$+$&  &  &  \\

33 & \scriptsize{$ \partial^{ \rho \sigma } Z^{ \mu } Z^{ \nu } \left( W^{ + }_{ \mu } \overleftrightarrow{ D }_{ \sigma } D_{ \nu } W^-_{ \rho } + \textrm{h.c.} \right) $} &$+$&  &  &  \\

34 & \scriptsize{$ \partial^{ \nu \sigma } Z^{ \mu } \partial^{ \rho } Z_{ \mu } \left( W^{ + }_{ \nu } \overleftrightarrow{ D }_{ \sigma } W^-_{ \rho } + \textrm{h.c.} \right) $} &$+$&  &  &  \\

35 & \scriptsize{$ \partial_{ \nu \rho \sigma } Z^{ \mu } \partial_{ \mu } Z^{ \nu } \left( W^{ + \rho } W^{ -  \sigma } \right) $} &$+$&  &  &  \\

36 & \scriptsize{$ \partial_{ \nu \rho } Z^{ \mu } \partial_{ \mu \sigma } Z^{ \nu } \left( W^{ + \rho } W^{ -  \sigma } \right) $} &$+$&  &  &  \\

37 & \scriptsize{$ \partial_{ \sigma } Z^{ \mu } \partial_{ \mu \rho } Z^{ \nu } \left( W^{ + \rho } D_{ \nu } W^{ - \sigma } + \textrm{h.c.} \right) $} &$+$&  &  &  \\

38 & \scriptsize{$ \partial_{ \rho \sigma } Z^{ \mu } Z^{ \nu } \left( W^{ + \rho } D_{ \mu \nu } W^{ -  \sigma } + \textrm{h.c.} \right) $} &$+$&  &  &  \\

39 & \scriptsize{$ \partial_{ \sigma } Z^{ \mu } \partial_{ \rho } Z^{ \nu } \left( W^{ + \rho } D_{ \mu \nu } W^{ -  \sigma } + \textrm{h.c.} \right) $} &$+$&  &  &  \\

\cline{5-5}

40 & \scriptsize{$ \varepsilon^{ \mu \nu \rho \sigma } \partial_{ \mu \alpha } Z^{ \beta } Z_{ \nu } \left( i W^{ + }_{ \rho } \overleftrightarrow{ D }^{ \alpha } D_{ \beta } W^{ - } _{ \sigma } + \textrm{h.c.} \right) $} &$+$&  & \multirow{2}{*}{ \scriptsize{$ \varepsilon D^8 H^4 $} } &  \\

41 & \scriptsize{$ \varepsilon^{ \mu \nu \rho \sigma } \partial_{ \mu \beta } Z^{ \alpha } \partial_{ \alpha } Z_{ \nu } \left( i W^+_{ \rho } D_{ \sigma } W^{ -  \beta } + \textrm{h.c.} \right) $} &$+$&  &  &  \\

\cline{5-5}

42 & \scriptsize{$ \partial_{ \sigma } Z^{ \mu } \partial_{ \mu \rho } Z^{ \nu } \left( i W^{ + \rho } \overleftrightarrow{ D }^{ \sigma } W^{ - } _{ \nu } + \textrm{h.c.} \right) $} &$-$&  & \multirow{8}{*}{ \scriptsize{$ D^8 H^4 $} } &  \\

43 & \scriptsize{$ \partial^{ \sigma } Z^{ \mu } \partial^{ \rho } Z^{ \nu } \left( i W^{ + }_{ \mu } \overleftrightarrow{ D }_{ \sigma } D_{ \nu } W^{ - } _{ \rho } + \textrm{h.c.} \right) $} &$-$&  &  &  \\

44 & \scriptsize{$ \partial^{ \rho \sigma } Z^{ \mu } Z^{ \nu } \left( i W^{ + }_{ \mu } \overleftrightarrow{ D }_{ \sigma } D_{ \nu } W^{ - } _{ \rho } + \textrm{h.c.} \right) $} &$-$&  &  &  \\

45 & \scriptsize{$ \partial^{ \rho \sigma } Z^{ \mu } \partial_{ \mu } Z^{ \nu } \left( i W^{ + }_{ \nu } \overleftrightarrow{ D }_{ \sigma } W^{ - } _{ \rho } + \textrm{h.c.} \right) $} &$-$&  &  &  \\

46 & \scriptsize{$ \partial^{ \nu \rho \sigma } Z^{ \mu } Z_{ \mu } \left( i W^{ + }_{ \nu } \overleftrightarrow{ D }_{ \sigma } W^{ - } _{ \rho } + \textrm{h.c.} \right) $} &$-$&  &  &  \\

47 & \scriptsize{$ \partial^{ \rho \sigma } Z^{ \mu } \partial_{ \mu } Z^{ \nu } \left( i W^+_{ \rho } D_{ \nu } W^-_{ \sigma } + \textrm{h.c.} \right) $} &$-$&  &  &  \\

48 & \scriptsize{$ \partial^{ \sigma } Z^{ \mu } \partial_{ \mu \rho } Z^{ \nu } \left( i W^{ + \rho } D_{ \nu } W^-_{ \sigma } + \textrm{h.c.} \right) $} &$-$&  &  &  \\

49 & \scriptsize{$ Z^{ \mu } \partial_{ \mu \rho \sigma } Z^{ \nu } \left( i W^{ + \rho } D_{ \nu } W^{ - \sigma } + \textrm{h.c.} \right) $} &$-$&  &  &  \\

\cline{5-5}

50 & \scriptsize{$ \varepsilon^{ \mu \nu \rho \sigma } \partial_{ \mu \alpha } Z^{ \beta } \partial_{ \beta } Z_{ \nu } \left( W^+_{ \rho } \overleftrightarrow{ D }^{ \alpha } W^-_{ \sigma } \right) $} &$-$&  & \multirow{5}{*}{ \scriptsize{$ \varepsilon D^8 H^4 $} } &  \\

51 & \scriptsize{$ \varepsilon^{ \mu \nu \rho \sigma } \partial_{ \mu \alpha } Z^{ \beta } \partial_{ \nu } Z_{ \beta } \left( W^+_{ \rho } \overleftrightarrow{ D }^{ \alpha } W^-_{ \sigma }  \right) $} &$-$&  &  &  \\

52 & \scriptsize{$ \varepsilon^{ \mu \nu \rho \sigma } \partial_{ \alpha } Z_{ \mu } \partial_{ \nu } Z^{ \beta } \left( W^+_{ \rho } \overleftrightarrow{ D }^{ \alpha } D_{ \sigma } W^-_{ \beta }  + \textrm{h.c.} \right) $} &$-$&  &  &  \\

53 & \scriptsize{$ \varepsilon^{ \mu \nu \rho \sigma } \partial_{ \mu } Z^{ \alpha } \partial_{ \nu \alpha } Z^{ \beta } \left( W^{ + }_{ \rho } D_{ \beta } W^{ - } _{ \sigma } + \textrm{h.c.} \right) $} &$-$&  &  &  \\

54 & \scriptsize{$ \varepsilon^{ \mu \nu \rho \sigma } \partial_{ \mu \beta } Z^{ \alpha } Z_{ \nu } \left( W^{ + }_{ \rho } D_{ \sigma \alpha } W^{ -  \beta } + \textrm{h.c.} \right) $} &$-$&  &  &  \\

\hline 

55 & \scriptsize{$ \partial^{ \sigma } Z^{ \mu } \partial_{ \mu } Z^{ \nu \rho } \left( W^{ + }_{ \nu } \overleftrightarrow{ D }_{ \sigma } W^-_{ \rho } \right) $} &$+$& \multirow{16}{*}{8} & \scriptsize{$ D^6 H^4 W^{ a }_{ \mu \nu } $} & \multirow{16}{*}{ $ \frac{0.001}{E_\text{TeV}^7}, \frac{0.004}{E_\text{TeV}^8} $ }  \\

\cline{5-5}

56 & \scriptsize{$ \partial_{ \alpha \beta } \widetilde{ Z }^{ \mu \nu } Z_{ \mu } \left( i W^{ + \beta } \overleftrightarrow{ D }^{ \alpha } W^{ - } _{ \nu } + \textrm{h.c.} \right) $} &$+$&  & \multirow{14}{*}{ \scriptsize{$ D^6 H^4 \widetilde{ W }^{ a }_{ \mu \nu } $} } &  \\

57 & \scriptsize{$ \partial_{ \alpha } Z^{ \beta } \widetilde{ Z }^{ \mu \nu } \left( i W^{ + }_{ \mu } \overleftrightarrow{ D }^{ \alpha } D_{ \beta } W^{ - } _{ \nu } + \textrm{h.c.} \right) $} &$+$&  &  &  \\

58 & \scriptsize{$ \partial_{ \mu \alpha } Z^{ \beta } \widetilde{ Z }^{ \mu \nu } \left( i W^{ + }_{ \beta } \overleftrightarrow{ D }^{ \alpha } W^{ - } _{ \nu } + \textrm{h.c.} \right) $} &$+$&  &  &  \\

59 & \scriptsize{$ \partial_{ \mu \beta } Z^{ \alpha } \widetilde{ Z }^{ \mu \nu } \left( i W^+_{ \nu } D_{ \alpha } W^{ -  \beta } + \textrm{h.c.} \right) $} &$+$&  &  &  \\

60 & \scriptsize{$ \partial_{ \alpha } Z^{ \beta } \partial_{ \beta } Z_{ \mu } \left( i W^{ + }_{ \nu } \overleftrightarrow{ D }^{ \alpha } \widetilde{ W }^{ - \mu \nu } + \textrm{h.c.} \right) $} &$+$&  &  &  \\

61 & \scriptsize{$ \partial_{ \mu \alpha } Z^{ \beta } Z_{ \beta } \left( i W^{ + }_{ \nu } \overleftrightarrow{ D }^{ \alpha } \widetilde{ W }^{ - \mu \nu } + \textrm{h.c.} \right) $} &$+$&  &  &  \\

62 & \scriptsize{$ \partial_{ \mu } Z^{ \alpha } \partial_{ \alpha } Z^{ \beta } \left( i W^+_{ \nu } D_{ \beta } \widetilde{ W }^{ - \mu \nu } + \textrm{h.c.} \right) $} &$+$&  &  &  \\

63 & \scriptsize{$ Z^{ \alpha } \partial_{ \mu \alpha } Z^{ \beta } \left( i W^+_{ \nu } D_{ \beta } \widetilde{ W }^{ - \mu \nu } + \textrm{h.c.} \right) $} &$+$&  &  &  \\

64 & \scriptsize{$ \partial^{ \alpha } \widetilde{ Z }^{ \mu \nu } Z^{ \beta } \left( W^+_{ \mu } \overleftrightarrow{ D }_{ \alpha } D_{ \nu } W^-_{ \beta }  + \textrm{h.c.} \right) $} &$-$&  &  &  \\

65 & \scriptsize{$ \partial^{ \alpha } Z^{ \rho } \partial_{ \rho } \widetilde{ Z }^{ \mu \nu } \left( W^+_{ \mu } \overleftrightarrow{ D }_{ \alpha } W^-_{ \nu } \right) $} &$-$&  &  &  \\

66 & \scriptsize{$ \partial_{ \nu \rho } Z^{ \mu } \partial_{ \mu } \widetilde{ Z }^{ \nu \sigma } \left( W^{ + \rho } W^{ - } _{ \sigma } + \textrm{h.c.} \right) $} &$-$&  &  &  \\

67 & \scriptsize{$ \partial^{ \alpha } Z^{ \mu } Z^{ \beta } \left( W^{ + \nu } \overleftrightarrow{ D }_{ \alpha } D_{ \beta } \widetilde{ W }^{ - }_{ \mu \nu } + \textrm{h.c.}  \right) $} &$-$&  &  &  \\

68 & \scriptsize{$ \partial_{ \mu \beta } Z^{ \alpha } \partial_{ \alpha } Z^{ \beta } \left( W^{ + }_{ \nu } \widetilde{ W }^{ - \mu \nu } + \textrm{h.c.} \right) $} &$-$&  &  &  \\

69 & \scriptsize{$ \partial_{ \mu } Z^{ \alpha } Z^{ \beta } \left( W^{ + }_{ \nu } D_{ \alpha \beta } \widetilde{ W }^{ - \mu \nu } + \textrm{h.c.} \right) $} &$-$&  &  &  \\

\hline

\end{tabular}
\begin{minipage}{5.7in}
\medskip
\caption{\label{tab:zzww2} \footnotesize Primary dimension 8 operators for the $Z Z W^+ W^-$ interaction.}
\end{minipage}
\end{center}
\end{adjustwidth}
\end{table}

\begin{table}[p]
\begin{adjustwidth}{-.5in}{-.5in}  
\begin{center}
\footnotesize
\centering
\renewcommand{\arraystretch}{0.9}
\setlength{\tabcolsep}{6pt}
\begin{tabular}{|c|c|c|c|c|c|}

\hline
\multirow{2}{*}{$i$} & \multirow{2}{*}{$\mathcal{O}_i^{ZZWW}$}   & \multirow{2}{*}{CP} & \multirow{2}{*}{$d_{\mathcal{O}_i}$}& SMEFT & $c$ Unitarity  \\
 & & & & Operator Form & Bound \\

\hline 

70 & \scriptsize{$ \partial_{ \nu \rho \alpha } Z^{ \mu } \partial_{ \mu \sigma } Z^{ \nu } \left( W^{ + \rho } \overleftrightarrow{ D }^{ \alpha } W^{ - \sigma } + \textrm{h.c.} \right) $} & $+$ & \multirow{11}{*}{10} & \multirow{4}{*}{ \scriptsize{$ D^{ 10 } H^4 $} } & \multirow{11}{*}{ $ \frac{2 \times 10^{ -5 }}{E_\text{TeV}^{10}} $ }  \\

71 & \scriptsize{$ \partial_{ \nu \alpha } Z^{ \mu } \partial_{ \rho \sigma } Z^{ \nu } \left( i W^{ + \rho } \overleftrightarrow{ D }^{ \alpha } D_{ \mu } W^{ - \sigma } + \textrm{h.c.} \right) $} &$+$&  &  &  \\

72 & \scriptsize{$ \partial_{ \nu \rho \sigma \alpha } Z^{ \mu } Z^{ \nu } \left( W^{ + \rho } \overleftrightarrow{ D }^{ \alpha } D_{ \mu } W^{ - \sigma } + \textrm{h.c.} \right) $} &$+$&  &  &  \\

73 & \scriptsize{$ \partial_{ \rho \alpha } Z^{ \mu } \partial_{ \sigma } Z^{ \nu } \left( W^{ + \rho } \overleftrightarrow{ D }^{ \alpha } D_{ \mu \nu } W^{ - \sigma } + \textrm{h.c.} \right) $} &$+$&  &  &  \\

\cline{5-5}

74 & \scriptsize{$ \varepsilon^{ \mu \nu \rho \sigma } \partial_{ \mu \tau } Z^{ \alpha } \partial_{ \nu \alpha } Z^{ \beta } \left( i W^+_{ \rho } \overleftrightarrow{ D }^{ \tau } D_{ \beta } W^{ - } _{ \sigma } + \textrm{h.c.} \right) $} &$+$&  & \multirow{2}{*}{ \scriptsize{$ \varepsilon D^{ 10 } H^4 $} } &  \\

75 & \scriptsize{$ \varepsilon^{ \mu \nu \rho \sigma } \partial_{ \mu \beta \tau } Z^{ \alpha } Z_{ \nu } \left( i W^+_{ \rho } \overleftrightarrow{ D }^{ \tau } D_{ \sigma \alpha } W^{ - \beta } \right) $} &$+$&  &  &  \\

\cline{5-5}

76 & \scriptsize{$ \partial_{ \nu \rho \sigma \alpha } Z^{ \mu } \partial_{ \mu } Z^{ \nu } \left( i W^{ + \rho } \overleftrightarrow{ D }^{ \alpha } W^{ - \sigma } \right) $} &$-$&  & \multirow{3}{*}{ \scriptsize{$ D^{ 10 } H^4 $} } &  \\

77 & \scriptsize{$ \partial_{ \sigma \alpha } Z^{ \mu } \partial_{ \mu \rho } Z^{ \nu } \left( i W^{ + \rho } \overleftrightarrow{ D }^{ \alpha } D_{ \nu } W^{ - \sigma } + \textrm{h.c.} \right) $} &$-$&  &  &  \\

78 & \scriptsize{$ \partial_{ \rho \sigma \alpha } Z^{ \mu } Z^{ \nu } \left( i W^{ + \rho } \overleftrightarrow{ D }^{ \alpha } D_{ \mu \nu } W^{ - \sigma } + \textrm{h.c.} \right) $} &$-$&  &  &  \\

\cline{5-5}

79 & \scriptsize{$ \varepsilon^{ \mu \nu \rho \sigma } \partial_{ \mu \beta \tau } Z^{ \alpha } \partial_{ \nu \alpha } Z^{ \beta } \left( W^+_{ \rho } \overleftrightarrow{ D }^{ \tau } W^{ - } _{ \sigma } \right) $} &$-$&  & \multirow{2}{*}{ \scriptsize{$ \varepsilon D^{ 10 } H^4 $} } &  \\

80 & \scriptsize{$ \varepsilon^{ \mu \nu \rho \sigma } \partial_{ \mu \tau } Z^{ \alpha } \partial_{ \nu } Z^{ \beta } \left( W^+_{ \rho } \overleftrightarrow{ D }^{ \tau } D_{ \alpha \beta } W^{ - } _{ \sigma } + \textrm{h.c.} \right) $} &$-$&  &  &  \\

\hline

81 & \scriptsize{$ \partial_{ \nu \rho \alpha } Z^{ \mu } \partial_{ \mu } Z^{ \nu } \left( i W^+_{ \sigma } \overleftrightarrow{ D }^{ \alpha } \widetilde{ W }^{ - \rho \sigma } + \textrm{h.c.} \right) $} &$+$& \multirow{3}{*}{10} & \multirow{3}{*}{ \scriptsize{$ D^{ 8 } H^4 \widetilde{ W }^{ a }_{ \mu \nu } $} } & \multirow{3}{*}{ $ \frac{8 \times 10^{ -5 }}{E_\text{TeV}^9}, \frac{3 \times 10^{ -4 }}{E_\text{TeV}^{10}} $ } \\

82 & \scriptsize{$ \partial_{ \rho \alpha } Z^{ \mu } Z^{ \nu } \left( i W^+_{ \sigma } \overleftrightarrow{ D }^{ \alpha } D_{ \mu \nu } \widetilde{ W }^{ - \rho \sigma } + \textrm{h.c.} \right) $} &$+$&  &  &  \\

83 & \scriptsize{$ \partial_{ \nu \rho \alpha } Z^{ \mu } Z^{ \nu } \left( W^+_{ \sigma } \overleftrightarrow{ D }^{ \alpha } D_{ \mu } \widetilde{ W }^{ - \rho \sigma } + \textrm{h.c.} \right) $} &$-$&  &  &  \\

\hline

\end{tabular}
\begin{minipage}{5.7in}
\medskip
\caption{\label{tab:zzww3} \footnotesize Primary dimension 10 operators for the $Z Z W^+ W^-$ interaction. Two redundancies, which are descendants of dimension 8 operators, appear at dimension 12. To form an independent set of operators, $s^n (t-u)^{ 2 m } \mathcal{ O }^{ ZZWW }_{ 39 }$ and $s^n (t-u)^{ 2 m } \mathcal{ O }^{ ZZWW }_{ 54 }$, with $n \geq 2$, $m \geq 0$, should be omitted. }
\end{minipage}
\end{center}
\end{adjustwidth}
\end{table}

\begin{table}[p]
\begin{adjustwidth}{-.75in}{-.75in}  
\begin{center}
\scriptsize
\centering
\renewcommand{\arraystretch}{0.9}
\setlength{\tabcolsep}{6pt}
\begin{tabular}{|c|c|c|c|c|c|}

\hline
\multirow{2}{*}{$i$} & \multirow{2}{*}{$\mathcal{O}_i^{ZZZZ}$}   & \multirow{2}{*}{CP} & \multirow{2}{*}{$d_{\mathcal{O}_i}$}& SMEFT & $c$ Unitarity  \\
 & & & & Operator Form & Bound \\

\hline 
1 & \scriptsize{$ Z^{ \mu } Z_{ \mu } Z^{ \nu } Z_{ \nu } $} & $+$ & 4 & \scriptsize{$ D^4 H^4$} & $\frac{ 0.09 }{ E_\text{TeV}^4}$ \\

\hline
2 & \scriptsize{$ \partial^{ \rho } Z^{ \mu } Z^{ \nu } \left( Z_{ \mu } \overleftrightarrow{ \partial_{ \rho } } Z_{ \nu } \right) $} & $+$ & \multirow{3}{*}{6} & \multirow{3}{*}{ \scriptsize{$ D^6 H^4 $} } & \multirow{3}{*}{ $ \frac{0.006}{E_\text{TeV}^6} $ } \\

3 & \scriptsize{$ \partial_{ \nu } Z^{ \mu } \partial_{ \mu } Z^{ \nu } Z^{ \rho } Z_{ \rho } $} & $+$ &  & & \\

4 & \scriptsize{$ Z^{ \mu } Z^{ \nu } \partial_{ \mu \nu }  Z^{ \rho } Z_{ \rho } $} &$+$&  &  &  \\

\hline
5 & \scriptsize{$ \widetilde{ Z }^{ \mu \nu } Z^{ \rho } \partial_{ \rho } Z_{ \mu } Z_{ \nu } $} &$-$& 6 & \scriptsize{$ D^4 H^4 \widetilde{ W }^{ a }_{ \mu \nu } $} & $ \frac{0.02}{E_\text{TeV}^5} $, $ \frac{0.07}{E_\text{TeV}^6} $ \\

\hline 
6 & \scriptsize{$ \partial_{ \sigma } Z^{ \mu } \partial_{ \mu } Z^{ \rho } \left( \partial_{ \rho } Z^{ \nu } \overleftrightarrow{ \partial^{ \sigma } } Z_{ \nu } \right) $} & $+$ & \multirow{6}{*}{8} & \multirow{4}{*}{ \scriptsize{$ D^8 H^4 $} } & \multirow{6}{*}{ $ \frac{3 \times 10^{-4}}{E_\text{TeV}^8} $ }  \\

7 & \scriptsize{$ \partial_{ \nu \sigma } Z^{ \mu } \partial_{ \mu } Z^{ \rho } \left( Z^{ \nu } \overleftrightarrow{ \partial^{ \sigma } } Z_{ \rho } \right) $} & $+$ &  &  &  \\

8 & \scriptsize{$ \partial_{ \nu \rho } Z^{ \mu } \partial_{ \mu \sigma } Z^{ \nu } Z^{ \rho } Z^{ \sigma } $} & $+$ &  &  &  \\

9 & \scriptsize{$ \partial^{ \rho \sigma } Z^{ \mu } \partial_{ \sigma } Z^{ \nu } \left( Z_{ \mu } \overleftrightarrow{ \partial_{ \rho } } Z_{ \nu } \right) $} & $+$ &  &  &  \\

\cline{5-5}

10 & \scriptsize{$ \varepsilon^{ \mu \nu \rho \sigma } \partial_{ \alpha \sigma } Z^{ \beta } \partial_{ \beta } Z_{ \mu } \left( Z_{ \nu } \overleftrightarrow{ \partial^{ \alpha } } Z_{ \rho } \right) $} & $-$ &  & \multirow{2}{*}{ \scriptsize{$ \varepsilon D^8 H^4 $} } &  \\

11 & \scriptsize{$ \varepsilon^{ \mu \nu \rho \sigma } \partial_{ \alpha \rho } Z^{ \beta } \partial_{ \sigma } Z_{ \beta } \left( Z_{ \mu } \overleftrightarrow{ \partial^{ \alpha } } Z_{ \nu } \right) $} & $-$ &  &  &  \\ 

\hline

12 & \scriptsize{$ \partial^{ \sigma } Z^{ \mu } \partial_{ \mu } Z^{ \nu \rho } \left( Z_{ \nu } \overleftrightarrow{ \partial_{ \sigma } } Z_{ \rho } \right) $} & $+$ & \multirow{2}{*}{ 8 } & \scriptsize{$ D^6 H^4 W^{ a }_{ \mu \nu } $} & \multirow{2}{*}{ $ \frac{0.001}{E_\text{TeV}^7}, \frac{0.004}{E_\text{TeV}^8} $ } \\

\cline{5-5}

13 & \scriptsize{$ \partial_{ \mu \beta } Z^{ \alpha } \partial_{ \alpha } Z^{ \beta } \widetilde{ Z }^{ \mu \nu } Z_{ \nu } $} & $-$ & & \scriptsize{$ D^6 H^4 \widetilde{ W }^{ a }_{ \mu \nu } $} & \\

\hline 
14 & \scriptsize{$ \partial_{ \nu \rho \alpha } Z^{ \mu } \partial_{ \mu \sigma } Z^{ \nu } \left( Z^{ \rho } \overleftrightarrow{ \partial^{ \alpha } } Z^{ \sigma } \right) $} & $+$ & \multirow{10}{*}{10} & \multirow{5}{*}{ \scriptsize{$ D^{ 10 } H^4 $} } & \multirow{10}{*}{ $ \frac{2 \times 10^{ -5 }}{E_\text{TeV}^{10}} $ } \\

15 & \scriptsize{$ \partial_{ \rho \sigma \alpha } Z^{ \mu } \partial_{ \mu } Z^{ \nu } \left( \partial_{ \nu } Z^{ \rho } \overleftrightarrow{ \partial^{ \alpha } } Z^{ \sigma } \right) $} & $+$ &  & &  \\

16 & \scriptsize{$ \partial_{ \alpha } Z^{ \mu } \partial_{ \mu \rho \sigma } Z^{ \nu } \left( \partial_{ \nu } Z^{ \rho } \overleftrightarrow{ \partial^{ \alpha } } Z^{ \sigma } \right) $} & $+$ &  & &  \\

17 & \scriptsize{$ \partial^{ \alpha \beta } Z^{ \mu } \partial_{ \mu \alpha } Z^{ \nu } \left( \partial_{ \nu } Z^{ \rho } \overleftrightarrow{ \partial_{ \beta } } Z_{ \rho } \right) $} & $+$ &  & &  \\

18 & \scriptsize{$ \partial^{ \alpha \rho \sigma } Z^{ \mu } \partial_{ \mu \alpha } Z^{ \nu } \left( Z_{ \rho } \overleftrightarrow{ \partial_{ \sigma } } Z_{ \nu } \right) $} & $+$ &  & &  \\

\cline{5-5}

19 & \scriptsize{$ \varepsilon^{ \mu \nu \rho \sigma } \partial_{ \mu \beta \tau } Z^{ \alpha } \partial_{ \nu \alpha } Z^{ \beta } \left( Z_{ \rho } \overleftrightarrow{ \partial^{ \tau } } Z_{ \sigma } \right) $} & $-$ &  & \multirow{4}{*}{ \scriptsize{$ \varepsilon D^{ 10 } H^4 $} } & \\

20 & \scriptsize{$ \varepsilon^{ \mu \nu \rho \sigma } \partial_{ \mu \tau } Z^{ \alpha } \partial_{ \nu } Z^{ \beta } \left( \partial_{ \alpha \beta } Z_{ \rho } \overleftrightarrow{ \partial^{ \tau } } Z_{ \sigma } \right) $} & $-$ &  & &  \\

21 & \scriptsize{$ \varepsilon^{ \mu \nu \rho \sigma } \partial_{ \sigma }^{ \enspace \alpha \tau } Z^{ \beta } \partial_{ \beta \tau } Z_{ \mu } \left( Z_{ \nu } \overleftrightarrow{ \partial_{ \alpha } } Z_{ \rho } \right) $} & $-$ &  & &  \\

22 & \scriptsize{$ \varepsilon^{ \mu \nu \rho \sigma } \partial_{ \rho }^{ \enspace \alpha \tau } Z^{ \beta } \partial_{ \sigma \tau } Z_{ \beta } \left( Z_{ \mu } \overleftrightarrow{ \partial_{ \alpha } } Z_{ \nu } \right) $} & $-$ &  & &  \\

\hline
23 & \scriptsize{$ \partial^{ \alpha \sigma } Z^{ \mu } \partial_{ \mu \alpha } Z^{ \nu \rho } \left( Z_{ \rho } \overleftrightarrow{ \partial_{ \sigma } } Z_{ \nu } \right) $} & $+$ & \multirow{2}{*}{ 10 } & \scriptsize{$ D^8 H^4 W^{ a }_{ \mu \nu } $} & \multirow{2}{*}{ $ \frac{8 \times 10^{ -5 }}{E_\text{TeV}^9}, \frac{3 \times 10^{ -4 }}{E_\text{TeV}^{10}} $ } \\

\cline{5-5}

24 & \scriptsize{$ \partial_{ \tau } Z^{ \alpha } \partial_{ \mu \alpha } Z^{ \beta } \left( \partial_{ \beta } \widetilde{Z}^{ \mu \sigma } \overleftrightarrow{ \partial^{ \tau } } Z_{ \sigma } \right) $} & $-$ & & \scriptsize{$ D^8 H^4 \widetilde{ W }^{ a }_{ \mu \nu } $} & \\

\hline

25 & \scriptsize{$ \partial^{ \nu \rho \alpha \beta } Z^{ \mu } \partial_{ \mu \sigma \beta } Z_{ \nu } \left( Z_{ \rho } \overleftrightarrow{ \partial_{ \alpha } } Z^{ \sigma } \right) $} & $+$ & \multirow{6}{*}{12} & \multirow{3}{*}{ \scriptsize{$ D^{ 12 } H^4 $} } & \multirow{6}{*}{ $ \frac{1 \times 10^{ -6 }}{E_\text{TeV}^{12}} $ } \\

26 & \scriptsize{$ \partial^{ \rho \sigma \alpha \beta } Z^{ \mu } \partial_{ \mu \beta } Z^{ \nu } \left( \partial_{ \nu } Z_{ \rho } \overleftrightarrow{ \partial_{ \alpha } } Z_{ \sigma } \right) $} & $+$ &  &  &  \\

27 & \scriptsize{$ \partial^{ \alpha \beta } Z^{ \mu } \partial_{ \mu \rho \sigma \beta } Z^{ \nu } \left( \partial_{ \nu } Z^{ \rho } \overleftrightarrow{ \partial_{ \alpha } } Z^{ \sigma } \right) $} & $+$ &  & &  \\

\cline{5-5}

28 & \scriptsize{$ \varepsilon^{ \mu \nu \rho \sigma } \partial_{ \mu }^{ \enspace \beta \tau \pi } Z^{ \alpha } \partial_{ \nu \alpha \pi } Z_{ \beta } \left( Z_{ \rho } \overleftrightarrow{ \partial_{ \tau } } Z_{ \sigma } \right) $} & $-$ &  & \multirow{2}{*}{ \scriptsize{$ \varepsilon D^{ 12 } H^4 $} } &  \\

29 & \scriptsize{$ \varepsilon^{ \mu \nu \rho \sigma } \partial_{ \rho }^{ \enspace \alpha \tau \pi } Z^{ \beta } \partial_{ \sigma \tau \pi } Z_{ \beta } \left( Z_{ \mu } \overleftrightarrow{ \partial_{ \alpha } } Z_{ \nu } \right) $} & $-$ &  & &  \\

\hline
\end{tabular}
\begin{minipage}{5.7in}
\medskip
\caption{\label{tab:zzzz} \footnotesize Primary operators for the $ZZZZ$ interaction up to dimension 12. 
Two redundancies, which are descendants of dimension 10 operators, appear at dimension 14. To form an independent set of operators,  $x^n y^m \mathcal{ O }^{ ZZZZ }_{ 22 }$ and $x^n y^m \mathcal{ O }^{ ZZZZ }_{ 23 }$, where $x = s^2 + t^2 + u^2$ and $y = s t u$, with $n \geq 1$ and $m \geq 0$ should be removed.}
\end{minipage}
\end{center}
\end{adjustwidth}
\end{table}

\begin{table}[p]
\begin{adjustwidth}{-.5in}{-.5in}  
\begin{center}
\footnotesize
\centering
\renewcommand{\arraystretch}{0.9}
\setlength{\tabcolsep}{6pt}
\begin{tabular}{|c|c|c|c|c|c|}

\hline
\multirow{2}{*}{$i$} & \multirow{2}{*}{$\mathcal{O}_i^{ W W Z \gamma }$}   & \multirow{2}{*}{CP} & \multirow{2}{*}{$d_{\mathcal{O}_i}$}& SMEFT & $c$ Unitarity  \\
 & & & & Operator Form & Bound \\

\hline 

1 & \scriptsize{$ \left( W^{ + \mu } D_{ \nu } W^{ - }_{ \mu } + \textrm{h.c.} \right) F^{ \nu \rho } Z_{ \rho } $} & $+$ & \multirow{18}{*}{6} & \multirow{5}{*}{ \scriptsize{$ D^{ 4 } H^4 B^{ \mu \nu } $} } & \multirow{18}{*}{ $ \frac{0.02}{E_\text{TeV}^5} $, $ \frac{0.07}{E_\text{TeV}^6} $ } \\

2 & \scriptsize{$ \left( W^{ + }_{ \mu } D_{ \rho } W^{ - }_{ \nu } + \textrm{h.c.} \right) F^{ \rho \mu } Z^{ \nu } $} & $+$ &  &  &  \\

3 & \scriptsize{$ \left( W^{ + \mu } D_{ \mu } W^{ - }_{ \nu } + \textrm{h.c.} \right) F^{ \nu \rho } Z_{ \rho } $} & $+$ &  &  &  \\

4 & \scriptsize{$ \left( W^{ + \mu } W^{ - \nu } + \textrm{h.c.} \right) \partial_{ \nu } F_{ \mu \rho } Z^{ \rho } $} & $+$ &  &  &  \\

5 & \scriptsize{$ \left( W^{ + }_{ \mu } D_{ \rho } W^{ - }_{ \nu } + \textrm{h.c.} \right) F^{ \mu \nu } Z^{ \rho } $} & $+$ &  &  &  \\

\cline{5-5}

6 & \scriptsize{$ \left( i W^{ + \mu } D_{ \mu } W^{ - }_{ \nu } + \textrm{h.c.} \right) \widetilde{ F }^{ \nu \rho } Z_{ \rho } $} & $+$ &  & \multirow{2}{*}{ \scriptsize{$ D^{ 4 } H^4 \widetilde{ B }^{ \mu \nu } $} } &  \\

7 & \scriptsize{$ \left( i W^{ + \mu } D_{ \nu } W^{ - }_{ \mu } + \textrm{h.c.} \right) \widetilde{ F }_{ \nu \rho } Z_{ \rho } $} & $+$ &  &  &  \\

\cline{5-5}

8 & \scriptsize{$ \varepsilon^{ \mu \nu \rho \sigma } \left( i W^{ + }_{ \mu } D^{ \alpha } W^{ - }_{ \nu } + \textrm{h.c.} \right) F_{ \rho \alpha } Z_{ \sigma } $} & $+$ &  & \multirow{2}{*}{ \scriptsize{$ \varepsilon D^{ 4 } H^4 B^{ \mu \nu } $} } &  \\

9 & \scriptsize{$ \varepsilon^{ \mu \nu \rho \sigma } \left( i W^{ + }_{ \mu } D_{ \nu } W^{ - \alpha } + \textrm{h.c.} \right) F_{ \rho \alpha } Z_{ \sigma } $} & $+$ &  &  &  \\

\cline{5-5}

10 & \scriptsize{$ \left( i W^{ + \mu } \overleftrightarrow{ D }^{ \rho } W^{ - }_{ \mu } \right) F_{ \rho \sigma } Z^{ \sigma } $} & $-$ &  & \multirow{4}{*}{ \scriptsize{$ D^{ 4 } H^4 B^{ \mu \nu } $} } &  \\

11 & \scriptsize{$ \left( i W^{ + \mu } D^{ \rho } W^{ - \nu } + \textrm{h.c.} \right) F_{ \mu \rho } Z_{ \nu } $} & $-$ &  &  &  \\

12 & \scriptsize{$ \left( i W^{ + \mu } W^{ - \nu } + \textrm{h.c.} \right) \partial_{ \nu } F_{ \mu \rho } Z^{ \rho } $} & $-$ &  &  &  \\

13 & \scriptsize{$ \left( i W^{ + \mu } D_{ \mu } W^{ - \nu } + \textrm{h.c.} \right) F_{ \nu \rho } Z^{ \rho } $} & $-$ &  &  &  \\

\cline{5-5}

14 & \scriptsize{$ \left( W^{ + \mu } D_{ \mu } W^{ - }_{ \nu } + \textrm{h.c.} \right) \widetilde{ F }^{ \nu \rho } Z_{ \rho } $} & $-$ &  & \multirow{3}{*}{ \scriptsize{$ D^{ 4 } H^4 \widetilde{ B }^{ \mu \nu } $} } &  \\

15 & \scriptsize{$ \left( W^{ + \mu } W^{ - }_{ \nu } + \textrm{h.c.} \right) \partial_{ \mu } \widetilde{ F }^{ \nu \rho } Z_{ \rho } $} & $-$ &  &  &  \\

16 & \scriptsize{$ \left( W^{ + \mu } D_{ \nu } W^{ - }_{ \mu } + \textrm{h.c.} \right) \widetilde{ F }^{ \nu \rho } Z_{ \rho } $} & $-$ &  &  &  \\

\cline{5-5}

17 & \scriptsize{$ \varepsilon^{ \mu \nu \rho \sigma } \left( W^{ + }_{ \mu } D^{ \alpha } W^{ - }_{ \nu } + \textrm{h.c.} \right) F_{ \rho \alpha } Z_{ \sigma } $} & $-$ &  & \multirow{2}{*}{ \scriptsize{$ \varepsilon D^{ 4 } H^4 B^{ \mu \nu } $} } &  \\

18 & \scriptsize{$ \varepsilon^{ \mu \nu \rho \sigma } \left( W^{ + }_{ \mu } D_{ \nu } W^{ - \alpha } + \textrm{h.c.} \right) F_{ \rho \alpha } Z_{ \sigma } $} & $-$ &  &  &  \\

\hline

19 & \scriptsize{$ \left( W^{ + }_{ \mu } W^{ - }_{ \nu \rho } + \textrm{h.c.} \right) F^{ \nu \rho } Z^{ \mu } $} & $+$ & \multirow{4}{*}{6} & \multirow{1}{*}{ \scriptsize{$ D^{ 2 } H^2 W^a_{ \mu \nu } B^{ \mu \nu } $} } & \multirow{4}{*}{ $ \frac{0.09}{E_\text{TeV}^4} $ } \\

\cline{5-5}

20 & \scriptsize{$ \left( i W^{ + \mu } \widetilde{ W }^{ - \nu \rho } + \textrm{h.c.} \right) F_{ \nu \mu } Z_{ \rho } $} & $+$ &  & \multirow{1}{*}{ \scriptsize{$ D^{ 2 } H^2 \widetilde{ W }^a_{ \mu \nu } B^{ \mu \nu } $} } &  \\

\cline{5-5}

21 & \scriptsize{$ \left( i W^{ + \mu } W^{ - \nu \rho } + \textrm{h.c.} \right) F_{ \nu \rho } Z_{ \mu } $} & $-$ &  & \multirow{1}{*}{ \scriptsize{$ D^{ 2 } H^2 W^a_{ \mu \nu } B^{ \mu \nu } $} } &  \\

\cline{5-5}

22 & \scriptsize{$ \left( W^{ + \mu } \widetilde{ W }^{ - \nu \rho } + \textrm{h.c.} \right) F_{ \nu \mu } Z_{ \rho } $} & $-$ &  & \multirow{1}{*}{ \scriptsize{$ D^{ 2 } H^2 \widetilde{ W }^a_{ \mu \nu } B^{ \mu \nu } $} } &  \\

\hline

\end{tabular}
\begin{minipage}{5.7in}
\medskip
\caption{\label{tab:wwzgam1} \footnotesize Primary dimension 6 operators for the $W^+ W^- Z \gamma$ interaction. }
\end{minipage}
\end{center}
\end{adjustwidth}
\end{table}

\begin{table}[p]
\begin{adjustwidth}{-.5in}{-.5in}  
\begin{center}
\scriptsize
\centering
\renewcommand{\arraystretch}{0.9}
\setlength{\tabcolsep}{6pt}
\begin{tabular}{|c|c|c|c|c|c|}

\hline
\multirow{2}{*}{$i$} & \multirow{2}{*}{$\mathcal{O}_i^{ W W Z \gamma }$}   & \multirow{2}{*}{CP} & \multirow{2}{*}{$d_{\mathcal{O}_i}$}& SMEFT & $c$ Unitarity  \\
 & & & & Operator Form & Bound \\

\hline 

23 & \scriptsize{$ \left( D_{ \nu } W^{ + \mu } D_{ \mu \rho } W^{ - \nu } + \textrm{h.c.} \right) F^{ \rho \sigma } Z_{ \sigma } $} & $+$ & \multirow{23}{*}{8} & \multirow{9}{*}{ \scriptsize{$ D^{ 6 } H^4 B^{ \mu \nu } $} } & \multirow{23}{*}{ $ \frac{0.001}{E_\text{TeV}^7}, \frac{0.004}{E_\text{TeV}^8} $ } \\

24 & \scriptsize{$ \left( D^{ \rho } W^{ + \mu } D_{ \mu } W^{ - \nu } + \textrm{h.c.} \right) \partial_{ \nu } F_{ \rho \sigma } Z^{ \sigma } $} & $+$ &  &  &  \\

25 & \scriptsize{$ \left( W^{ + }_{ \mu } D^{ \mu \rho } W^{ - \nu } + \textrm{h.c.} \right) \partial_{ \nu } F_{ \rho \sigma } Z^{ \sigma } $} & $+$ &  &  &  \\

26 & \scriptsize{$ \left( W^{ + \mu } D^{ \rho } W^{ - \nu } + \textrm{h.c.} \right) \partial_{ \mu \nu } F_{ \rho \sigma } Z^{ \sigma } $} & $+$ &  &  &  \\

27 & \scriptsize{$ \left( D_{ \sigma } W^{ + \mu } D_{ \mu \rho } W^{ - \nu } + \textrm{h.c.} \right) F_{ \rho \sigma } Z_{ \nu } $} & $+$ &  &  &  \\

28 & \scriptsize{$ \left( D_{ \rho \sigma } W^{ + \mu } D_{ \mu } W^{ - }_{ \nu } + \textrm{h.c.} \right) F^{ \nu \rho } Z^{ \sigma } $} & $+$ &  &  &  \\

29 & \scriptsize{$ \left( W^{ + \mu } D^{ \rho \sigma } W^{ - \nu } + \textrm{h.c.} \right) \partial_{ \nu } F_{ \mu \rho } Z_{ \sigma } $} & $+$ &  &  &  \\

30 & \scriptsize{$ \left( D_{ \nu \sigma } W^{ + \mu } D_{ \rho } W^{ - }_{ \mu } + \textrm{h.c.} \right) F^{ \nu \rho } Z^{ \sigma } $} & $+$ &  &  &  \\

\cline{5-5}

31 & \scriptsize{$ \varepsilon^{ \mu \nu \rho \sigma } \left( i D_{ \mu }^{ \enspace \alpha } W^{ + \beta } D_{ \beta } W^{ - }_{ \nu } + \textrm{h.c.} \right) F_{ \rho \alpha } Z_{ \sigma } $} & $+$ &  & \multirow{3}{*}{ \scriptsize{$ \varepsilon D^{ 6 } H^4 B^{ \mu \nu } $} } &  \\

32 & \scriptsize{$ \varepsilon^{ \mu \nu \rho \sigma } \left( i D_{ \mu } W^{ + }_{ \beta } D^{ \alpha \beta } W^{ - }_{ \nu } + \textrm{h.c.} \right) F_{ \rho \alpha } Z_{ \sigma } $} & $+$ &  &  &  \\

33 & \scriptsize{$ \varepsilon^{ \mu \nu \rho \sigma } \left( i W^{ + }_{ \mu } D_{ \nu }^{ \enspace \alpha } W^{ - \beta } + \textrm{h.c.} \right) \partial_{ \beta } F_{ \rho \alpha } Z_{ \sigma } $} & $+$ &  &  &  \\

\cline{5-5}

34 & \scriptsize{$ \left( i D_{ \nu } W^{ + \mu } D_{ \mu \rho } W^{ - \nu } + \textrm{h.c.} \right) F^{ \rho \sigma } Z_{ \sigma } $} & $-$ &  & \multirow{8}{*}{ \scriptsize{$ D^{ 6 } H^4 B^{ \mu \nu } $} } &  \\

35 & \scriptsize{$ \left( i D^{ \rho } W^{ + \mu } D_{ \mu } W^{ - \nu } + \textrm{h.c.} \right) \partial_{ \nu } F_{ \rho \sigma } Z^{ \sigma } $} & $-$ &  &  &  \\

36 & \scriptsize{$ \left( i W^{ + }_{ \mu } D^{ \mu \rho } W^{ - \nu } + \textrm{h.c.} \right) \partial_{ \nu } F_{ \rho \sigma } Z^{ \sigma } $} & $-$ &  &  &  \\

37 & \scriptsize{$ \left( i W^{ + \mu } D^{ \rho } W^{ - \nu } + \textrm{h.c.} \right) \partial_{ \mu \nu } F_{ \rho \sigma } Z^{ \sigma } $} & $-$ &  &  &  \\

38 & \scriptsize{$ \left( i D_{ \sigma } W^{ + \mu } D_{ \mu \rho } W^{ - \nu } + \textrm{h.c.} \right) F^{ \rho \sigma } Z_{ \nu } $} & $-$ &  &  &  \\

39 & \scriptsize{$ \left( i D_{ \rho \sigma } W^{ + \mu } D_{ \mu } W^{ - }_{ \nu } + \textrm{h.c.} \right) F^{ \nu \rho } Z^{ \sigma } $} & $-$ &  &  &  \\

40 & \scriptsize{$ \left( i W^{ + \mu } D^{ \rho \sigma } W^{ - \nu } + \textrm{h.c.} \right) \partial_{ \nu } F_{ \mu \rho } Z_{ \sigma } $} & $-$ &  &  &  \\

\cline{5-5}

41 & \scriptsize{$ \varepsilon^{ \mu \nu \rho \sigma } \left( D_{ \mu }^{ \enspace \alpha } W^{ + \beta } D_{ \beta } W^{ - }_{ \nu } + \textrm{h.c.} \right) F_{ \rho \alpha } Z_{ \sigma } $} & $-$ &  & \multirow{3}{*}{ \scriptsize{$ \varepsilon D^{ 6 } H^4 B^{ \mu \nu } $} } &  \\

42 & \scriptsize{$ \varepsilon^{ \mu \nu \rho \sigma } \left( D_{ \mu } W^{ + }_{ \beta } D^{ \alpha \beta } W^{ - }_{ \nu } + \textrm{h.c.} \right) F_{ \rho \alpha } Z_{ \sigma } $} & $-$ &  &  &  \\

43 & \scriptsize{$ \varepsilon^{ \mu \nu \rho \sigma } \left( W^{ + }_{ \mu } D_{ \nu }^{ \enspace \alpha } W^{ - \beta } + \textrm{h.c.} \right) \partial_{ \beta } F_{ \rho \alpha } Z_{ \sigma } $} & $-$ &  &  &  \\

\hline

44 & \scriptsize{$ \left( D_{ \sigma } W^{ + \mu } D_{ \mu } W^{ - }_{ \nu \rho } + \textrm{h.c.} \right) F^{ \nu \rho } Z^{ \sigma } $} & $+$ & \multirow{13}{*}{8} & \multirow{1}{*}{ \scriptsize{$ D^4 H^2 W^{ a }_{ \mu \nu } B^{ \mu \nu } $} } & \multirow{13}{*}{ $ \frac{0.006}{E_\text{TeV}^6} $ } \\

\cline{5-5}

45 & \scriptsize{$ \left( i D^{ \mu } W^{ + \nu } D_{ \nu } \widetilde{ W }^{ - \rho \sigma } + \textrm{h.c.} \right) F_{ \rho \mu } Z_{ \sigma } $} & $+$ &  & \multirow{4}{*}{ \scriptsize{$ D^4 H^2 \widetilde{ W }^{ a }_{ \mu \nu } B^{ \mu \nu } $} } &  \\

46 & \scriptsize{$ \left( i W^{ + }_{ \nu } D^{ \nu \mu } \widetilde{ W }^{ - \rho \sigma } + \textrm{h.c.} \right) F_{ \rho \nu } Z_{ \sigma } $} & $+$ &  &  &  \\

47 & \scriptsize{$ \left( i D^{ \rho } W^{ + \mu } \widetilde{ W }^{ - \nu \sigma } + \textrm{h.c.} \right) \partial_{ \mu } F_{ \nu \rho } Z_{ \sigma } $} & $+$ &  &  &  \\

48 & \scriptsize{$ \left( i W^{ + \mu } D^{ \rho } \widetilde{ W }^{ - \nu \sigma } + \textrm{h.c.} \right) \partial_{ \mu } F_{ \nu \rho } Z_{ \sigma } $} & $+$ &  &  &  \\

\cline{5-5}

49 & \scriptsize{$ \left( i D_{ \sigma } W^{ + \mu } D_{ \mu } W^{ - }_{ \nu \rho } + \textrm{h.c.} \right) F^{ \nu \rho } Z^{ \sigma } $} & $-$ &  & \multirow{1}{*}{ \scriptsize{$ D^4 H^2 W^{ a }_{ \mu \nu } B^{ \mu \nu } $} } &  \\

\cline{5-5}

50 & \scriptsize{$ \left( D^{ \mu } W^{ + \nu } D_{ \nu } \widetilde{ W }^{ - \rho \sigma } + \textrm{h.c.} \right) F_{ \rho \mu } Z_{ \sigma } $} & $-$ &  & \multirow{5}{*}{ \scriptsize{$ D^4 H^2 \widetilde{ W }^{ a }_{ \mu \nu } B^{ \mu \nu } $} } &  \\

51 & \scriptsize{$ \left( W^{ + }_{ \nu } D^{ \nu \mu } \widetilde{ W }^{ - \rho \sigma } + \textrm{h.c.} \right) F_{ \rho \nu } Z_{ \sigma } $} & $-$ &  &  &  \\

52 & \scriptsize{$ \left( W^{ + \mu } D^{ \rho } \widetilde{ W }^{ - \nu \sigma } + \textrm{h.c.} \right) \partial_{ \mu } F_{ \nu \rho } Z_{ \sigma } $} & $-$ &  &  &  \\

53 & \scriptsize{$ \left( D^{ \rho } W^{ + \mu } \widetilde{ W }^{ - \nu \sigma } + \textrm{h.c.} \right) \partial_{ \mu } F_{ \nu \rho } Z_{ \sigma } $} & $-$ &  &  &  \\

54 & \scriptsize{$ \left( D_{ \nu } W^{ + }_{ \mu } D_{ \rho } \widetilde{ W }^{ - \nu \sigma } + \textrm{h.c.} \right) F^{ \mu \rho } Z_{ \sigma } $} & $-$ &  &  &  \\
  
\hline

55 & \scriptsize{$ \left( i D_{ \nu } W^{ + }_{ \mu \rho } \widetilde{ W }^{ - \nu \sigma } + \textrm{h.c.} \right) F^{ \mu \rho } Z_{ \sigma } $} & $+$ & \multirow{2}{*}{8} & \multirow{2}{*}{ \scriptsize{$ D^2 H^2 W^{ a }_{ \mu \nu } \widetilde{ W }^{ a }_{ \mu \nu } B^{ \mu \nu } $} } & \multirow{2}{*}{ $ \frac{0.02}{E_\text{TeV}^5}, \frac{0.07}{E_\text{TeV}^6} $ } \\

56 & \scriptsize{$ \left( D_{ \nu } W^{ + }_{ \mu \rho } \widetilde{ W }^{ - \nu \sigma } + \textrm{h.c.} \right) F^{ \mu \rho } Z_{ \sigma } $} & $-$ &  &  &  \\

\hline

57 & \scriptsize{$ \left( D_{ \nu \rho \sigma } W^{ + \mu } D_{ \mu \alpha } W^{ - \nu } + \textrm{h.c.} \right) F^{ \rho \alpha } Z^{ \sigma } $} & $+$ & \multirow{1}{*}{10} & \multirow{1}{*}{ \scriptsize{$ D^8 H^4 B^{ \mu \nu } $} } & \multirow{1}{*}{ $ \frac{8 \times 10^{ -5 }}{E_\text{TeV}^9}, \frac{3 \times 10^{ -4 }}{E_\text{TeV}^{10}} $ } \\[1.1 ex]

\hline

58 & \scriptsize{$ \left( D_{ \nu \rho } W^{ + \mu } D_{ \mu \alpha } \widetilde{ W }^{ - \nu \sigma } + \textrm{h.c.} \right) F^{ \rho \alpha } Z_{ \sigma } $} & $-$ & \multirow{1}{*}{10} & \multirow{1}{*}{ \scriptsize{$ D^6 H^2 \widetilde{ W }^{ a }_{ \mu \nu } B^{ \mu \nu } $} } & \multirow{1}{*}{ $ \frac{3 \times 10^{ -4 }}{E_\text{TeV}^8} $ } \\[1.1 ex]

\hline

\end{tabular}
\begin{minipage}{5.7in}
\medskip
\caption{\label{tab:wwzgam2} \footnotesize Primary dimension 8 and 10 operators for the $W^+ W^- Z \gamma$ interaction. At dimension 10 there are 4 redundancies, which are descendants of dimension 8 operators. In order to form a set of independent operators, $s^n t^m \mathcal{ O }^{ WW\gamma Z }_{ 44 }$, $s^n t^m \mathcal{ O }^{ WW\gamma Z }_{ 49 }$, $s^n t^m \mathcal{ O }^{ WW\gamma Z }_{ 54 }$, and $s^n t^m \mathcal{ O }^{ WW\gamma Z }_{ 55 }$, with $n \geq 1$, $m \geq 0$, should be omitted. }
\end{minipage}
\end{center}
\end{adjustwidth}
\end{table}

\begin{table}[p]
\begin{adjustwidth}{-.5in}{-.5in}  
\begin{center}
\scriptsize
\centering
\renewcommand{\arraystretch}{0.9}
\setlength{\tabcolsep}{6pt}
\begin{tabular}{|c|c|c|c|c|c|}

\hline
\multirow{2}{*}{$i$} & \multirow{2}{*}{$\mathcal{O}_i^{ Z Z Z \gamma }$}   & \multirow{2}{*}{CP} & \multirow{2}{*}{$d_{\mathcal{O}_i}$}& SMEFT & $c$ Unitarity  \\
 & & & & Operator Form & Bound \\

\hline 

1 & \scriptsize{$ Z^{ \mu } \partial_{ \nu } Z_{ \mu } Z_{ \rho } F^{ \nu \rho } $} & $+$ & \multirow{3}{*}{6} & \multirow{2}{*}{ \scriptsize{$ D^{ 4 } H^4 B^{ \mu \nu } $} } & \multirow{3}{*}{ $ \frac{0.02}{E_\text{TeV}^5} $, $ \frac{0.07}{E_\text{TeV}^6} $ } \\

2 & \scriptsize{$ Z^{ \mu } \partial_{ \mu } Z_{ \nu } Z_{ \rho } F^{ \nu \rho } $} & $+$ &  &  &  \\

\cline{5-5}

3 & \scriptsize{$ Z^{ \mu } \partial_{ \mu } Z_{ \nu } Z_{ \rho } \widetilde{ F }^{ \nu \rho } $} & $-$ &  & \multirow{1}{*}{ \scriptsize{$ D^{ 4 } H^4 \widetilde{ B }^{ \mu \nu } $} } &  \\

\hline

4 & \scriptsize{$ Z^{ \mu } \widetilde{ Z }^{ \nu \rho } Z_{ \rho } F_{ \nu \mu } $} & $-$ & \multirow{1}{*}{6} & \multirow{1}{*}{ \scriptsize{$ D^{ 2 } H^2 \widetilde{ W }^a_{ \mu \nu } B^{ \mu \nu } $} } & \multirow{1}{*}{ $ \frac{0.09}{E_\text{TeV}^4} $ } \\[1.1ex]

\hline

5 & \scriptsize{$ \partial_{ \nu } Z^{ \mu } \partial_{ \mu \rho } Z^{ \nu } Z_{ \sigma } F^{ \rho \sigma } $} & $+$ & \multirow{10}{*}{8} & \multirow{7}{*}{ \scriptsize{$ D^{ 6 } H^4 B^{ \mu \nu } $} } & \multirow{10}{*}{ $ \frac{0.001}{E_\text{TeV}^7}, \frac{0.004}{E_\text{TeV}^8} $ } \\

6 & \scriptsize{$ \partial^{ \rho } Z^{ \mu } \partial_{ \mu } Z^{ \nu } Z^{ \sigma } \partial_{ \nu } F_{ \rho \sigma } $} & $+$ &  &  &  \\

7 & \scriptsize{$ Z_{ \mu } \partial^{ \mu \rho } Z^{ \nu } Z^{ \sigma } \partial_{ \nu } F_{ \rho \sigma } $} & $+$ &  &  &  \\

8 & \scriptsize{$ Z^{ \mu } \partial^{ \rho } Z^{ \nu } Z^{ \sigma } \partial_{ \mu \nu } F_{ \rho \sigma } $} & $+$ &  &  &  \\

9 & \scriptsize{$ \partial_{ \sigma } Z^{ \mu } \partial_{ \mu \rho } Z^{ \nu } Z_{ \nu } F^{ \rho \sigma } $} & $+$ &  &  &  \\

10 & \scriptsize{$ \partial^{ \sigma } Z^{ \mu } \partial_{ \nu \sigma } Z_{ \mu } Z_{ \rho } F^{ \nu \rho } $} & $+$ &  &  &  \\

11 & \scriptsize{$ \partial^{ \sigma } Z_{ \mu } \partial_{ \rho \sigma } Z_{ \nu } Z^{ \nu } F^{ \rho \mu } $} & $+$ &  &  &  \\

\cline{5-5}

12 & \scriptsize{$ \partial^{ \sigma } Z^{ \mu } \partial_{ \nu \sigma } Z_{ \mu } Z_{ \rho } \widetilde{ F }^{ \nu \rho } $} & $-$ &  & \multirow{1}{*}{ \scriptsize{$ D^{ 6 } H^4 \widetilde{ B }^{ \mu \nu } $} } &  \\

\cline{5-5}

13 & \scriptsize{$ \varepsilon^{ \mu \nu \rho \sigma } \partial_{ \mu }^{ \enspace \alpha } Z^{ \beta } \partial_{ \beta } Z_{ \nu } Z_{ \sigma } F_{ \rho \alpha } $} & $-$ &  & \multirow{2}{*}{ \scriptsize{$ \varepsilon D^{ 6 } H^4 B^{ \mu \nu } $} } &  \\

14 & \scriptsize{$ \varepsilon^{ \mu \nu \rho \sigma } \partial_{ \mu } Z_{ \beta } \partial^{ \alpha \beta } Z_{ \nu } Z_{ \sigma } F_{ \rho \alpha } $} & $-$ &  &  &  \\

\hline

15 & \scriptsize{$ \partial^{ \mu } Z^{ \nu } \partial_{ \nu } \widetilde{ Z }^{ \rho \sigma } Z_{ \sigma } F_{ \rho \mu } $} & $-$ & \multirow{3}{*}{8} & \multirow{3}{*}{ \scriptsize{$ D^4 H^2 \widetilde{ W }^{ a }_{ \mu \nu } B^{ \mu \nu } $} } & \multirow{3}{*}{ $ \frac{0.006}{E_\text{TeV}^6} $ } \\

16 & \scriptsize{$ Z_{ \mu } \partial^{ \mu \nu } \widetilde{ Z }^{ \rho \sigma } Z_{ \sigma } F_{ \rho \nu } $} & $-$ &  &  &  \\

17 & \scriptsize{$ \partial_{ \nu } Z_{ \mu } \partial_{ \rho } \widetilde{ Z }^{ \nu \sigma } Z_{ \sigma } F^{ \mu \rho } $} & $-$ &  &  &  \\

\hline

18 & \scriptsize{$ \partial_{ \nu } Z_{ \mu \rho } \widetilde{ Z }^{ \nu \sigma } Z_{ \sigma } F^{ \mu \rho } $} & $-$ & \multirow{1}{*}{8} & \multirow{1}{*}{ \scriptsize{$ D^2 H^2 W^{ a }_{ \mu \nu } \widetilde{ W }^{ a }_{ \mu \nu } B^{ \mu \nu } $} } & \multirow{1}{*}{ $ \frac{0.02}{E_\text{TeV}^5}, \frac{0.07}{E_\text{TeV}^6} $ } \\[1.1ex]

\hline

19 & \scriptsize{$ \partial_{ \nu \rho \sigma } Z^{ \mu } \partial_{ \mu \alpha } Z^{ \nu } Z^{ \sigma } F^{ \rho \alpha } $} & $+$ & \multirow{14}{*}{10} & \multirow{10}{*}{ \scriptsize{$ D^8 H^4 B^{ \mu \nu } $} } & \multirow{14}{*}{ $ \frac{8 \times 10^{ -5 }}{E_\text{TeV}^9}, \frac{3 \times 10^{ -4 }}{E_\text{TeV}^{10}} $ } \\

20 & \scriptsize{$ \partial_{ \nu }^{ \enspace \alpha } Z^{ \mu } \partial_{ \mu \rho \alpha } Z^{ \nu } Z_{ \sigma } F^{ \rho \sigma } $} & $+$ &  &  &  \\

21 & \scriptsize{$ \partial^{ \rho \alpha } Z^{ \mu } \partial_{ \mu \alpha } Z^{ \nu } Z^{ \sigma } \partial_{ \nu } F_{ \rho \sigma } $} & $+$ &  &  &  \\

22 & \scriptsize{$ \partial_{ \alpha } Z_{ \mu } \partial^{ \mu \rho \alpha } Z^{ \nu } Z^{ \sigma } \partial_{ \nu } F_{ \rho \sigma } $} & $+$ &  &  &  \\

23 & \scriptsize{$ \partial_{ \alpha } Z^{ \mu } \partial^{ \rho \alpha } Z^{ \nu } Z^{ \sigma } \partial_{ \mu \nu } F_{ \rho \sigma } $} & $+$ &  &  &  \\

24 & \scriptsize{$ \partial_{ \sigma }^{ \enspace \alpha } Z^{ \mu } \partial_{ \mu \rho \alpha } Z^{ \nu } Z_{ \nu } F^{ \rho \sigma } $} & $+$ &  &  &  \\

25 & \scriptsize{$ \partial_{ \rho \sigma \alpha } Z^{ \mu } \partial_{ \mu }^{ \enspace \alpha } Z_{ \nu } Z^{ \sigma } F^{ \nu \rho } $} & $+$ &  &  &  \\

26 & \scriptsize{$ \partial_{ \alpha } Z^{ \mu } \partial^{ \rho \sigma \alpha } Z^{ \nu } Z_{ \sigma } \partial_{ \nu } F_{ \mu \rho } $} & $+$ &  &  &  \\

27 & \scriptsize{$ \partial_{ \nu \sigma \alpha } Z^{ \mu } \partial_{ \rho }^{ \enspace \alpha } Z_{ \mu } Z^{ \sigma } F^{ \nu \rho } $} & $+$ &  &  &  \\

28 & \scriptsize{$ \partial^{ \sigma \alpha } Z^{ \mu } \partial_{ \nu \sigma \alpha } Z_{ \mu } Z_{ \rho } F^{ \nu \rho } $} & $+$ &  &  &  \\

\cline{5-5}

29 & \scriptsize{$ \partial^{ \rho \alpha } Z^{ \mu } \partial_{ \nu \rho \alpha } Z_{ \mu } Z_{ \sigma } \widetilde{ F }^{ \nu \sigma } $} & $-$ &  & \multirow{1}{*}{ \scriptsize{$ D^8 H^4 \widetilde{ B }^{ \mu \nu } $} } &  \\

\cline{5-5}

30 & \scriptsize{$ \varepsilon^{ \mu \nu \rho \sigma } \partial_{ \mu }^{ \enspace \alpha \tau } Z^{ \beta } \partial_{ \beta \tau } Z_{ \nu } Z_{ \sigma } F_{ \rho \alpha } $} & $-$ &  & \multirow{3}{*}{ \scriptsize{$ \varepsilon D^{ 8 } H^4 B^{ \mu \nu } $} } &  \\

31 & \scriptsize{$ \varepsilon^{ \mu \nu \rho \sigma } \partial_{ \mu \tau } Z_{ \beta } \partial^{ \alpha \beta \tau } Z_{ \nu } Z_{ \sigma } F_{ \rho \alpha } $} & $-$ &  &  &  \\

32 & \scriptsize{$ \varepsilon^{ \mu \nu \rho \sigma } \partial_{ \tau } Z_{ \mu } \partial_{ \nu }^{ \enspace \alpha \tau } Z^{ \beta } Z_{ \sigma } \partial_{ \beta } F_{ \rho \alpha } $} & $-$ &  &  &  \\

\hline

33 & \scriptsize{$ \partial_{ \sigma }^{ \enspace \alpha } Z^{ \mu } \partial_{ \mu \alpha } Z_{ \nu \rho } Z^{ \sigma } F^{ \nu \rho } $} & $+$ & \multirow{8}{*}{10} & \multirow{1}{*}{ \scriptsize{$ D^6 H^2 W^{ a }_{ \mu \nu } B^{ \mu \nu } $} } & \multirow{8}{*}{ $ \frac{3 \times 10^{ -4 }}{E_\text{TeV}^8} $ } \\

\cline{5-5}

34 & \scriptsize{$ \partial_{ \nu \rho } Z^{ \mu } \partial_{ \mu \alpha } \widetilde{ Z }^{ \nu \sigma } Z_{ \sigma } F^{ \rho \alpha } $} & $-$ &  & \multirow{6}{*}{ \scriptsize{$ D^6 H^4 \widetilde{ W }^{ a }_{ \mu \nu } B^{ \mu \nu } $} } &  \\

35 & \scriptsize{$ \partial^{ \mu \alpha } Z^{ \nu } \partial_{ \nu \alpha } \widetilde{ Z }^{ \rho \sigma } Z_{ \sigma } F_{ \rho \mu } $} & $-$ &  &  &  \\

36 & \scriptsize{$ \partial_{ \alpha } Z_{ \mu } \partial^{ \mu \nu \alpha } \widetilde{ Z }^{ \rho \sigma } Z_{ \sigma } F_{ \rho \nu } $} & $-$ &  &  &  \\

37 & \scriptsize{$ \partial^{ \rho \alpha } Z^{ \mu } \partial_{ \alpha } \widetilde{ Z }^{ \nu \sigma } Z_{ \sigma } \partial_{ \mu } F_{ \nu \rho } $} & $-$ &  &  &  \\

38 & \scriptsize{$ \partial_{ \alpha } Z^{ \mu } \partial^{ \rho \alpha } \widetilde{ Z }^{ \nu \sigma } Z_{ \sigma } \partial_{ \mu } F_{ \nu \rho } $} & $-$ &  &  &  \\

39 & \scriptsize{$ \partial_{ \nu }^{ \enspace \alpha } Z_{ \mu } \partial_{ \rho \alpha } \widetilde{ Z }^{ \nu \sigma } Z_{ \sigma } F^{ \mu \rho } $} & $-$ &  &  &  \\

\hline

40 & \scriptsize{$ \partial_{ \nu \alpha } Z_{ \mu \rho } \partial^{ \alpha } \widetilde{ Z }^{ \nu \sigma } Z_{ \sigma } F^{ \mu \rho } $} & $-$ & \multirow{1}{*}{10} & \multirow{1}{*}{ \scriptsize{$ D^4 H^2 W^{ a }_{ \mu \nu } \widetilde{ W }^{ a }_{ \mu \nu } B^{ \mu \nu } $} } & \multirow{1}{*}{ $ \frac{0.001}{E_\text{TeV}^7}, \frac{0.004}{E_\text{TeV}^{8}} $ } \\[1.1ex]

\hline

\end{tabular}
\begin{minipage}{5.7in}
\medskip
\caption{\label{tab:zzzgam1} \footnotesize Primary dimension 6, 8, and 10 operators for the $ZZZ \gamma$ interaction. }
\end{minipage}
\end{center}
\end{adjustwidth}
\end{table}

\begin{table}[p]
\begin{adjustwidth}{-.5in}{-.5in}  
\begin{center}
\footnotesize
\centering
\renewcommand{\arraystretch}{0.9}
\setlength{\tabcolsep}{6pt}
\begin{tabular}{|c|c|c|c|c|c|}

\hline
\multirow{2}{*}{$i$} & \multirow{2}{*}{$\mathcal{O}_i^{ Z Z Z \gamma }$}   & \multirow{2}{*}{CP} & \multirow{2}{*}{$d_{\mathcal{O}_i}$}& SMEFT & $c$ Unitarity  \\
 & & & & Operator Form & Bound \\

\hline 

41 & \scriptsize{$ \partial_{ \nu \rho \sigma \beta } Z^{ \mu } \partial_{ \mu \alpha }^{ \quad \beta } Z^{ \nu } Z^{ \sigma } F^{ \rho \alpha } $} & $+$ & \multirow{9}{*}{12} & \multirow{6}{*}{ \scriptsize{$ D^{ 10 } H^4 B^{ \mu \nu } $} } & \multirow{9}{*}{ $ \frac{5 \times 10^{ -6 }}{E_\text{TeV}^{11}}, \frac{2 \times 10^{ -5 }}{E_\text{TeV}^{12}} $ } \\

42 & \scriptsize{$ \partial_{ \nu }^{ \enspace \alpha \beta } Z^{ \mu } \partial_{ \mu \rho \alpha \beta } Z^{ \nu } Z_{ \sigma } F^{ \rho \sigma } $} & $+$ &  &  &  \\

43 & \scriptsize{$ \partial^{ \rho \alpha \beta } Z^{ \mu } \partial_{ \mu \alpha \beta } Z^{ \nu } Z^{ \sigma } \partial_{ \nu } F_{ \rho \sigma } $} & $+$ &  &  &  \\

44 & \scriptsize{$ \partial_{ \alpha \beta } Z_{ \mu } \partial^{ \mu \rho \alpha \beta } Z^{ \nu } Z^{ \sigma } \partial_{ \nu } F_{ \rho \sigma } $} & $+$ &  &  &  \\

45 & \scriptsize{$ \partial_{ \alpha \beta } Z^{ \mu } \partial^{ \rho \alpha \beta } Z^{ \nu } Z^{ \sigma } \partial_{ \mu \nu } F_{ \rho \sigma } $} & $+$ &  &  &  \\

46  & \scriptsize{$ \partial_{ \nu \sigma }^{ \quad \alpha \beta } Z^{ \mu } \partial_{ \rho \alpha \beta } Z_{ \mu } Z^{ \sigma } F^{ \nu \rho } $} & $+$ &  &  &  \\

\cline{5-5}

47 & \scriptsize{$ \varepsilon^{ \mu \nu \rho \sigma } \partial_{ \mu }^{ \enspace \alpha \tau \pi } Z^{ \beta } \partial_{ \beta \tau \pi } Z_{ \nu } Z_{ \sigma } F_{ \rho \alpha } $} & $-$ &  & \multirow{3}{*}{ \scriptsize{$ \varepsilon D^{ 10 } H^4 B^{ \mu \nu } $} } &  \\

48 & \scriptsize{$ \varepsilon^{ \mu \nu \rho \sigma } \partial_{ \mu \tau \pi } Z_{ \beta } \partial^{ \alpha \beta \tau \pi } Z_{ \nu } Z_{ \sigma } F_{ \rho \alpha } $} & $-$ &  &  &  \\

49 & \scriptsize{$ \varepsilon^{ \mu \nu \rho \sigma } \partial_{ \tau \pi } Z_{ \mu } \partial_{ \nu }^{ \enspace \alpha \tau \pi } Z^{ \beta } Z_{ \sigma } \partial_{ \beta } F_{ \rho \alpha } $} & $-$ &  &  &  \\

\hline

50 & \scriptsize{$ \partial_{ \nu \rho }^{ \quad \beta } Z^{ \mu } \partial_{ \mu \alpha \beta } \widetilde{ Z }^{ \nu \sigma } Z_{ \sigma } F^{ \rho \alpha } $} & $-$ & \multirow{2}{*}{12} & \multirow{2}{*}{ \scriptsize{$ D^8 H^2 \widetilde{ W }^{ a }_{ \mu \nu } B^{ \mu \nu } $} } & \multirow{2}{*}{ $ \frac{2 \times 10^{ -5 }}{E_\text{TeV}^{10}} $ } \\

51 & \scriptsize{$ \partial^{ \mu \alpha \beta } Z^{ \nu } \partial_{ \nu \alpha \beta } \widetilde{ Z }^{ \rho \sigma } Z_{ \sigma } F_{ \rho \mu } $} & $-$ &  &  &  \\

\hline

52 & \scriptsize{$ \partial_{ \nu \alpha \beta } Z_{ \mu \rho } \partial^{ \alpha \beta } \widetilde{ Z }^{ \nu \sigma } Z_{ \sigma } F^{ \mu \rho } $} & $-$ &\multirow{1}{*}{12}  & \multirow{1}{*}{ \scriptsize{$ D^6 H^2 \widetilde{ W }^{ a }_{ \mu \nu } B^{ \mu \nu } B^{ \mu \nu } $} } & \multirow{1}{*}{ $ \frac{8 \times 10^{ -5 }}{E_\text{TeV}^{9}}, \frac{3 \times 10^{ -4 }}{E_\text{TeV}^{10}} $ } \\[1.1ex]

\hline

53 & \scriptsize{$ \partial_{ \nu \rho \sigma \beta \tau } Z^{ \mu } \partial_{ \mu \alpha }^{ \quad \beta \tau } Z^{ \nu } Z^{ \sigma } F^{ \rho \alpha } $} & $+$ & \multirow{3}{*}{14} & \multirow{2}{*}{ \scriptsize{$ D^{ 12 } H^4 B^{ \mu \nu } $} } & \multirow{3}{*}{ $ \frac{3 \times 10^{ -7 }}{E_\text{TeV}^{13}}, \frac{9 \times 10^{ -7 }}{E_\text{TeV}^{14}} $ } \\

54 & \scriptsize{$ \left( \partial_{ \nu \alpha \beta } Z^{ \mu } \overleftrightarrow{ \partial }^{ \tau } \partial_{ \mu \rho }^{ \quad \alpha \beta } Z^{ \nu } \right) Z_{ \sigma } \partial_{ \tau } F^{ \rho \sigma } $} & $+$ &  &  &  \\

\cline{5-5}

55 & \scriptsize{$ \varepsilon^{ \mu \nu \rho \sigma } \left( \partial_{ \mu }^{ \enspace \alpha \tau \pi } Z^{ \beta } \overleftrightarrow{ \partial }^{ \delta } \partial_{ \beta \tau \pi } Z_{ \nu } \right) Z_{ \sigma } \partial_{ \delta } F_{ \rho \alpha } $} & $-$ &  & \multirow{1}{*}{ \scriptsize{$ \varepsilon D^{ 12 } H^4 B^{ \mu \nu } $} } &  \\
 
\hline

56 & \scriptsize{$ \partial_{ \nu \rho \beta \tau } Z^{ \mu } \partial_{ \mu \alpha }^{ \quad \beta \tau } \widetilde{ Z }^{ \nu \sigma } Z_{ \sigma } F^{ \rho \alpha } $} & $-$ & \multirow{1}{*}{14} & \multirow{1}{*}{ \scriptsize{$ D^{ 10 } H^2 \widetilde{ W }^{ a }_{ \mu \nu } B^{ \mu \nu } $} } & \multirow{1}{*}{ $ \frac{1 \times 10^{ -6 }}{E_\text{TeV}^{12}} $ } \\[1.5 ex]
 
 \hline

\end{tabular}
\begin{minipage}{5.7in}
\medskip
\caption{\label{tab:zzzgam2} \footnotesize Primary dimension 12 and 14 operators for the $ZZZ \gamma$ interaction. At dimension 14, two descendant operators become redundant. To form a set of independent operators, $x^n y^m \mathcal{ O }^{ ZZZ \gamma }_{ 28 }$ and $x^n y^m \mathcal{ O }^{ ZZZ \gamma }_{ 29 }$, with $x = s^2 + t^2 + u^2$, $y = s t u$, $n \geq 1$ and $m \geq 0$, should be omitted. }
\end{minipage}
\end{center}
\end{adjustwidth}
\end{table}

\begin{table}[p]
\begin{adjustwidth}{-.5in}{-.5in}  
\begin{center}
\scriptsize
\centering
\renewcommand{\arraystretch}{0.9}
\setlength{\tabcolsep}{6pt}
\begin{tabular}{|c|c|c|c|c|c|}

\hline
\multirow{2}{*}{$i$} & \multirow{2}{*}{$\mathcal{O}_i^{ W W \gamma \gamma }$}   & \multirow{2}{*}{CP} & \multirow{2}{*}{$d_{\mathcal{O}_i}$}& SMEFT & $c$ Unitarity  \\
 & & & & Operator Form & Bound \\

\hline 

1 & \scriptsize{$ W^{ + \mu } W^{ - }_{ \mu } F^{ \nu \rho } F_{ \nu \rho } $} & $+$ & \multirow{3}{*}{6} & \multirow{2}{*}{ \scriptsize{$ D^{ 2 } H^2 B^{ \mu \nu } B^{ \mu \nu } $} } & \multirow{3}{*}{ $ \frac{0.09}{E_\text{TeV}^{4}} $ } \\

2 & \scriptsize{$ W^{ + \mu } W^{ - }_{ \nu } F_{ \mu \rho } F^{ \nu \rho } $} & $+$ &  &  &  \\

\cline{5-5}

3 & \scriptsize{$ \left( W^{ + \mu } W^{ - }_{ \nu } + \textrm{h.c.} \right) F_{ \mu \rho } \widetilde{ F }^{ \nu \rho } $} & $-$ &  & \multirow{1}{*}{ \scriptsize{$ D^{ 2 } H^2 B^{ \mu \nu } \widetilde{ B }^{ \mu \nu } $} } &  \\

\hline

4 & \scriptsize{$ W^{ + \mu } W^{ - \nu } F^{ \rho \sigma } \partial_{ \mu \nu } F_{ \rho \sigma } $} & $+$ & \multirow{14}{*}{8} & \multirow{6}{*}{ \scriptsize{$ D^4 H^2 B^{ \mu \nu } B^{ \mu \nu } $} } & \multirow{14}{*}{ $ \frac{0.006}{E_\text{TeV}^{6}} $ } \\

5 & \scriptsize{$ W^{ + \mu } W^{ - \nu } \partial_{ \mu } F^{ \rho \sigma } \partial_{ \nu } F_{ \rho \sigma } $} & $+$ &  &  &  \\

6 & \scriptsize{$ \left( W^{ + \mu } D_{ \sigma } W^{ - \nu } + \textrm{h.c.} \right) F^{ \rho \sigma } \partial_{ \mu } F_{ \nu \rho } $} & $+$ &  &  &  \\

7 & \scriptsize{$ \left( W^{ + \mu } D^{ \sigma } W^{ - }_{ \nu } + \textrm{h.c.} \right) F^{ \nu \rho } \partial_{ \mu } F_{ \rho \sigma } $} & $+$ &  &  &  \\

8 & \scriptsize{$ \left( W^{ + \mu } D_{ \rho }^{ \enspace \sigma } W^{ - }_{ \mu } + \textrm{h.c.} \right) F^{ \nu \rho } F_{ \nu \sigma } $} & $+$ &  &  &  \\

9 & \scriptsize{$ \left( W^{ + }_{ \mu } \overleftrightarrow{ D }^{ \sigma } W^{ - \nu } + \textrm{h.c.} \right) \partial_{ \sigma } F^{ \mu \rho } F_{ \rho \nu } $} & $+$ &  &  &  \\

\cline{5-5}

10 & \scriptsize{$ \left( i W^{ + }_{ \mu } D^{ \rho } W^{ - \nu } + \textrm{h.c.} \right) \widetilde{ F }^{ \mu \sigma } \partial_{ \nu } F_{ \sigma \rho } $} & $+$ &  & \multirow{1}{*}{ \scriptsize{$ D^4 H^2 B^{ \mu \nu } \widetilde{ B }^{ \mu \nu } $} } &  \\

\cline{5-5}

11 & \scriptsize{$ \varepsilon^{ \mu \nu \rho \sigma } \left( i W^{ + }_{ \mu } D_{ \nu } W^{ - \alpha } + \textrm{h.c.} \right) F_{ \rho }^{ \enspace \beta } \partial_{ \alpha } F_{ \sigma \beta } $} & $+$ &  & \multirow{1}{*}{ \scriptsize{$ \varepsilon D^4 H^2 B^{ \mu \nu } B^{ \mu \nu } $} } &  \\

\cline{5-5}

12 & \scriptsize{$ \left( i W^{ + \mu } D_{ \sigma } W^{ - \nu } + \textrm{h.c.} \right) F^{ \rho \sigma } \partial_{ \mu } F_{ \nu \rho } $} & $-$ &  & \multirow{2}{*}{ \scriptsize{$ D^4 H^2 B^{ \mu \nu } B^{ \mu \nu } $} } &  \\

13 & \scriptsize{$ \left( i W^{ + \mu } D^{ \sigma } W^{ - }_{ \nu } + \textrm{h.c.} \right) F^{ \nu \rho } \partial_{ \mu } F_{ \rho \sigma } $} & $-$ &  &  &  \\

\cline{5-5}

14 & \scriptsize{$ \left( W^{ + \mu } \overleftrightarrow{ D }^{ \sigma } W^{ - }_{ \nu } + \textrm{h.c.} \right) \partial_{ \sigma } F_{ \mu \rho } \widetilde{ F }^{ \nu \rho } $} & $-$ &  & \multirow{2}{*}{ \scriptsize{$ D^4 H^2 B^{ \mu \nu } \widetilde{ B }^{ \mu \nu } $} } &  \\

15 & \scriptsize{$ \left( W^{ + }_{ \mu } W^{ - \nu } + \textrm{h.c.} \right) \partial_{ \nu } F_{ \rho \sigma } \partial^{ \sigma } \widetilde{ F }^{ \mu \rho } $} & $-$ &  &  &  \\

\cline{5-5}

16 & \scriptsize{$ \varepsilon^{ \mu \nu \rho \sigma } \left( W^{ + }_{ \mu } D_{ \beta } W^{ - }_{ \nu } + \textrm{h.c.} \right) F_{ \rho \alpha } \partial_{ \sigma } F^{ \alpha \beta } $} & $-$ &  & \multirow{2}{*}{ \scriptsize{$ \varepsilon D^4 H^2 B^{ \mu \nu } B^{ \mu \nu } $} } &  \\
 
17 & \scriptsize{$ \varepsilon^{ \mu \nu \rho \sigma } \left( W^{ + }_{ \mu } \overleftrightarrow{ D }^{ \beta } W^{ - }_{ \nu } + \textrm{h.c.} \right) \partial_{ \beta } F^{ \alpha }_{ \enspace \rho } F_{ \alpha \sigma } $} & $-$ &  &  &  \\
 
\hline

18 & \scriptsize{$ \left( W^{ + \mu } D^{ \sigma } W^{ - }_{ \nu \rho } + \textrm{h.c.} \right) F^{ \nu \rho } F_{ \sigma \mu } $} & $+$ & \multirow{5}{*}{8} & \multirow{1}{*}{ \scriptsize{$ D^{ 2 } H^2 W^{ a }_{ \mu \nu } B^{ \mu \nu } B^{ \mu \nu } $} } & \multirow{5}{*}{ $ \frac{0.02}{E_\text{TeV}^{5}}, \frac{0.07}{E_\text{TeV}^{6}} $ } \\
 
\cline{5-5}

19 & \scriptsize{$ \left( i W^{ + }_{ \mu } D_{ \rho } \widetilde{ W }^{ - \mu \sigma } + \textrm{h.c.} \right) F^{ \nu \rho } F_{ \sigma \nu } $} & $+$ &  & \multirow{1}{*}{ \scriptsize{$ D^{ 2 } H^2 \widetilde{ W }^{ a }_{ \mu \nu } B^{ \mu \nu } B^{ \mu \nu } $} } &  \\

\cline{5-5}
 
20 & \scriptsize{$ \left( W^{ + }_{ \mu } D_{ \rho } W^{ - }_{ \nu \sigma } + \textrm{h.c.} \right) \widetilde{ F }^{ \mu \rho } F^{ \nu \sigma } $} & $-$ &  & \multirow{1}{*}{ \scriptsize{$ D^{ 2 } H^2 W^{ a }_{ \mu \nu } B^{ \mu \nu } \widetilde{ B }^{ \mu \nu } $} } &  \\

\cline{5-5}
 
21 & \scriptsize{$ \left( W^{ + }_{ \mu } \widetilde{ W }^{ - \mu \nu } + \textrm{h.c.} \right) F^{ \rho \sigma } \partial_{ \sigma } F_{ \nu \rho } $} & $-$ &  & \multirow{2}{*}{ \scriptsize{$ D^{ 2 } H^2 \widetilde{ W }^{ a }_{ \mu \nu } B^{ \mu \nu } B^{ \mu \nu } $} } &  \\

22 & \scriptsize{$ \left( W^{ + }_{ \mu } D_{ \sigma } \widetilde{ W }^{ - \mu \nu } + \textrm{h.c.} \right) F^{ \rho \sigma } F_{ \nu \rho } $} & $-$ &  &  &  \\
 
\hline

23 & \scriptsize{$ \left( W^{ + \mu } \overleftrightarrow{ D }^{ \alpha } W^{ - \nu } + \textrm{h.c.} \right) \partial_{ \mu \alpha } F^{ \rho \sigma } \partial_{ \nu } F_{ \rho \sigma } $} & $+$ &  \multirow{9}{*}{10} & \multirow{2}{*}{ \scriptsize{$ D^6 H^2 B^{ \mu \nu } B^{ \mu \nu } $} } & \multirow{9}{*}{ $ \frac{3 \times 10^{ -4 }}{E_\text{TeV}^{8}} $ } \\

24 & \scriptsize{$ \left( W^{ + \mu } \overleftrightarrow{ D }^{ \alpha } D_{ \sigma } W^{ - \nu } + \textrm{h.c.} \right) \partial_{ \alpha } F^{ \rho \sigma } \partial_{ \mu } F_{ \nu \rho } $} & $+$ &  &  &  \\

\cline{5-5}

25 & \scriptsize{$ \varepsilon^{ \mu \nu \rho \sigma } \left( i W^{ + }_{ \mu } \overleftrightarrow{ D }^{ \tau } D_{ \beta } W^{ - }_{ \nu } + \textrm{h.c.} \right) \partial_{ \tau } F_{ \rho \alpha } \partial_{ \sigma } F^{ \alpha \beta } $} & $+$ &  & \multirow{1}{*}{ \scriptsize{$ \varepsilon D^6 H^2 B^{ \mu \nu } B^{ \mu \nu } $} } &  \\

\cline{5-5}
 
26 & \scriptsize{$ \left( i W^{ + \mu } \overleftrightarrow{ D }^{ \alpha } W^{ - \nu } + \textrm{h.c.} \right) \partial_{ \alpha } F^{ \rho \sigma } \partial_{ \mu \nu } F_{ \rho \sigma } $} & $-$ &  & \multirow{3}{*}{ \scriptsize{$ D^6 H^2 B^{ \mu \nu } B^{ \mu \nu } $} } &  \\

27 & \scriptsize{$ \left( i W^{ + \mu } \overleftrightarrow{ D }^{ \alpha } D_{ \sigma } W^{ - \nu } + \textrm{h.c.} \right) \partial_{ \alpha } F^{ \rho \sigma } \partial_{ \mu } F_{ \nu \rho } $} & $-$ &  &  &  \\

28 & \scriptsize{$ \left( i W^{ + \mu } \overleftrightarrow{ D }_{ \alpha } D^{ \rho } W^{ - \nu } + \textrm{h.c.} \right) \partial^{ \alpha } F_{ \nu \sigma } \partial^{ \sigma } F_{ \mu \rho } $} & $-$ &  &  &  \\

\cline{5-5}

29 & \scriptsize{$ \varepsilon^{ \mu \nu \rho \sigma } \left( W^{ + }_{ \mu } \overleftrightarrow{ D }_{ \tau } W^{ - }_{ \nu } \right) \partial^{ \alpha \tau } F_{ \rho \beta } \partial^{ \beta } F_{ \sigma \alpha } $} & $-$ &  & \multirow{2}{*}{ \scriptsize{$ \varepsilon D^6 H^2 B^{ \mu \nu } B^{ \mu \nu } $} } &  \\

30 & \scriptsize{$ \varepsilon^{ \mu \nu \rho \sigma } \left( W^{ + }_{ \mu } \overleftrightarrow{ D }^{ \tau } D^{ \alpha \beta } W^{ - }_{ \nu } + \textrm{h.c.} \right) \partial_{ \tau } F_{ \rho \beta } F_{ \sigma \alpha } $} & $-$ &  &  &  \\

\hline 

31 & \scriptsize{$ \left( W^{ + }_{ \mu } \overleftrightarrow{ D }_{ \alpha } D_{ \sigma } W^{ - }_{ \nu \rho } + \textrm{h.c.} \right) \partial^{ \alpha } F^{ \nu \rho } F^{ \sigma \mu } $} & $+$ & \multirow{7}{*}{10} & \multirow{2}{*}{ \scriptsize{$ D^4 H^2 W^{ a }_{ \mu \nu } B^{ \mu \nu } B^{ \mu \nu } $} } & \multirow{7}{*}{ $ \frac{0.001}{E_\text{TeV}^{7}}, \frac{0.004}{E_\text{TeV}^{8}} $ } \\

32 & \scriptsize{$ \left( W^{ + \mu } D_{ \sigma } W^{ - }_{ \alpha \nu } + \textrm{h.c.} \right) F^{ \rho \sigma } \partial_{ \mu \rho } F^{ \alpha \nu } $} & $+$ &  &  &  \\

\cline{5-5}

33 & \scriptsize{$ \left( i W^{ + }_{ \mu } \overleftrightarrow{ D }_{ \alpha } D_{ \rho } W^{ - }_{ \nu \sigma } + \textrm{h.c.} \right) \partial^{ \alpha } F^{ \nu \sigma } \widetilde{ F }^{ \mu \rho } $} & $+$ &  & \multirow{1}{*}{ \scriptsize{$ D^4 H^2 W^{ a }_{ \mu \nu } B^{ \mu \nu } \widetilde{ B }^{ \mu \nu } $} } &  \\

\cline{5-5}

34 & \scriptsize{$ \left( i W^{ + }_{ \mu } \overleftrightarrow{ D }_{ \alpha } \widetilde{ W }^{ - \mu \nu } + \textrm{h.c.} \right) \partial^{ \alpha } F^{ \rho \sigma } \partial_{ \sigma } F_{ \nu \rho } $} & $+$ &  & \multirow{3}{*}{ \scriptsize{$ D^4 H^2 \widetilde{ W }^{ a }_{ \mu \nu } B^{ \mu \nu } B^{ \mu \nu } $} } &  \\

35 & \scriptsize{$ \left( i W^{ + }_{ \mu } \overleftrightarrow{ D }_{ \alpha } D_{ \sigma } \widetilde{ W }^{ - \mu \nu } + \textrm{h.c.} \right) \partial^{ \alpha } F_{ \nu \rho } F^{ \rho \sigma } $} & $+$ &  &  &  \\

36 & \scriptsize{$ \left( W^{ + \mu } D^{ \rho }_{ \enspace \alpha } \widetilde{ W }^{ - }_{ \mu \nu } + \textrm{h.c.} \right) F^{ \sigma \alpha } \partial^{ \nu } F_{ \rho \sigma } $} & $-$ &  &  &  \\
 
\hline
 
37 & \scriptsize{$ \left( i W^{ + \mu } \overleftrightarrow{ D }^{ \beta } D_{ \rho }^{ \enspace \sigma } W^{ - \nu } + \textrm{h.c.} \right) \partial_{ \beta } F^{ \alpha \rho } \partial_{ \mu \nu } F_{ \alpha \sigma } $} & $-$ & \multirow{1}{*}{12} & \multirow{1}{*}{ \scriptsize{$ D^8 H^2 B^{ \mu \nu } B^{ \mu \nu } $} } & \multirow{1}{*}{ $ \frac{2 \times 10^{ -5 }}{E_\text{TeV}^{10}} $ } \\[1.1ex]

\hline

38 & \scriptsize{$ \left( i W^{ + }_{ \mu } \overleftrightarrow{ D }^{ \beta } D_{ \enspace \alpha }^{ \sigma } \widetilde{ W }^{ - \mu \nu } + \textrm{h.c.} \right) \partial_{ \beta } F_{ \rho \sigma } \partial_{ \nu } F^{ \rho \alpha } $} & $+$ & \multirow{1}{*}{12} & \multirow{1}{*}{ \scriptsize{$ D^6 H^2 \widetilde{ W }^{ a }_{ \mu \nu } B^{ \mu \nu } B^{ \mu \nu } $} } & \multirow{1}{*}{ $ \frac{8 \times 10^{ -5 }}{E_\text{TeV}^{9}}, \frac{3 \times 10^{ -4 }}{E_\text{TeV}^{10}} $ } \\[1.1ex]
 
\hline

\end{tabular}
\begin{minipage}{5.7in}
\medskip
\caption{\label{tab:wwgamgam} \footnotesize Operators up to dimension 12 for the $W^+ W^- \gamma \gamma $ interaction. At dimension 12 there are 2 redundancies such that in order to form a set of independent operators, $s^n (t-u)^{ 2 m } \mathcal{ O }^{ WW \gamma \gamma }_{ 9 }$ and $s^n (t-u)^{ 2 m } \mathcal{ O }^{ WW \gamma \gamma }_{ 17 }$, with $n \geq 0$, and $m \geq 1$, should be omitted. Operators for $W^+ W^- g g$ interactions can be obtained by replacing $F_{ \mu \nu }$'s with $G^A_{ \mu \nu }$'s contracted with $\delta_{AB}$. The same redundancies apply to the corresponding operators.}
\end{minipage}
\end{center}
\end{adjustwidth}
\end{table}

\begin{table}[p]
\begin{adjustwidth}{-.5in}{-.5in}  
\begin{center}
\footnotesize
\centering
\renewcommand{\arraystretch}{0.9}
\setlength{\tabcolsep}{6pt}
\begin{tabular}{|c|c|c|c|c|c|}

\hline
\multirow{2}{*}{$i$} & \multirow{2}{*}{$\mathcal{O}_i^{ Z Z \gamma \gamma }$}   & \multirow{2}{*}{CP} & \multirow{2}{*}{$d_{\mathcal{O}_i}$}& SMEFT & $c$ Unitarity  \\
 & & & & Operator Form & Bound \\

\hline 

1 & \scriptsize{$ Z^{ \mu } Z_{ \mu } F^{ \nu \rho } F_{ \nu \rho } $} & $+$ & \multirow{3}{*}{6} & \multirow{2}{*}{ \scriptsize{$ D^{ 2 } H^2 B^{ \mu \nu } B^{ \mu \nu } $} } & \multirow{3}{*}{ $ \frac{0.09}{E_\text{TeV}^{4}} $ } \\

2 & \scriptsize{$ Z^{ \mu } Z_{ \nu } F_{ \mu \rho } F^{ \nu \rho } $} & $+$ &  &  &  \\

\cline{5-5}

3 & \scriptsize{$ Z^{ \mu } Z_{ \nu } F_{ \mu \rho } \widetilde{ F }^{ \nu \rho } $} & $-$ &  & \multirow{1}{*}{ \scriptsize{$ D^{ 2 } H^2 B^{ \mu \nu } \widetilde{ B }^{ \mu \nu } $} } &  \\

\hline

4 & \scriptsize{$ Z^{ \mu } Z^{ \nu } F^{ \rho \sigma } \partial_{ \mu \nu } F_{ \rho \sigma } $} & $+$ & \multirow{9}{*}{8} & \multirow{6}{*}{ \scriptsize{$ D^4 H^2 B^{ \mu \nu } B^{ \mu \nu } $} } & \multirow{9}{*}{ $ \frac{0.006}{E_\text{TeV}^{6}} $ } \\

5 & \scriptsize{$ Z^{ \mu } Z^{ \nu } \partial_{ \mu } F^{ \rho \sigma } \partial_{ \nu } F_{ \rho \sigma } $} & $+$ &  &  &  \\

6 & \scriptsize{$ Z^{ \mu } \partial_{ \sigma } Z^{ \nu } F^{ \rho \sigma } \partial_{ \mu } F_{ \nu \rho } $} & $+$ &  &  &  \\

7 & \scriptsize{$ Z^{ \mu } \partial^{ \sigma } Z_{ \nu } F^{ \nu \rho } \partial_{ \mu } F_{ \rho \sigma } $} & $+$ &  &  &  \\

8 & \scriptsize{$ Z^{ \mu } \partial_{ \rho }^{ \enspace \sigma } Z_{ \mu } F^{ \nu \rho } F_{ \nu \sigma } $} & $+$ &  &  &  \\

9 & \scriptsize{$ \left( Z_{ \mu } \overleftrightarrow{ \partial }^{ \sigma } Z^{ \nu } \right)\partial_{ \sigma } F^{ \mu \rho } F_{ \rho \nu } $} & $+$ &  &  &  \\

\cline{5-5}

10 & \scriptsize{$ Z_{ \mu } Z^{ \nu } \partial_{ \nu } F_{ \rho \sigma } \partial^{ \sigma } \widetilde{ F }^{ \mu \rho } $} & $-$ &  & \multirow{1}{*}{ \scriptsize{$ D^4 H^2 B^{ \mu \nu } \widetilde{ B }^{ \mu \nu } $} } &  \\

\cline{5-5}

11 & \scriptsize{$ \varepsilon^{ \mu \nu \rho \sigma } Z_{ \mu } \partial_{ \beta } Z_{ \nu } F_{ \rho \alpha } \partial_{ \sigma } F^{ \alpha \beta } $} & $-$ &  & \multirow{2}{*}{ \scriptsize{$ \varepsilon D^4 H^2 B^{ \mu \nu } B^{ \mu \nu } $} } &  \\
 
12 & \scriptsize{$ \varepsilon^{ \mu \nu \rho \sigma } \left( Z_{ \mu } \overleftrightarrow{ \partial }^{ \beta } Z_{ \nu } \right) \partial_{ \beta } F^{ \alpha }_{ \enspace \rho } F_{ \alpha \sigma } $} & $-$ &  &  &  \\
 
\hline

13 & \scriptsize{$ Z^{ \mu } \partial^{ \sigma } Z_{ \nu \rho } F^{ \nu \rho } F_{ \sigma \mu } $} & $+$ & \multirow{4}{*}{8} & \multirow{1}{*}{ \scriptsize{$ D^{ 2 } H^2 B^{ \mu \nu } B^{ \mu \nu } B^{ \mu \nu } $} } & \multirow{4}{*}{ $ \frac{0.02}{E_\text{TeV}^{5}}, \frac{0.07}{E_\text{TeV}^{6}} $ } \\

\cline{5-5}

14 & \scriptsize{$ Z_{ \mu } \partial_{ \rho } Z_{ \nu \sigma } \widetilde{ F }^{ \mu \rho } F^{ \nu \sigma } $} & $-$ &  & \multirow{3}{*}{ \scriptsize{$ D^{ 2 } H^2 B^{ \mu \nu } B^{ \mu \nu } \widetilde{ B }^{ \mu \nu } $} } &  \\
 
15 & \scriptsize{$ Z_{ \mu } \widetilde{ Z }^{ \mu \nu } F^{ \rho \sigma } \partial_{ \sigma } F_{ \nu \rho } $} & $-$ &  &  &  \\

16 & \scriptsize{$ Z_{ \mu } \partial_{ \sigma } \widetilde{ Z }^{ \mu \nu } F^{ \rho \sigma } F_{ \nu \rho } $} & $-$ &  &  &  \\
 
\hline

17 & \scriptsize{$ \left( Z^{ \mu } \overleftrightarrow{ \partial }^{ \alpha } Z^{ \nu } \right) \partial_{ \mu \alpha } F^{ \rho \sigma } \partial_{ \nu } F_{ \rho \sigma } $} & $+$ & \multirow{4}{*}{10} & \multirow{2}{*}{ \scriptsize{$ D^6 H^2 B^{ \mu \nu } B^{ \mu \nu } $} } & \multirow{4}{*}{ $ \frac{3 \times 10^{ -4 }}{E_\text{TeV}^{8}} $ } \\

18 & \scriptsize{$ \left( Z^{ \mu } \overleftrightarrow{ \partial }^{ \alpha } \partial_{ \sigma } Z^{ \nu } \right) \partial_{ \alpha } F^{ \rho \sigma } \partial_{ \mu } F_{ \nu \rho } $} & $+$ &  &  &  \\

\cline{5-5}

19 & \scriptsize{$ \varepsilon^{ \mu \nu \rho \sigma } \left( Z_{ \mu } \overleftrightarrow{ \partial }_{ \tau } Z_{ \nu } \right) \partial^{ \alpha \tau } F_{ \rho \beta } \partial^{ \beta } F_{ \sigma \alpha } $} & $-$ &  & \multirow{2}{*}{ \scriptsize{$ \varepsilon D^6 H^2 B^{ \mu \nu } B^{ \mu \nu } $} } &  \\

20 & \scriptsize{$ \varepsilon^{ \mu \nu \rho \sigma } \left( Z_{ \mu } \overleftrightarrow{ \partial }^{ \tau } \partial^{ \alpha \beta } Z_{ \nu } \right) \partial_{ \tau } F_{ \rho \beta } F_{ \sigma \alpha } $} & $-$ &  &  &  \\
 
\hline

21 & \scriptsize{$ Z^{ \mu } \partial_{ \rho } Z^{ \nu \alpha } F^{ \rho \sigma } \partial_{ \mu \sigma } F_{ \nu \alpha } $} & $+$ & \multirow{3}{*}{10} & \multirow{2}{*}{ \scriptsize{$ D^4 H^2 B^{ \mu \nu } B^{ \mu \nu } B^{ \mu \nu } $} } & \multirow{3}{*}{ $ \frac{0.001}{E_\text{TeV}^{7}}, \frac{0.004}{E_\text{TeV}^{8}} $ } \\

22 & \scriptsize{$ \left( Z_{ \mu } \overleftrightarrow{ \partial }_{ \alpha } \partial_{ \sigma } Z_{ \nu \rho } \right) \partial^{ \alpha } F^{ \nu \rho } F^{ \sigma \mu } $} & $+$ &  &  &  \\

\cline{5-5}

23 & \scriptsize{$ Z_{ \mu } \partial_{ \rho }^{ \enspace \alpha } \widetilde{ Z }^{ \mu \nu } F_{ \sigma \alpha } \partial_{ \nu } F^{ \rho \sigma } $} & $-$ &  & \multirow{1}{*}{ \scriptsize{$ D^4 H^2 B^{ \mu \nu } B^{ \mu \nu } \widetilde{ B }^{ \mu \nu } $} } &  \\

\hline

\end{tabular}
\begin{minipage}{5.7in}
\medskip
\caption{\label{tab:zzgamgam} \footnotesize Operators up to dimension 10 for the $Z Z \gamma \gamma $ interaction. At dimension 12 there are 2 redundancies such that in order to form a set of independent operators, $s^n (t-u)^{ 2 m } \mathcal{ O }^{ ZZ \gamma \gamma }_{ 9 }$ and $s^n (t-u)^{ 2 m } \mathcal{ O }^{ ZZ \gamma \gamma }_{ 12 }$, with $n \geq 0$ and $m \geq 1$, should be omitted. Operators for $Z Z g g$ interactions can be obtained by replacing $F_{ \mu \nu }$'s with $G^A_{ \mu \nu }$'s contracted with $\delta_{AB}$. The same redundancies apply to the corresponding operators. }
\end{minipage}
\end{center}
\end{adjustwidth}
\end{table}

\begin{table}[p]
\begin{adjustwidth}{-.5in}{-.5in}  
\begin{center}
\footnotesize
\centering
\renewcommand{\arraystretch}{0.9}
\setlength{\tabcolsep}{6pt}
\begin{tabular}{|c|c|c|c|c|c|}

\hline
\multirow{2}{*}{$i$} & \multirow{2}{*}{$\mathcal{O}_i^{ Z \gamma g g }$}   & \multirow{2}{*}{CP} & \multirow{2}{*}{$d_{\mathcal{O}_i}$}& SMEFT & $c$ Unitarity  \\
 & & & & Operator Form & Bound \\
 
 \hline

1 & \scriptsize{$ G^{ \mu \nu } D_{ \sigma } G_{ \mu \rho } F_{ \nu }^{ \enspace \rho } Z^{ \sigma } $} & $+$ & \multirow{12}{*}{8} & \multirow{6}{*}{ \scriptsize{$ D^{ 2 } H^2 G^{ \mu \nu } G^{ \mu \nu } B^{ \mu \nu } $} } & \multirow{12}{*}{ $ \frac{0.02}{E_\text{TeV}^{5}}, \frac{0.07}{E_\text{TeV}^{6}} $ } \\

2 & \scriptsize{$ G^{ \mu \nu } D_{ \sigma } G_{ \nu \rho } F^{ \rho \sigma } Z_{ \mu } $} & $+$ &  &  &  \\

3 & \scriptsize{$ G^{ \mu \nu } D^{ \sigma } G_{ \mu \rho } F_{ \nu \sigma } Z^{ \rho }$} & $+$ &  &  &  \\

4 & \scriptsize{$ G^{ \mu \nu } D_{ \rho } G_{ \mu \nu } F^{ \rho \sigma } Z_{ \sigma } $} & $+$ &  &  &  \\

5 & \scriptsize{$ G^{ \mu \nu } G^{ \rho \sigma } \partial_{ \rho } F_{ \mu \nu } Z_{ \sigma } $} & $+$ &  &  &  \\

6 & \scriptsize{$ G^{ \mu \nu } G_{ \mu \rho } \partial^{ \rho } F_{ \nu \sigma } Z^{ \sigma } $} & $+$ &  &  &  \\

\cline{5-5}

7 & \scriptsize{$ D_{ \rho } G_{ \mu \nu } G^{ \nu \rho } \widetilde{ F }^{ \mu \sigma } Z_{ \sigma } $} & $-$ &  & \multirow{1}{*}{ \scriptsize{$ D^{ 2 } H^2 G^{ \mu \nu } G^{ \mu \nu } \widetilde{ B }^{ \mu \nu } $} } &  \\

\cline{5-5}

8 & \scriptsize{$ G^{ \mu \nu } \widetilde{ G }^{ \rho \sigma } \partial_{ \rho } F_{ \mu \nu } Z_{ \sigma } $} & $-$ &  & \multirow{2}{*}{ \scriptsize{$ D^{ 2 } H^2 G^{ \mu \nu } \widetilde{ G }^{ \mu \nu } B^{ \mu \nu } $} } &  \\

9 & \scriptsize{$ G^{ \mu \nu } D_{ \nu } \widetilde{ G }^{ \rho \sigma } F_{ \mu \rho } Z_{ \sigma } $} & $-$ &  &  &  \\

\cline{5-5}

10 & \scriptsize{$ \varepsilon^{ \mu \nu \rho \sigma } G_{ \mu \alpha } D_{ \nu } G^{ \alpha \beta } F_{ \rho \beta } Z_{ \sigma } $} & $-$ &  & \multirow{3}{*}{ \scriptsize{$ \varepsilon D^{ 2 } H^2 G^{ \mu \nu } G^{ \mu \nu } B^{ \mu \nu } $} } &  \\

11 & \scriptsize{$ \varepsilon^{ \mu \nu \rho \sigma } G_{ \mu \alpha } G^{ \alpha \beta } \partial_{ \nu } F_{ \rho \beta } Z_{ \sigma } $} & $-$ &  &  &  \\

12 & \scriptsize{$ \varepsilon^{ \mu \nu \rho \sigma } G_{ \mu \alpha } D^{ \beta } G_{ \nu }^{ \enspace \alpha } F_{ \rho \beta } Z_{ \sigma } $} & $-$ &  &  &  \\
 
\hline

13 & \scriptsize{$ G^{ \mu \nu } D_{ \mu \alpha } G^{ \rho \sigma } \partial_{ \nu } F_{ \rho \sigma } Z^{ \alpha } $} & $+$ & \multirow{12}{*}{10} & \multirow{6}{*}{ \scriptsize{$ D^4 H^2 G^{ \mu \nu } G^{ \mu \nu } B^{ \mu \nu }$} } & \multirow{12}{*}{ $ \frac{0.001}{E_\text{TeV}^{7}}, \frac{0.004}{E_\text{TeV}^{8}} $ } \\

14 & \scriptsize{$ D_{ \rho } G^{ \mu \nu } D_{ \sigma \alpha } G_{ \mu \nu } F^{ \rho \sigma } Z^{ \alpha } $} & $+$ &  &  &  \\

15 & \scriptsize{$ \left( G^{ \mu \nu } \overleftrightarrow{ D }^{ \alpha } D_{ \sigma } G_{ \mu \rho } \right) \partial_{ \alpha } F_{ \nu }^{ \enspace \rho } Z^{ \sigma } $} & $+$ &  &  &  \\

16 & \scriptsize{$ \left( G^{ \mu \nu } \overleftrightarrow{ D }^{ \alpha } D_{ \nu } G^{ \rho \sigma } \right) \partial_{ \alpha } F_{ \mu \rho } Z_{ \sigma } $} & $+$ &  &  &  \\

17 & \scriptsize{$ \left( G^{ \mu \nu } \overleftrightarrow{ D }^{ \alpha } D_{ \sigma } G_{ \nu \rho } \right) \partial_{ \alpha } F^{ \rho \sigma } Z_{ \mu } $} & $+$ &  &  &  \\

18 & \scriptsize{$ \left( G^{ \mu \nu } \overleftrightarrow{ D }^{ \alpha } G^{ \rho \sigma } \right) \partial_{ \nu \alpha } F_{ \mu \rho } Z_{ \sigma } $} & $+$ & &  &  \\


\cline{5-5}

19 & \scriptsize{$ G^{ \mu \nu } D_{ \rho } G_{ \mu \alpha } \partial_{ \nu }^{ \enspace \alpha } \widetilde{ F }^{ \rho \sigma } Z_{ \sigma } $} & $-$ &  & \multirow{3}{*}{ \scriptsize{$ D^4 H^2 G^{ \mu \nu } G^{ \mu \nu } \widetilde{ B }^{ \mu \nu }$} } &  \\

20 & \scriptsize{$ \left( G^{ \mu \nu } \overleftrightarrow{ D }^{ \alpha } D_{ \nu } G_{ \mu \rho } \right) \partial_{ \alpha } \widetilde{ F }^{ \rho \sigma } Z_{ \sigma } $} & $-$ &  &  &  \\
 
21 & \scriptsize{$ \left( G_{ \mu \nu } \overleftrightarrow{ D }^{ \alpha } G^{ \nu \rho } \right) \partial_{ \rho \alpha } \widetilde{ F }^{ \mu \sigma } Z_{ \sigma } $} & $-$ &  &  &  \\

\cline{5-5}
 
22 & \scriptsize{$ D^{ \alpha } G_{ \mu \nu } D^{ \nu \rho } \widetilde{ G }^{ \mu \sigma } F_{ \rho \alpha } Z_{ \sigma } $} & $-$ &  & \multirow{1}{*}{ \scriptsize{$ D^4 H^2 G^{ \mu \nu } \widetilde{ G }^{ \mu \nu } B^{ \mu \nu }$} } &  \\

\cline{5-5}

23 & \scriptsize{$ \varepsilon^{ \mu \nu \rho \sigma } \left( G_{ \mu \alpha } \overleftrightarrow{ D }^{ \tau } D_{ \nu } G^{ \alpha \beta } \right) \partial_{ \tau } F_{ \rho \beta } Z_{ \sigma } $} & $-$ &  & \multirow{2}{*}{ \scriptsize{$ \varepsilon D^4 H^2 G^{ \mu \nu } G^{ \mu \nu } B^{ \mu \nu }$} } &  \\

24 & \scriptsize{$ \varepsilon^{ \mu \nu \rho \sigma } \left( G_{ \mu \alpha } \overleftrightarrow{ D }_{ \tau } G_{ \nu \beta } \right) \partial^{ \beta \tau } F_{ \rho }^{ \enspace \alpha } Z_{ \sigma } $} & $-$ &  &  &  \\

\hline

25 & \scriptsize{$ \left( G^{ \mu \nu } \overleftrightarrow{ D }_{ \beta } D^{ \sigma \alpha } G_{ \mu \rho } \right) \partial^{ \rho \beta } F_{ \nu \sigma } Z_{ \alpha } $} & $+$ & \multirow{2}{*}{12} & \multirow{1}{*}{ \scriptsize{$ D^6 H^2 G^{ \mu \nu } G^{ \mu \nu } B^{ \mu \nu }$} } & \multirow{2}{*}{ $ \frac{8 \times 10^{ -5 }}{E_\text{TeV}^{9}}, \frac{3 \times 10^{ -4 }}{E_\text{TeV}^{10}} $ } \\

26 & \scriptsize{$ \left( G^{ \mu \nu } \overleftrightarrow{ D }_{ \beta } D_{ \alpha } G_{ \mu \rho } \right) \partial_{ \nu }^{ \enspace \rho \beta } \widetilde{ F }^{ \alpha \sigma } Z_{ \sigma } $} & $-$ &  & \multirow{1}{*}{ \scriptsize{$ D^6 H^2 G^{ \mu \nu } G^{ \mu \nu } \widetilde{ B }^{ \mu \nu }$} } &  \\

\hline

\end{tabular}
\begin{minipage}{5.7in}
\medskip
\caption{\label{tab:zgamgg} \footnotesize Primary operators up to dimension 12 for the $Z\gamma g g $ interaction. There are two redundancies that both appear at dimension 12. To form a set of operators which are independent, $s^n (t-u)^{ 2 m } \mathcal{ O }^{ Z \gamma g g }_{ 4 }, s^n (t-u)^{ 2 m } \mathcal{ O }^{ Z \gamma g g }_{ 9 }$, with $n \geq 0$, $m \geq 1$, should be omitted. }
\end{minipage}
\end{center}
\end{adjustwidth}
\end{table}

\begin{table}[p]
\begin{adjustwidth}{-.5in}{-.5in}  
\begin{center}
\footnotesize
\centering
\renewcommand{\arraystretch}{0.9}
\setlength{\tabcolsep}{6pt}
\begin{tabular}{|c|c|c|c|c|c|}

\hline
\multirow{2}{*}{$i$} & \multirow{2}{*}{$\mathcal{O}_i^{ Z \gamma \gamma \gamma }$}   & \multirow{2}{*}{CP} & \multirow{2}{*}{$d_{\mathcal{O}_i}$}& SMEFT & $c$ Unitarity  \\
 & & & & Operator Form & Bound \\

\hline 

1 & \scriptsize{$ F^{ \mu \nu } \partial_{ \sigma } F_{ \nu \rho } F^{ \rho \sigma } Z_{ \mu } $} & $+$ & \multirow{4}{*}{8} & \multirow{2}{*}{ \scriptsize{$ D^{ 2 } H^2 B^{ \mu \nu } B^{ \mu \nu } B^{ \mu \nu } $} } & \multirow{4}{*}{ $ \frac{0.02}{E_\text{TeV}^{5}}, \frac{0.07}{E_\text{TeV}^{6}} $ } \\

2 & \scriptsize{$ F^{ \mu \nu } \partial^{ \sigma } F_{ \mu \rho } F_{ \nu \sigma } Z^{ \rho } $} & $+$ &  &  &  \\

\cline{5-5}

3 & \scriptsize{$ \partial_{ \rho } F_{ \mu \nu } F^{ \nu \rho } \widetilde{ F }^{ \sigma \mu } Z_{ \sigma } $} & $-$ &  & \multirow{1}{*}{ \scriptsize{$ D^{ 2 } H^2 B^{ \mu \nu } B^{ \mu \nu } \widetilde{ B }^{ \mu \nu } $} } &  \\

\cline{5-5}  

4 & \scriptsize{$ \varepsilon^{ \mu \nu \rho \sigma } F_{ \mu \alpha } F^{ \alpha \beta } \partial_{ \nu } F_{ \beta \rho } Z_{ \sigma } $} & $-$ &  & \multirow{1}{*}{ \scriptsize{$ \varepsilon D^{ 2 } H^2 B^{ \mu \nu } B^{ \mu \nu } B^{ \mu \nu } $} } &  \\

\hline

5 & \scriptsize{$ \partial_{ \rho \alpha } F^{ \mu \nu } \partial_{ \sigma } F_{ \mu \nu } F^{ \rho \sigma } Z^{ \alpha } $} & $+$ & \multirow{10}{*}{10} & \multirow{5}{*}{ \scriptsize{$ D^4 H^2 B^{ \mu \nu } B^{ \mu \nu } B^{ \mu \nu } $} } & \multirow{10}{*}{ $ \frac{0.001}{E_\text{TeV}^{7}}, \frac{0.004}{E_\text{TeV}^{8}} $ } \\

6 & \scriptsize{$ \partial_{ \sigma \alpha } F^{ \mu \nu } \partial^{ \alpha } F_{ \nu \rho } F^{ \rho }_{ \enspace \mu } Z^{ \sigma } $} & $+$ &  &  &  \\

7 & \scriptsize{$ \partial^{ \alpha } F^{ \mu \nu } \partial_{ \sigma \alpha } F_{ \nu \rho } F^{ \rho \sigma } Z_{ \mu } $} & $+$ &  &  &  \\

8 & \scriptsize{$ \partial_{ \alpha } F^{ \mu \nu } \partial^{ \sigma \alpha } F_{ \mu \rho } F_{ \nu \sigma } Z^{ \rho } $} & $+$ &  &  &  \\

9 & \scriptsize{$ \partial^{ \alpha } F^{ \mu \nu } \partial_{ \alpha } F^{ \rho \sigma } \partial_{ \rho } F_{ \mu \nu } Z_{ \sigma } $} & $+$ &  &  &  \\

\cline{5-5}

10 & \scriptsize{$ F^{ \mu \nu } \partial_{ \rho } F_{ \mu \sigma } \partial_{ \nu }^{ \enspace \sigma } \widetilde{ F }^{ \rho \alpha } Z_{ \alpha } $} & $-$ &  & \multirow{3}{*}{ \scriptsize{$ D^4 H^2 B^{ \mu \nu } B^{ \mu \nu } \widetilde{ B }^{ \mu \nu }$} } &  \\

11 & \scriptsize{$ \partial_{ \rho } F_{ \mu \nu } \partial_{ \sigma }^{ \enspace \nu } \widetilde{ F }^{ \mu \alpha } F^{ \rho \sigma } Z_{ \alpha } $} & $-$ &  &  &  \\

12 & \scriptsize{$ \partial_{ \rho \alpha } F_{ \mu \nu } \partial^{ \alpha } F^{ \nu \rho } \widetilde{ F }^{ \mu \sigma } Z_{ \sigma } $} & $-$ &  &  &  \\

\cline{5-5}

13 & \scriptsize{$ \varepsilon^{ \mu \nu \rho \sigma } \partial^{ \tau } F_{ \mu \alpha } \partial_{ \nu \tau } F^{ \alpha \beta } F_{ \rho \beta } Z_{ \sigma } $} & $-$ &  & \multirow{2}{*}{ \scriptsize{$ \varepsilon D^4 H^2 B^{ \mu \nu } B^{ \mu \nu } B^{ \mu \nu } $} } &  \\

14 & \scriptsize{$ \varepsilon^{ \mu \nu \rho \sigma } \partial^{ \tau } F_{ \mu \alpha } \partial_{ \tau } F^{ \alpha \beta } \partial_{ \nu } F_{ \rho \beta } Z_{ \sigma } $} & $-$ &  &  &  \\
 
\hline

15 & \scriptsize{$ \partial_{ \rho \alpha \beta } F^{ \mu \nu } \partial_{ \sigma }^{ \enspace \beta } F_{ \mu \nu } F^{ \rho \sigma } Z^{ \alpha } $} & $+$ & \multirow{8}{*}{12} & \multirow{4}{*}{ \scriptsize{$ D^6 H^2 B^{ \mu \nu } B^{ \mu \nu } B^{ \mu \nu }$} } & \multirow{8}{*}{ $ \frac{8 \times 10^{ -5 }}{E_\text{TeV}^{9}}, \frac{3 \times 10^{ -4 }}{E_\text{TeV}^{10}} $ } \\

16 & \scriptsize{$ \partial_{ \sigma \alpha \beta } F^{ \mu \nu } \partial^{ \alpha \beta } F_{ \nu \rho } F^{ \rho }_{ \enspace \mu } Z^{ \sigma } $} & $+$ &  &  &  \\

17 & \scriptsize{$ \partial^{ \alpha \beta } F^{ \mu \nu } \partial_{ \sigma \alpha \beta } F_{ \nu \rho } F^{ \rho \sigma } Z_{ \mu } $} & $+$ &  &  &  \\

18 & \scriptsize{$ \left( F^{ \mu \nu } \overleftrightarrow{ \partial }_{ \beta } \partial^{ \alpha \sigma } F_{ \mu \rho } \right) \partial^{ \rho \beta } F_{ \nu \sigma } Z_{ \alpha } $} & $+$ & &  &  \\

\cline{5-5}

19 & \scriptsize{$ \partial^{ \beta } F^{ \mu \nu } \partial_{ \rho \beta } F_{ \mu \sigma } \partial_{ \nu }^{ \enspace \sigma } \widetilde{ F }^{ \rho \alpha } Z_{ \alpha } $} & $-$ &  & \multirow{4}{*}{ \scriptsize{$ D^6 H^2 B^{ \mu \nu } B^{ \mu \nu } \widetilde{ B }^{ \mu \nu }$} } &  \\

20 & \scriptsize{$ \partial_{ \rho }^{ \enspace \beta } F_{ \mu \nu } \partial_{ \sigma \beta }^{ \quad \nu } \widetilde{ F }^{ \mu \alpha } F^{ \rho \sigma } Z_{ \alpha } $} & $-$ &  &  &  \\
 
21 & \scriptsize{$ \partial_{ \rho \alpha \beta } F_{ \mu \nu } \partial^{ \alpha \beta } F^{ \nu \rho } \widetilde{ F }^{ \mu \sigma } Z_{ \sigma } $} & $-$ &  &  &  \\
 
22 & \scriptsize{$ \left( F^{ \mu \nu } \overleftrightarrow{ \partial }_{ \beta } \partial_{ \rho } F_{ \mu \alpha } \right) \partial^{ \alpha \beta }_{ \quad \nu } \widetilde{ F }^{ \sigma \rho } Z_{ \sigma } $} & $-$ &  &  &  \\
 
\hline

23 & \scriptsize{$ \partial^{ \beta \tau } F^{ \mu \nu } \partial_{ \mu \alpha \beta \tau } F^{ \rho \sigma } \partial_{ \nu } F_{ \rho \sigma } Z^{ \alpha } $} & $+$ & \multirow{4}{*}{14} & \multirow{2}{*}{ \scriptsize{$ D^8 H^2 B^{ \mu \nu } B^{ \mu \nu } B^{ \mu \nu } $} } & \multirow{4}{*}{ $ \frac{5 \times 10^{ -6 }}{E_\text{TeV}^{11}}, \frac{2 \times 10^{ -5 }}{E_\text{TeV}^{12}} $ } \\

24 & \scriptsize{$ \partial_{ \rho \alpha \beta \tau } F^{ \mu \nu } \partial_{ \sigma }^{ \enspace \beta \tau } F_{ \mu \nu } F^{ \rho \sigma } Z^{ \alpha } $} & $+$ &  &  &  \\

\cline{5-5}
 
25 & \scriptsize{$ \partial^{ \beta \tau } F^{ \mu \nu } \partial_{ \rho \beta \tau } F_{ \mu \sigma } \partial_{ \nu }^{ \enspace \sigma } \widetilde{ F }^{ \rho \alpha } Z_{ \alpha } $} & $-$ &  & \multirow{2}{*}{ \scriptsize{$ D^8 H^2 B^{ \mu \nu } B^{ \mu \nu } \widetilde{ B }^{ \mu \nu }$} } &  \\

26 & \scriptsize{$ \partial_{ \rho }^{ \enspace \beta \tau } F_{ \mu \nu } \partial_{ \sigma \beta \tau }^{ \quad \enspace \nu } \widetilde{ F }^{ \mu \alpha } F^{ \rho \sigma } Z_{ \alpha } $} & $-$ &  &  &  \\

\hline

\end{tabular}
\begin{minipage}{5.7in}
\medskip
\caption{\label{tab:zgamgamgam} \footnotesize Primary operators up to dimension 14 for the $Z \gamma \gamma \gamma$ interaction. At dimension 14, two operators become redundant to operators at dimension 10. To form a set of independent operators, $x^n y^m \mathcal{ O }^{ Z \gamma \gamma \gamma }_{ 9 }$ and $x^n y^m \mathcal{ O }^{ Z \gamma \gamma \gamma }_{ 13 }$, with $x = s^2 + t^2 + u^2$, $y = s t u$, $n \geq 1$, and $m \geq 0$, should be omitted. Replacing all $F_{ \mu \nu }$'s with $G^A_{ \mu \nu }$'s contracted with $d_{ABC}$ yields the symmetric (in exchange of gluon kinematics) primary operators up to dimension 14 for the $Zggg$ interaction. The same redundancies apply to the corresponding operators.}
\end{minipage}
\end{center}
\end{adjustwidth}
\end{table}

\begin{table}[p]
\begin{adjustwidth}{-.5in}{-.5in}  
\begin{center}
\footnotesize
\centering
\renewcommand{\arraystretch}{0.9}
\setlength{\tabcolsep}{6pt}
\begin{tabular}{|c|c|c|c|c|c|}

\hline
\multirow{2}{*}{$i$} & \multirow{2}{*}{$\mathcal{O}_i^{ Z g g g, A }$} & \multirow{2}{*}{CP} & \multirow{2}{*}{$d_{\mathcal{O}_i}$} & SMEFT & $c$ Unitarity  \\
 & & & & Operator Form & Bound \\
 
\hline

1 & \scriptsize{$ G^{ \mu \nu } D_{ \sigma } G_{ \nu \rho } G^{ \rho \sigma } Z_{ \mu } $} & $+$ & \multirow{2}{*}{8} & \multirow{1}{*}{ \scriptsize{$ D^{ 2 } H^2 G^{ \mu \nu } G^{ \mu \nu } G^{ \mu \nu } $} } & \multirow{2}{*}{ $ \frac{0.02}{E_\text{TeV}^{5}}, \frac{0.07}{E_\text{TeV}^{6}} $ } \\

\cline{5-5}

2 & \scriptsize{$ G^{ \mu \nu } D_{ \nu } G_{ \mu \rho } \widetilde{ G }^{ \rho \sigma } Z_{ \sigma } $} & $-$ &  & \multirow{1}{*}{ \scriptsize{$ D^{ 2 } H^2 G^{ \mu \nu } G^{ \mu \nu } \widetilde{ G }^{ \mu \nu } $} } &  \\

\hline

3 & \scriptsize{$ D^{ \alpha } G^{ \mu \nu } D_{ \sigma \alpha } G_{ \mu \rho } G_{ \nu }^{ \enspace \rho } Z^{ \sigma } $} & $+$ & \multirow{8}{*}{10} & \multirow{4}{*}{ \scriptsize{$ D^{ 4 } H^2 G^{ \mu \nu } G^{ \mu \nu } G^{ \mu \nu } $} } & \multirow{8}{*}{ $ \frac{0.001}{E_\text{TeV}^{7}}, \frac{0.004}{E_\text{TeV}^{8}} $ } \\

4 & \scriptsize{$ D^{ \alpha } G^{ \mu \nu } D_{ \nu \alpha } G_{ \rho \sigma } G_{ \mu }^{ \enspace \rho } Z^{ \sigma } $} & $+$ &  &  &  \\

5 & \scriptsize{$ D^{ \alpha } G^{ \mu \nu } D_{ \sigma \alpha } G_{ \nu \rho } G^{ \rho \sigma } Z_{ \mu } $} & $+$ &  &  &  \\

6 & \scriptsize{$ G^{ \mu \nu } D^{ \sigma }_{ \enspace \alpha } G_{ \mu \rho } D^{ \rho } G_{ \nu \sigma } Z^{ \alpha } $} & $+$ &  &  &  \\

\cline{5-5}

7 & \scriptsize{$ D^{ \alpha } G^{ \mu \nu } D_{ \nu \alpha } G_{ \mu \rho } \widetilde{ G }^{ \rho \sigma } Z_{ \sigma } $} & $-$ &  & \multirow{3}{*}{ \scriptsize{$ D^{ 4 } H^2 G^{ \mu \nu } G^{ \mu \nu } \widetilde{ G }^{ \mu \nu } $} } &  \\

8 & \scriptsize{$ D^{ \alpha } G_{ \mu \nu } D_{ \alpha } G^{ \nu \rho } D_{ \rho } \widetilde{ G }^{ \mu \sigma } Z_{ \sigma } $} & $-$ &  &  &  \\

9 & \scriptsize{$ G^{ \mu \nu } D_{ \alpha } G_{ \mu \rho } D_{ \nu }^{ \enspace \rho } \widetilde{ G }^{ \sigma \alpha } Z_{ \sigma } $} & $-$ &  &  &  \\

\cline{5-5}

10 & \scriptsize{$ \varepsilon^{ \mu \nu \rho \sigma } D^{ \tau } G_{ \mu \alpha } D_{ \nu \tau } G^{ \alpha \beta } G_{ \rho \beta } Z_{ \sigma } $} & $-$ &  & \multirow{1}{*}{ \scriptsize{$ \varepsilon D^{ 4 } H^2 G^{ \mu \nu } G^{ \mu \nu } G^{ \mu \nu } $} } &  \\

\hline

11 & \scriptsize{$ D^{ \alpha \beta } G^{ \mu \nu } D_{ \nu \alpha \beta } G_{ \rho \sigma } G_{ \mu }^{ \enspace \rho } Z^{ \sigma } $} & $+$ & \multirow{8}{*}{12} & \multirow{4}{*}{ \scriptsize{$ D^6 H^2 G^{ \mu \nu } G^{ \mu \nu } G^{ \mu \nu }$} } & \multirow{8}{*}{ $ \frac{8 \times 10^{ -5 }}{E_\text{TeV}^{9}}, \frac{3 \times 10^{ -4 }}{E_\text{TeV}^{10}} $ } \\

12 & \scriptsize{$ D^{ \beta } G^{ \mu \nu } D^{ \sigma }_{ \enspace \alpha \beta } G_{ \mu \rho } D^{ \rho } G_{ \nu \sigma } Z^{ \alpha } $} & $+$ &  &  &  \\

13 & \scriptsize{$ \left( G^{ \mu \nu } \overleftrightarrow{ D }^{ \beta } D_{ \mu \alpha } G^{ \rho \sigma } \right) D_{ \nu \beta } G_{ \rho \sigma } Z^{ \alpha } $} & $+$ &  &  &  \\

14 & \scriptsize{$ \left( D_{ \rho \alpha } G^{ \mu \nu } \overleftrightarrow{ D }^{ \beta } D_{ \sigma } G_{ \mu \nu } \right) D_{ \beta } G^{ \rho \sigma } Z^{ \alpha } $} & $+$ &  &  &  \\

\cline{5-5}

15 & \scriptsize{$ D^{ \alpha \beta } G^{ \mu \nu } D_{ \nu \alpha \beta } G_{ \mu \rho } \widetilde{ G }^{ \rho \sigma } Z_{ \sigma } $} & $-$ &  & \multirow{4}{*}{ \scriptsize{$ D^6 H^2 G^{ \mu \nu } G^{ \mu \nu } \widetilde{ G }^{ \mu \nu }$} } &  \\

16 & \scriptsize{$ D^{ \beta } G^{ \mu \nu } D_{ \alpha \beta } G_{ \mu \rho } D_{ \nu }^{ \enspace \rho } \widetilde{ G }^{ \sigma \alpha } Z_{ \sigma } $} & $-$ &  &  &  \\

17 & \scriptsize{$ \left( G^{ \mu \nu } \overleftrightarrow{ D }^{ \beta } D_{ \rho } G_{ \mu \alpha } \right) D_{ \nu \beta }^{ \quad \alpha } \widetilde{ G }^{ \rho \sigma } Z_{ \sigma } $} & $-$ &  &  &  \\

18 & \scriptsize{$ \left(  D_{ \rho } G^{ \mu \nu } \overleftrightarrow{ D }^{ \beta } D_{ \nu \alpha } \widetilde{ G }_{ \mu \sigma } \right) D_{ \beta } G^{ \rho \alpha } Z^{ \sigma } $} & $-$ &  &  &  \\

\hline

19 & \scriptsize{$ D^{ \beta \tau } G^{ \mu \nu } D^{ \sigma }_{ \enspace \alpha \beta \tau } G_{ \mu \rho } D^{ \rho } G_{ \nu \sigma } Z^{ \alpha } $} & $+$ & \multirow{4}{*}{14} & \multirow{2}{*}{ \scriptsize{$ D^8 H^2 G^{ \mu \nu } G^{ \mu \nu } 
G^{ \mu \nu }$} } & \multirow{4}{*}{ $ \frac{5 \times 10^{ -6 }}{E_\text{TeV}^{11}}, \frac{2 \times 10^{ -5 }}{E_\text{TeV}^{12}} $ } \\

20 & \scriptsize{$ \left( D^{ \tau } G^{ \mu \nu } \overleftrightarrow{ D }^{ \beta } D_{ \mu \alpha \tau } G^{ \rho \sigma } \right) D_{ \nu \beta } G_{ \rho \sigma } Z^{ \alpha } $} & $+$ &  &  &  \\

\cline{5-5}

21 & \scriptsize{$ D^{ \beta \tau } G^{ \mu \nu } D_{ \alpha \beta \tau } G_{ \mu \rho } D_{ \nu }^{ \enspace \rho } \widetilde{ G }^{ \sigma \alpha } Z_{ \sigma } $} & $-$ &  & \multirow{2}{*}{ \scriptsize{$ D^8 H^2 G^{ \mu \nu } G^{ \mu \nu } \widetilde{ G }^{ \mu \nu }$} } &  \\
 
22 & \scriptsize{$ \left( D^{ \tau } G^{ \mu \nu } \overleftrightarrow{ D }^{ \beta } D_{ \rho \tau } G_{ \mu \alpha } \right) D_{ \nu \beta }^{ \quad \alpha } \widetilde{ G }^{ \rho \sigma } Z_{ \sigma } $} & $-$ &  &  &  \\
 
\hline

23 & \scriptsize{$ \left( D^{ \tau \pi } G^{ \mu \nu } \overleftrightarrow{ D }^{ \beta } D_{ \mu \alpha \tau \pi } G^{ \rho \sigma } \right) D_{ \nu \beta } G_{ \rho \sigma } Z^{ \alpha } $} & $+$ & \multirow{2}{*}{16} & \multirow{1}{*}{ \scriptsize{$ D^{ 10 } H^2 G^{ \mu \nu } G^{ \mu \nu } G^{ \mu \nu }$} } & \multirow{2}{*}{ $ \frac{3 \times 10^{ -7 }}{E_\text{TeV}^{13}}, \frac{9 \times 10^{ -7 }}{E_\text{TeV}^{14}} $ } \\

24 & \scriptsize{$ \left( D^{ \tau \pi } G^{ \mu \nu } \overleftrightarrow{ D }^{ \beta } D_{ \rho \tau \pi } G_{ \mu \alpha } \right) D_{ \nu \beta }^{ \quad \alpha } \widetilde{ G }^{ \rho \sigma } Z_{ \sigma } $} & $-$ &  & \multirow{1}{*}{ \scriptsize{$ D^{ 10 } H^2 G^{ \mu \nu } G^{ \mu \nu } \widetilde{ G }^{ \mu \nu }$} } &  \\

\hline

\end{tabular}
\begin{minipage}{5.7in}
\medskip
\caption{\label{tab:zggg} \footnotesize Antisymmetric (in exchange of gluon kinematics) primary operators up to dimension 16 for the $Z g g g$ interaction. Note that $\textrm{SU}(3)$ indices are suppressed, where the $G^A_{ \mu \nu }$'s are contracted with the fully antisymmetric structure constant tensor $f_{ABC}$.}
\end{minipage}
\end{center}
\end{adjustwidth}
\end{table}

\begin{table}[p]
\begin{adjustwidth}{-.5in}{-.5in}  
\begin{center}
\footnotesize
\centering
\renewcommand{\arraystretch}{0.9}
\setlength{\tabcolsep}{6pt}
\begin{tabular}{|c|c|c|c|c|c|}

\hline
\multirow{2}{*}{$i$} & \multirow{2}{*}{$\mathcal{O}_i^{ \gamma \gamma \gamma \gamma }$}   & \multirow{2}{*}{CP} & \multirow{2}{*}{$d_{\mathcal{O}_i}$}& SMEFT & $c$ Unitarity  \\
 & & & & Operator Form & Bound \\
 
 \hline

1 & \scriptsize{$ F^{ \mu \nu } F_{ \mu \rho } F_{ \nu \sigma } F^{\sigma \rho } $} & $+$ & \multirow{3}{*}{8} & \multirow{2}{*}{ \scriptsize{$ B^{ \mu \nu } B^{ \mu \nu } B^{ \mu \nu } B^{ \mu \nu } $} } & \multirow{3}{*}{ $ \frac{0.1}{E_\text{TeV}^{4}}$ } \\

2 & \scriptsize{$ F^{ \mu \nu } F_{ \rho \sigma } F_{ \mu \nu } F^{\rho \sigma } $} & $+$ &  & & \\

\cline{5-5}

3 & \scriptsize{$ F^{ \nu \sigma } F_{ \sigma \rho } \widetilde{F}_{ \mu \nu } F^{\rho \mu } $} &$-$&  & {$ B^{ \mu \nu } B^{ \mu \nu } B^{ \mu \nu } \widetilde{B}^{ \mu \nu } $} &  \\

\hline

4 & \scriptsize{$ \partial_{\beta} F^{ \mu \nu }  \partial_{\alpha}F_{ \nu \rho } F^{ \rho }_{ \enspace \mu } F^{\alpha \beta } $} & $+$ & \multirow{5}{*}{10} & \multirow{3}{*}{ \scriptsize{$  D^2 B^{ \mu \nu } B^{ \mu \nu } B^{ \mu \nu } B^{ \mu \nu } $}} & \multirow{5}{*}{ $ \frac{0.006}{E_\text{TeV}^{6}} $ } \\

5 & \scriptsize{$ \partial_{\alpha} F^{ \mu \nu }  \partial_{\sigma}F_{ \nu \rho } F^{ \rho \sigma } F^{ \alpha }_{ \enspace \mu } $} & $+$ &  & &  \\

6 & \scriptsize{$ F^{ \mu \nu }  \partial_{ \alpha \sigma }F_{ \nu \rho } F^{ \rho \sigma } F^{\alpha}_{ \enspace \mu } $} & $+$ &  & &  \\

\cline{5-5}

7 & \scriptsize{$ \partial^{\nu}F^{ \sigma \mu }  \partial_{ \alpha }F_{ \mu \nu } \widetilde{F}_{ \rho \sigma } F^{\alpha \rho } $} & $-$ &  & \multirow{2}{*}{ \scriptsize{$ \varepsilon D^2 B^{ \mu \nu } B^{ \mu \nu } B^{ \mu \nu } B^{ \mu \nu } $}} &  \\

8 & \scriptsize{$ \varepsilon^{ \mu \nu \rho \sigma } \partial^{\tau}F_{ \rho \alpha }  F^{ \alpha \beta } \partial_{\sigma} F_{ \beta \mu } F_{\tau \nu } $} & $-$ &  & &  \\

\hline

9 & \scriptsize{$ \partial_{\alpha}^{\enspace \beta} F^{ \mu \nu }  \partial_{ \sigma \beta}F_{ \nu \rho } F^{ \rho \sigma } F^{ \alpha }_{ \enspace \mu } $} & $+$ & \multirow{1}{*}{12} & \multirow{1}{*}{ \scriptsize{$  D^4 B^{ \mu \nu } B^{ \mu \nu } B^{ \mu \nu } B^{ \mu \nu } $} } & \multirow{1}{*}{ $ \frac{4 \times 10^{ -4 }}{E_\text{TeV}^{8}} $ } \\[1.1ex]

\hline

\end{tabular}
\begin{minipage}{5.7in}
\medskip
\caption{\label{tab:gamgamgamgam} \scriptsize \footnotesize Operators up to dimension 12 for the $\gamma \gamma \gamma \gamma$ interaction. At dimension 14, there are two redundancies such that in order to have a set of independent operators, $x^n y^{m} \mathcal{ O }^{ \gamma \gamma \gamma \gamma }_{ 6 }$ and $x^{n} y^{m} \mathcal{ O }^{ \gamma \gamma \gamma \gamma }_{ 7 }$, with $x = s^2 + t^2 + u^2$, $y = s t u$, $n \geq 1$, and $m \geq 0$, should be omitted. }
\end{minipage}
\end{center}
\end{adjustwidth}
\end{table}

\begin{table}[p]
\begin{adjustwidth}{-.5in}{-.5in}  
\begin{center}
\scriptsize
\centering
\renewcommand{\arraystretch}{0.9}
\setlength{\tabcolsep}{6pt}
\begin{tabular}{|c|c|c|c|c|c|}

\hline
\multirow{2}{*}{$i$} & \multirow{2}{*}{$\mathcal{O}_i^{ \gamma \gamma g g }$}   & \multirow{2}{*}{CP} & \multirow{2}{*}{$d_{\mathcal{O}_i}$}& SMEFT & $c$ Unitarity  \\
 & & & & Operator Form & Bound \\
 
 \hline

1 & \scriptsize{$ F^{ \mu \nu } F_{ \mu \nu } G^{ \rho \sigma } G_{\rho \sigma } $} & $+$ & \multirow{7}{*}{8} & \multirow{4}{*}{ \scriptsize{$ B^{ \mu \nu } B^{ \mu \nu } G^{ \mu \nu } G^{ \mu \nu } $} } & \multirow{7}{*}{ $ \frac{0.1}{E_\text{TeV}^{4}}$ } \\

2 & \scriptsize{$ F^{ \mu \nu } F^{ \rho \sigma } G_{ \mu \nu }  G_{\rho \sigma } $} & $+$ &  &  &  \\

3 & \scriptsize{$ F^{ \mu \nu } F_{ \nu \rho } G^{ \rho \sigma } G_{\sigma \mu } $} & $+$ &  &  &  \\

4 & \scriptsize{$ F^{ \mu \nu } F^{  \rho \sigma } G_{ \nu \rho }  G_{\sigma \mu } $} & $+$ &  &  &  \\

\cline{5-5}

5 & \scriptsize{$ \widetilde{ F }^{ \mu \nu } F_{  \mu \nu } G^{ \rho \sigma }  G_{ \rho \sigma } $} & $-$ &  & \multirow{2}{*}{ \scriptsize{$ B^{ \mu \nu } \widetilde{ B }^{ \mu \nu } G^{ \mu \nu } G^{ \mu \nu }$} } &  \\

6 & \scriptsize{$ \widetilde{ F }^{ \mu \nu } F^{ \rho \sigma } G_{  \mu \nu } G_{ \rho \sigma } $} & $-$ &  &  &  \\

\cline{5-5}

7 & \scriptsize{$ F^{ \mu \nu } F_{  \mu \nu } \widetilde{ G }^{ \rho \sigma }  G_{ \rho \sigma } $} & $-$ &  &  \multirow{1}{*}{ \scriptsize{$  B^{ \mu \nu } B^{ \mu \nu } G^{ \mu \nu } \widetilde{ G }^{ \mu \nu } $} } &  \\
 
\hline

8 & \scriptsize{$ \partial^{ \alpha } F^{ \mu \nu } \partial^{ \sigma } F_{ \nu \rho } G^{ \rho }_{ \enspace \mu } G_{\sigma \alpha } $} & $+$ & \multirow{5}{*}{10} & \multirow{3}{*}{ \scriptsize{$ D^2  B^{ \mu \nu } B^{ \mu \nu } G^{ \mu \nu } G^{ \mu \nu }$} } & \multirow{5}{*}{ $ \frac{0.006}{E_\text{TeV}^{6}} $ } \\

9 & \scriptsize{$ F^{ \mu \nu } \partial^{ \alpha }_{ \enspace \sigma } F_{ \nu \rho } G^{ \rho \sigma } G_{\alpha \mu } $} & $+$ &  &  &  \\

10 & \scriptsize{$ F^{ \mu \nu } \partial^{\alpha}F^{ \rho \sigma } D_{\rho}G_{ \mu \nu } G_{\alpha \sigma } $} & $+$ &  &  &  \\

\cline{5-5}

11 & \scriptsize{$ \varepsilon^{ \mu \nu \rho \sigma } F_{ \rho \alpha } \partial^{\tau}_{\enspace \sigma} F^{ \alpha \beta } G_{ \beta \mu } G_{\tau \nu } $} & $-$ &  & \multirow{2}{*}{ \scriptsize{$ \varepsilon D^2 B^{ \mu \nu } B^{ \mu \nu } G^{ \mu \nu } G^{ \mu \nu } $}} &  \\

12 & \scriptsize{$ \varepsilon^{ \mu \nu \rho \sigma } \partial^{\tau} F_{ \rho \alpha } F^{ \alpha \beta } D_{\sigma} G_{ \beta \mu } G_{\tau \nu } $} & $-$ &  &  &  \\

\hline

\end{tabular}
\begin{minipage}{5.7in}
\medskip
\caption{\label{tab:gamgamgg} \scriptsize \footnotesize Operators up to dimension 10 for the $\gamma \gamma gg$ interaction. At dimension 12, there are two redundancies that should be omitted, along with their descendants, if we want to have a list of independent operators. The corresponding redundancies are given by $s^n (t-u)^{ 2 m } \mathcal{ O }^{ \gamma \gamma g g }_{ 4 }$ and $s^{m+1} (t-u)^{ 2 n } \mathcal{ O }^{ \gamma \gamma g g }_{ 6 }$, with $n \geq 0$, and $m \geq 1$. }
\end{minipage}
\end{center}
\end{adjustwidth}
\end{table}

\begin{table}[p]
\begin{adjustwidth}{-.5in}{-.5in}  
\begin{center}
\footnotesize
\centering
\renewcommand{\arraystretch}{0.9}
\setlength{\tabcolsep}{6pt}
\begin{tabular}{|c|c|c|c|c|c|c|}

\hline
\multirow{2}{*}{$i$} & \multirow{2}{*}{$\mathcal{O}_i^{ \gamma g g g }$} & \multirow{2}{*}{CP} & \multirow{2}{*}{$d_{\mathcal{O}_i}$} & \multirow{2}{*}{$ \textrm{SU}(3) $} & SMEFT & $c$ Unitarity  \\
 & & & & & Operator Form & Bound \\
 
\hline

1 & \scriptsize{$ G^{ \mu \nu } G_{ \mu \rho } G_{ \nu \sigma } F^{ \rho \sigma} $} & $+$ & \multirow{4}{*}{8} & \multirow{4}{*}{$ d_{ ABC } $} & \multirow{2}{*}{ \scriptsize{$  G^{ \mu \nu } G^{ \mu \nu } G^{ \mu \nu } B^{ \mu \nu } $} } & \multirow{4}{*}{ $ \frac{0.1}{E_\text{TeV}^{4}}$ } \\

2 & \scriptsize{$ G^{ \mu \nu } G_{ \mu \nu } G_{ \rho \sigma } F^{ \rho \sigma} $} & $+$ & & &  &  \\

\cline{6-6}

3 & \scriptsize{$ G^{ \mu \nu } G_{ \nu \rho } \widetilde{ G }_{ \mu \sigma } F^{ \rho \sigma } $} & $-$ &  &  & \multirow{1}{*}{ \scriptsize{$  G^{ \mu \nu } G^{ \mu \nu } \widetilde{G}^{ \mu \nu } B^{ \mu \nu } $}} &  \\

\cline{6-6}

4 & \scriptsize{$ G^{ \mu \nu } G_{ \nu \rho } G^{ \rho \sigma } \widetilde{F}_{ \sigma \mu } $} & $-$ &  &  &  \multirow{1}{*}{ \scriptsize{$  G^{ \mu \nu } G^{ \mu \nu } G^{ \mu \nu } \widetilde{B}^{ \mu \nu } $}} &  \\

\hline

5 & \scriptsize{$ D_{ \alpha } G^{ \mu \nu } D_{\sigma} G_{ \nu \rho } G^{ \rho }_{ \enspace \mu } F^{ \sigma \alpha } $} & $+$ & \multirow{12}{*}{10} & \multirow{4}{*}{$ d_{ ABC } $} & \multirow{6}{*}{ \scriptsize{$  D^2 G^{ \mu \nu } G^{ \mu \nu } G^{ \mu \nu } B^{ \mu \nu } $} } & \multirow{12}{*}{ $ \frac{0.006}{E_\text{TeV}^{6}} $ } \\

6 & \scriptsize{$ D^{ \alpha } G^{ \mu \nu } D_{ \sigma } G_{ \nu \rho } G^{ \rho \sigma } F_{  \alpha \mu } $} & $+$ &  &  &  &  \\

7 & \scriptsize{$ G^{ \mu \nu } D_{ \sigma }^{\enspace \alpha} G_{ \nu \rho } G^{ \rho \sigma } F_{  \alpha \mu } $} & $+$ &  &  &  &  \\

8 & \scriptsize{$ D_{\alpha} G^{ \mu \nu } D^{ \sigma } G_{ \mu \rho } G_{ \nu \sigma } F^{ \alpha \rho } $} & $+$ &  &  &  &  \\

\cline{5-5}

9 & \scriptsize{$ D_{ \alpha } G^{ \mu \nu } D_{\nu} G^{ \rho \sigma } G_{ \mu \rho } F^{ \alpha }_{\enspace \sigma } $} & $+$ &  & \multirow{2}{*}{$ f_{ ABC } $} &  &  \\

10 & \scriptsize{$ D^{ \alpha } G^{ \mu \nu } G^{ \rho \sigma } D_{ \nu }G_{ \rho \mu } F_{ \alpha \sigma } $} & $+$ &  &  &  &  \\

\cline{5-5}
\cline{6-6}

11 & \scriptsize{$ D^{ \rho \alpha } G^{ \mu \nu }  G_{ \nu \rho } \widetilde{ G }_{ \mu }^{ \enspace \sigma } F_{\sigma \alpha } $} & $-$ &  & \multirow{2}{*}{$ d_{ ABC } $} & \multirow{4}{*}{ \scriptsize{$ D^{ 2 } G^{ \mu \nu } G^{ \mu \nu } \widetilde{ G }^{ \mu \nu } B^{ \mu \nu }$} } &  \\

12 & \scriptsize{$ D_{ \rho } G_{ \mu \nu } D^{ \alpha } G^{ \nu \rho } \widetilde{ G }^{ \mu \sigma } F_{ \sigma \alpha} $} & $-$ &  &  &  &  \\

\cline{5-5}

13 & \scriptsize{$ D^{ \rho }G^{ \mu \nu } D_{ \alpha } G_{ \nu \rho } \widetilde{G}_{ \sigma \mu } F^{  \alpha \sigma } $} & $-$ &  & \multirow{2}{*}{$ f_{ ABC } $} &  &  \\

14 & \scriptsize{$ G^{ \mu \nu } D_{ \alpha } G_{ \nu \rho } D^{ \rho }\widetilde{G}_{ \sigma \mu } F^{  \alpha \sigma } $} & $-$ &  &  &  &  \\

\cline{5-6}

15 & \scriptsize{$ \varepsilon^{ \mu \nu \rho \sigma } D^{ \tau } G_{ \rho \alpha } G^{ \alpha \beta } D_{\sigma} G_{ \beta \mu } F_{ \tau \nu } $} & $-$ &  &  \multirow{2}{*}{$ d_{ ABC } $} &  \multirow{2}{*}{ \scriptsize{$ \varepsilon D^{ 4 } G^{ \mu \nu } G^{ \mu \nu } G^{ \mu \nu } B^{ \mu \nu } $} } &  \\

16 & \scriptsize{$ \varepsilon^{ \mu \nu \rho \sigma } G_{ \rho \alpha } D^{ \tau } G^{ \alpha \beta } D_{\sigma}G_{ \mu \beta } F_{ \tau \nu } $} & $-$ &  &  & &  \\

\hline

\end{tabular}
\begin{minipage}{5.7in}
\medskip
\caption{\label{tab:ggggamma1} \scriptsize \footnotesize Primary dimension 8, and 10 operators for the $\gamma ggg$ interaction. $G^{ A }_{\mu \nu}$'s are contracted with the fully symmetric structure constant tensor $d_{ABC}$ and the fully antisymmetric structure constant tensor $f_{ABC}$. }
\end{minipage}
\end{center}
\end{adjustwidth}
\end{table}

\begin{table}[p]
\begin{adjustwidth}{-.5in}{-.5in}  
\begin{center}
\footnotesize
\centering
\renewcommand{\arraystretch}{0.9}
\setlength{\tabcolsep}{6pt}
\begin{tabular}{|c|c|c|c|c|c|c|}

\hline
\multirow{2}{*}{$i$} & \multirow{2}{*}{$\mathcal{O}_i^{ \gamma g g g }$} & \multirow{2}{*}{CP} & \multirow{2}{*}{$d_{\mathcal{O}_i}$} & \multirow{2}{*}{$ \textrm{SU}(3) $} & SMEFT & $c$ Unitarity  \\
 & & & & & Operator Form & Bound \\
 
\hline

17 & \scriptsize{$ D^{ \alpha \beta } G^{ \mu \nu } D_{ \sigma \beta } G_{ \nu \rho } G^{ \rho \sigma } F_{  \alpha \mu } $} & $+$ & \multirow{8}{*}{12} & \multirow{2}{*}{$ d_{ ABC } $} & \multirow{4}{*}{ \scriptsize{$  D^4 G^{ \mu \nu } G^{ \mu \nu } G^{ \mu \nu } B^{ \mu \nu } $} } & \multirow{8}{*}{ $ \frac{4 \times 10^{ -4 }}{E_\text{TeV}^{8}} $ } \\

18 & \scriptsize{$ D_{\beta}G^{ \mu \nu } D_{ \sigma }^{\enspace \alpha \beta } G_{ \nu \rho } G^{ \rho \sigma } F_{  \alpha \mu } $} & $+$ &  &  &  &  \\

\cline{5-5}

19 & \scriptsize{$ D_{ \beta }G^{ \mu \nu } D_{\nu}^{\enspace \alpha \beta } G^{ \rho \sigma } G_{ \rho \mu } F_{  \alpha \sigma } $} & $+$ &  & \multirow{2}{*}{$ f_{ ABC } $} &  &  \\

20 & \scriptsize{$ D_{ \rho \beta } G^{ \mu \nu } D^{ \alpha \beta } G_{ \mu \nu }  G^{ \rho \sigma } F_{ \sigma \alpha} $} & $+$ &  &  &  &  \\

\cline{5-6}

21 & \scriptsize{$ D^{ \rho \alpha \beta } G^{ \mu \nu } D_{\beta}G_{ \nu \rho } \widetilde{ G }_{ \mu }^{ \enspace \sigma } F_{\sigma \alpha } $} & $-$ &  & \multirow{2}{*}{$ d_{ ABC } $} & \multirow{3}{*}{ \scriptsize{$  D^4 G^{ \mu \nu } G^{ \mu \nu } \widetilde{G}^{ \mu \nu } B^{ \mu \nu } $} } &  \\

22 & \scriptsize{$ D_{ \rho \beta } G_{ \mu \nu } D^{ \alpha \beta } G^{ \nu \rho } \widetilde{ G }^{ \mu \sigma } F_{ \sigma \alpha} $} & $-$ &  &  &  &  \\

\cline{5-5}

23 & \scriptsize{$ D^{ \rho \beta }G^{ \mu \nu } D_{ \alpha \beta } G_{ \nu \rho } \widetilde{G}_{ \sigma \mu } F^{  \alpha \sigma } $} & $-$ &  & \multirow{2}{*}{$ f_{ ABC } $} &  &  \\

\cline{6-6}

24 & \scriptsize{$ \varepsilon^{ \mu \nu \rho \sigma} D^{ \pi }G_{ \rho \alpha } D^{ \tau }_{\enspace \sigma \pi }G^{ \alpha \beta } G_{ \beta \mu } F_{\tau \nu } $} & $-$ &  &  & \multirow{1}{*}{ \scriptsize{$ \varepsilon D^4 G^{ \mu \nu } G^{ \mu \nu } G^{ \mu \nu } B^{ \mu \nu } $} } &  \\

\hline

25 & \scriptsize{$ D_{\beta \tau}G^{ \mu \nu } D_{ \sigma }^{\enspace \alpha \beta \tau } G_{ \nu \rho } G^{ \rho \sigma } F_{  \alpha \mu } $} & $+$ & \multirow{6}{*}{14} & \multirow{1}{*}{$ d_{ ABC } $} & \multirow{3}{*}{ \scriptsize{$ D^6 G^{ \mu \nu } G^{ \mu \nu } 
G^{ \mu \nu } B^{ \mu \nu }$} } & \multirow{6}{*}{ $ \frac{2 \times 10^{ -5 }}{E_\text{TeV}^{10}} $ } \\

\cline{5-5}

26 & \scriptsize{$ D_{ \alpha \beta \tau} G^{ \mu \nu } D_{\nu}^{\enspace \beta \tau} G^{ \rho \sigma } G_{ \mu \rho } F^{ \alpha }_{\enspace \sigma } $} & $+$ & & \multirow{2}{*}{$ f_{ ABC } $} &  &   \\

27 & \scriptsize{$ D_{ \beta \tau }G^{ \mu \nu } D_{\nu}^{\enspace \alpha \beta \tau } G^{ \rho \sigma } G_{ \rho \mu } F_{  \alpha \sigma } $} & $+$ &  &  &  &  \\

\cline{5-6}

28 & \scriptsize{$ D^{ \rho \alpha \beta \tau } G^{ \mu \nu } D_{\beta \tau}G_{ \nu \rho } \widetilde{ G }_{\mu}^{ \enspace \sigma } F_{\sigma \alpha } $} & $-$ &  & \multirow{1}{*}{$ d_{ ABC } $} & \multirow{2}{*}{ \scriptsize{$ D^6 G^{ \mu \nu } G^{ \mu \nu } \widetilde{G}^{ \mu \nu } B^{ \mu \nu }$} } &  \\

\cline{5-5}

29 & \scriptsize{$ D^{ \rho \beta \tau }G^{ \mu \nu } D_{ \alpha \beta \tau } G_{ \nu \rho } \widetilde{G}_{ \sigma \mu } F^{  \alpha \sigma } $} & $-$ &  & \multirow{2}{*}{$ f_{ ABC } $} &  &  \\

\cline{6-6}

30 & \scriptsize{$ \varepsilon^{ \mu \nu \rho \sigma} D^{ \pi \eta}G_{ \rho \alpha } D^{ \tau }_{\enspace \sigma \pi \eta }G^{ \alpha \beta } G_{ \beta \mu } F_{\tau \nu } $} & $-$ &  &  & \multirow{1}{*}{ \scriptsize{$ \varepsilon D^6 G^{ \mu \nu } G^{ \mu \nu }
G^{ \mu \nu } B^{ \mu \nu }$} } &  \\

\hline

31 & \scriptsize{$ \left( D^{ \tau \pi }_{\enspace \enspace \alpha} G^{ \mu \nu } \overleftrightarrow{ D }^{ \beta } D_{ \tau \pi \sigma } G_{ \nu \rho } \right) D_{ \beta } G^{ \rho }_{ \enspace \mu } F^{ \sigma \alpha } $} & $+$ & \multirow{4}{*}{16} & \multirow{4}{*}{$ f_{ ABC } $} &  \multirow{2}{*}{ \scriptsize{$ D^8 G^{ \mu \nu } G^{ \mu \nu } 
G^{ \mu \nu } B^{ \mu \nu }$} } & \multirow{4}{*}{ $ \frac{ 10^{ -6 }}{E_\text{TeV}^{12}} $ } \\

32 & \scriptsize{$ \left( D^{ \tau \pi } G^{ \mu \nu } \overleftrightarrow{ D }^{ \beta } D^{\alpha}_{ \enspace \sigma \tau \pi } G_{ \nu \rho } \right) D_{ \beta } G^{ \rho \sigma } F_{ \alpha \mu } $} & $+$ &  &  & &  \\
\cline{6-6}

33 & \scriptsize{$ \left( D^{ \rho \alpha \tau \pi } G^{ \mu \nu } \overleftrightarrow{ D }^{ \beta } D_{ \tau \pi } G_{ \nu \rho } \right) D_{ \beta } \widetilde{ G }_{ \mu }^{ \enspace \sigma } F_{\sigma \alpha } $} & $-$ & &  & \multirow{1}{*}{\scriptsize{ $D^8 G^{ \mu \nu } G^{ \mu \nu } 
\widetilde{G}^{ \mu \nu } B^{ \mu \nu }$}} & \\

\cline{6-6}

34 & \scriptsize{$ \varepsilon^{ \mu \nu \rho \sigma } \left( D^{ \tau \pi \eta } G_{ \rho \alpha } \overleftrightarrow{ D }^{ \xi } D_{ \pi \eta } G^{ \alpha \beta } \right) D_{ \sigma \xi } G_{ \beta \mu } F_{ \tau \nu } $} & $-$ &  &  & \multirow{1}{*}{ \scriptsize{$ \varepsilon D^8 G^{ \mu \nu } G^{ \mu \nu } 
G^{ \mu \nu } B^{ \mu \nu }$} } &  \\

\hline

\end{tabular}
\begin{minipage}{5.7in}
\medskip
\caption{\label{tab:ggggamma2} \scriptsize \footnotesize Primary dimension 12, 14 and 16 operators for the $\gamma ggg$ interaction. There are two
redundancies that appear at dimension 14. We can form a set of independent operators by removing the operators and descendants, $x^n y^m \mathcal{ O }^{ \gamma ggg}_{ 8 }$, $x^n y^m \mathcal{ O }^{ \gamma ggg}_{ 16 }$, with $x = s^2 + t^2 + u^2$, $y = s t u$, $n \geq 1$ and $m \geq 0$, which are redundant.}
\end{minipage}
\end{center}
\end{adjustwidth}
\end{table}

\begin{table}[p]
\begin{adjustwidth}{-.5in}{-.5in}  
\begin{center}
\footnotesize
\centering
\renewcommand{\arraystretch}{0.9}
\setlength{\tabcolsep}{6pt}
\begin{tabular}{|c|c|c|c|c|c|c|}

\hline
\multirow{2}{*}{$i$} & \multirow{2}{*}{$\mathcal{O}_i^{ g g g g }$} & \multirow{2}{*}{CP} & \multirow{2}{*}{$d_{\mathcal{O}_i}$} & \multirow{2}{*}{$ \textrm{SU(3) Trace} $} & SMEFT & $c$ Unitarity  \\
 & & & & & Operator Form & Bound \\
 
\hline

1 & \scriptsize{$ G^{ A \mu \nu } G^{B}_{ \enspace \mu \nu } G^{ C \rho \sigma } G^{ D }_{ \enspace \rho \sigma} $} & $+$ & \multirow{9}{*}{8} & \multirow{4}{*}{$ Tr (T^2) Tr (T^2) $} & \multirow{6}{*}{ \scriptsize{$  G^{ \mu \nu } G^{ \mu \nu } G^{ \mu \nu } G^{ \mu \nu } $} } & \multirow{9}{*}{ $ \frac{0.1}{E_\text{TeV}^{4}}$ } \\

2 & \scriptsize{$ G^{ A \mu \nu }  G^{ B \rho \sigma } G^{C}_{ \enspace \mu \nu } G^{ D }_{ \enspace \rho \sigma} $} & $+$ & & &  &  \\

3 & \scriptsize{$ G^{ A \mu \nu } G^{B}_{ \enspace \nu \rho } G^{ C \rho \sigma } G^{ D }_{ \enspace \sigma \mu} $} & $+$ & & &  &  \\

4 & \scriptsize{$ G^{ A \mu \nu } G^{ B \rho \sigma } G^{C}_{ \enspace \nu \rho } G^{ D }_{ \enspace \sigma \mu} $} & $+$ & & &  &  \\

\cline{5-5}

5 & \scriptsize{$ G^{ A \mu \nu } G^{ B \rho \sigma }  G^{C}_{ \enspace \mu \nu } G^{ D }_{ \enspace \rho \sigma} $} & $+$ & & \multirow{2}{*}{$ Tr(T^4) $} &  &  \\

6 & \scriptsize{$ G^{ A \mu \nu } G^{B}_{ \enspace \nu \rho } G^{ C }_{ \enspace \sigma \mu} G^{ D \rho \sigma } $} & $+$ & & &  &  \\

\cline{5-5}
\cline{6-6}

7 & \scriptsize{$ G^{ A \mu \nu } \widetilde{G}^{ B }_{ \enspace \mu \nu } G^{ C \rho \sigma } G^{ D }_{ \enspace \rho \sigma } $} & $-$ &  & \multirow{1}{*}{$ Tr (T^2) Tr (T^2) $} & \multirow{3}{*}{ \scriptsize{$  G^{ \mu \nu } G^{ \mu \nu } G^{ \mu \nu } \widetilde{G}^{ \mu \nu } $}} &  \\

\cline{5-5}

8 & \scriptsize{$ G^{ A \mu \nu } G^{ B \rho \sigma } \widetilde{G}^{ C }_{ \enspace \mu \nu }  G^{ D }_{ \enspace \rho \sigma } $} & $-$ &  & \multirow{2}{*}{$ Tr(T^4) $} & &  \\

9 & \scriptsize{$ \widetilde{G}^{ A \mu \nu } G^{ B }_{ \enspace \nu \rho } G^{ C \rho \sigma } G^{ D }_{ \enspace \sigma \mu } $} & $-$ &  & & &  \\

\hline

10 & \scriptsize{$ D^{ \alpha }G^{ A \mu \nu } D_{ \alpha}G^{B}_{ \enspace \mu \nu } G^{ C \rho \sigma } G^{ D }_{ \enspace \rho \sigma} $} & $+$ & \multirow{14}{*}{10} & \multirow{5}{*}{$ Tr (T^2) Tr (T^2) $} & \multirow{6}{*}{ \scriptsize{$  D^{2} G^{ \mu \nu } G^{ \mu \nu } G^{ \mu \nu } G^{ \mu \nu } $} } & \multirow{14}{*}{ $ \frac{0.006}{E_\text{TeV}^{6}} $ }  \\

11 & \scriptsize{$ D^{ \alpha }G^{ A \mu \nu } D_{ \alpha }G^{ B \rho \sigma } G^{C}_{ \enspace \mu \nu } G^{ D }_{ \enspace \rho \sigma} $} & $+$ & & &  &  \\

12 & \scriptsize{$ D^{ \alpha }G^{ A \mu \nu } D_{ \alpha }G^{B}_{ \enspace \nu \rho } G^{ C \rho \sigma } G^{ D }_{ \enspace \sigma \mu} $} & $+$ & & &  &  \\

13 & \scriptsize{$ D^{ \alpha }G^{ A \mu \nu } D_{ \alpha }G^{ B \rho \sigma } G^{C}_{ \enspace \nu \rho } G^{ D }_{ \enspace \sigma \mu} $} & $+$ & & &  &  \\

14 & \scriptsize{$ \left( G^{ A \mu \nu } \overleftrightarrow{ D }^{ \alpha } G^{ B \rho \sigma } \right) D_{ \alpha } G^{C}_{ \enspace \mu \nu } G^{ D }_{ \enspace \rho \sigma} $} & $+$ & & &  &  \\
\cline{5-5}

15 & \scriptsize{$ D^{ \alpha }G^{ A \mu \nu } G^{ B \rho \sigma }  D_{ \alpha }G^{C}_{ \enspace \mu \nu } G^{ D }_{ \enspace \rho \sigma} $} & $+$ & & \multirow{4}{*}{$ Tr(T^4) $} &  &  \\

16 & \scriptsize{$ D^{ \alpha }G^{ A \mu \nu } G^{B}_{ \enspace \nu \rho } D_{ \alpha }G^{ C }_{ \enspace \sigma \mu} G^{ D \rho \sigma } $} & $+$ & & &  &  \\

17 & \scriptsize{$ \left( G^{ A \mu \nu } \overleftrightarrow{ D }^{ \alpha } G^{ C \rho \sigma } \right) D_{ \alpha }G^{B}_{ \enspace \mu \nu } G^{ D }_{ \enspace \rho \sigma} $} & $+$ & &  &  &  \\

18 & \scriptsize{$ D_{ \alpha }G^{ A \mu \nu } D_{ \sigma }G^{B}_{ \enspace \nu \rho } G^{ C \rho }_{ \quad \mu } G^{ D \sigma \alpha} $} & $+$ & & & &  \\

\cline{5-6}

19 & \scriptsize{$ D^{\alpha} G^{ A \mu \nu } D_{ \alpha }\widetilde{ G }^{ B }_{ \enspace \mu \nu } G^{ C \rho \sigma } G^{ D }_{ \enspace \rho \sigma } $} & $-$ &  & \multirow{1}{*}{$ Tr (T^2) Tr (T^2) $} & \multirow{3}{*}{ \scriptsize{$  D^{2} G^{ \mu \nu } G^{ \mu \nu } G^{ \mu \nu } \widetilde{G}^{ \mu \nu } $}} &  \\

\cline{5-5}

20 & \scriptsize{$ D^{\alpha}G^{ A \mu \nu } G^{ B \rho \sigma } D_{ \alpha } \widetilde{G}^{ C }_{ \enspace \mu \nu }  G^{ D }_{ \enspace \rho \sigma } $} & $-$ &  & \multirow{2}{*}{$ Tr(T^4) $} & &  \\

21 & \scriptsize{$ D^{ \alpha } \widetilde{G}^{ A \mu \nu } G^{ B }_{ \enspace \nu \rho } D_{ \alpha }G^{ C \rho \sigma } G^{ D }_{ \enspace \sigma \mu } $} & $-$ &  & & &  \\

\cline{5-6}

22 & \scriptsize{$ \varepsilon^{ \mu \nu \rho \sigma } D^{ \tau } G^{ A }_{ \enspace \rho \alpha } G^{ B \alpha \beta } D_{\sigma} G^{ C }_{ \enspace \beta \mu } G^{ D }_{ \enspace \tau \nu } $} & $-$ &  & \multirow{1}{*}{$ Tr (T^2) Tr (T^2) $} & \multirow{2}{*}{ \scriptsize{$  \varepsilon D^{2} G^{ \mu \nu } G^{ \mu \nu } G^{ \mu \nu } G^{ \mu \nu } $}} &  \\
\cline{5-5}

23 & \scriptsize{$ \varepsilon^{ \mu \nu \rho \sigma } D^{ \tau } G^{ A }_{ \enspace \rho \alpha } G^{ B \alpha \beta } D_{\sigma} G^{ C }_{ \enspace \beta \mu } G^{ D }_{ \enspace \tau \nu } $} & $-$ &  &  \multirow{1}{*}{$ Tr(T^4) $} &  &  \\

\hline

\end{tabular}
\begin{minipage}{5.7in}
\medskip
\caption{\label{tab:ggggg} \scriptsize
\footnotesize Primary dimension 8 and 10 operators for the
$gggg$ interaction. $G^{A}_{\mu \nu}$'s are contacted with trace factors where $Tr (T^{2}) Tr (T^{2})$ represents $Tr(T^{A}T^{B})Tr(T^{C}T^{D})$, and $Tr (T^{4}) $ represents $Tr(T^{A}T^{B}T^{C}T^{D})$.}
\end{minipage}
\end{center}
\end{adjustwidth}
\end{table}

\begin{table}[p]
\begin{adjustwidth}{-.5in}{-.5in}  
\begin{center}
\scriptsize
\centering
\renewcommand{\arraystretch}{0.9}
\setlength{\tabcolsep}{6pt}
\begin{tabular}{|c|c|c|c|c|c|c|}

\hline
\multirow{2}{*}{$i$} & \multirow{2}{*}{$\mathcal{O}_i^{ g g g g }$} & \multirow{2}{*}{CP} & \multirow{2}{*}{$d_{\mathcal{O}_i}$} & \multirow{2}{*}{$ \textrm{SU(3) Trace} $} & SMEFT & $c$ Unitarity  \\
 & & & & & Operator Form & Bound \\

\hline

24 & \scriptsize{$ \left( G^{ A \mu \nu } \overleftrightarrow{D}^{ \alpha \beta } G^{B}_{ \enspace \mu \nu } \right) D_{ \alpha \beta }G^{ C \rho \sigma } G^{ D }_{ \enspace \rho \sigma} $} & $+$ & \multirow{16}{*}{12} & \multirow{5}{*}{$ Tr (T^2) Tr (T^2) $} & \multirow{10}{*}{ \scriptsize{$  D^{4} G^{ \mu \nu } G^{ \mu \nu } G^{ \mu \nu } G^{ \mu \nu } $} } & \multirow{16}{*}{ $ \frac{4 \times 10^{ -4 }}{E_\text{TeV}^{8}} $ }  \\

25 & \scriptsize{$ \left( G^{ A \mu \nu } \overleftrightarrow{D}^{ \alpha \beta} G^{ B \rho \sigma } \right) D_{ \alpha \beta }G^{C}_{ \enspace \mu \nu } G^{ D }_{ \enspace \rho \sigma} $} & $+$ & & &  &  \\

26 & \scriptsize{$ \left( G^{ A \mu \nu } \overleftrightarrow{D}^{ \alpha \beta } G^{B}_{ \enspace \nu \rho } \right) D_{ \alpha \beta }G^{ C \rho \sigma } G^{ D }_{ \enspace \sigma \mu} $} & $+$ & & &  &  \\

27 & \scriptsize{$ \left(D^{ \beta }G^{ A \mu \nu } \overleftrightarrow{ D }^{ \alpha } D_{ \beta }G^{ B \rho \sigma } \right) D_{ \alpha } G^{C}_{ \enspace \mu \nu } G^{ D }_{ \enspace \rho \sigma} $} & $+$ & & &  &  \\
\cline{5-5}

28 & \scriptsize{$ \left( G^{ A \mu \nu } \overleftrightarrow{D}^{ \alpha \beta }G^{C}_{ \enspace \mu \nu } \right) D_{ \alpha \beta }G^{ B \rho \sigma }   G^{ D }_{ \enspace \rho \sigma} $} & $+$ & & \multirow{6}{*}{$ Tr(T^4) $} &  &  \\

29 & \scriptsize{$ \left( G^{ A \mu \nu } \overleftrightarrow{D}^{ \alpha \beta }G^{ C }_{ \enspace \sigma \mu} \right) D_{ \alpha \beta }G^{B}_{ \enspace \nu \rho } G^{ D \rho \sigma } $} & $+$ & & &  &  \\

30 & \scriptsize{$ \left( D^{ \beta }G^{ A \mu \nu } \overleftrightarrow{ D }^{ \alpha } D_{ \beta }G^{ C \rho \sigma } \right) D_{ \alpha }G^{B}_{ \enspace \mu \nu } G^{ D }_{ \enspace \rho \sigma} $} & $+$ & &  &  &  \\

31 & \scriptsize{$ D_{ \alpha \beta } G^{ A \mu \nu } D_{ \sigma }G^{B}_{ \enspace \nu \rho } D^{ \beta }G^{ C \rho }_{ \quad \mu } G^{ D \sigma \alpha} $} & $+$ & & & &  \\

32 & \scriptsize{$ \left( D_{ \alpha }G^{ A \mu \nu } \overleftrightarrow{D}^{ \beta } G^{ C \rho }_{ \quad \mu } \right)D_{ \sigma \beta }G^{B}_{ \enspace \nu \rho }  G^{ D \sigma \alpha} $} & $+$ & & & &  \\

33 & \scriptsize{$ D_{ \alpha \beta }G^{ A \mu \nu } D_{ \sigma }^{ \enspace \beta } G^{B}_{ \enspace \nu \rho } G^{ C \rho }_{ \quad \mu } G^{ D \sigma \alpha} $} & $+$ & & & &  \\

\cline{5-6}

34 & \scriptsize{$ \left( G^{ A \mu \nu } \overleftrightarrow{D}^{ \alpha \beta }\widetilde{ G }^{ B }_{ \enspace \mu \nu } \right) D_{ \alpha \beta}G^{ C \rho \sigma } G^{ D }_{ \enspace \rho \sigma } $} & $-$ &  & \multirow{1}{*}{$ Tr (T^2) Tr (T^2) $} & \multirow{3}{*}{ \scriptsize{$  D^{4} G^{ \mu \nu } G^{ \mu \nu } G^{ \mu \nu } \widetilde{G}^{ \mu \nu } $}} &  \\

\cline{5-5}

35 & \scriptsize{$ \left( G^{ A \mu \nu } \overleftrightarrow{D}^{ \alpha \beta }\widetilde{G}^{ C }_{ \enspace \mu \nu } \right) D_{ \alpha \beta}G^{ B \rho \sigma } G^{ D }_{ \enspace \rho \sigma } $} & $-$ &  & \multirow{2}{*}{$ Tr(T^4) $} & &  \\

36 & \scriptsize{$ \left( D^{ \rho}_{ \enspace \alpha }G^{ A \mu \nu } \overleftrightarrow{D}^{ \beta } \widetilde{G}^{ C }_{\enspace \mu \sigma } \right) D_{ \beta }G^{B}_{ \enspace \nu \rho } G^{ D \sigma \alpha} $} & $-$ &  &   &  &  \\

\cline{5-6}

37 & \scriptsize{$ \varepsilon^{ \mu \nu \rho \sigma } D^{ \tau \pi } G^{ A }_{ \enspace \rho \alpha } D_{ \pi }G^{ B \alpha \beta } D_{\sigma} G^{ C }_{ \enspace \beta \mu } G^{ D }_{ \enspace \tau \nu } $} & $-$ &  & \multirow{1}{*}{$ Tr (T^2) Tr (T^2) $} & \multirow{3}{*}{ \scriptsize{$  \varepsilon D^{4} G^{ \mu \nu } G^{ \mu \nu } G^{ \mu \nu } G^{ \mu \nu } $}} &  \\
\cline{5-5}

38 & \scriptsize{$ \varepsilon^{ \mu \nu \rho \sigma } D^{ \tau \pi } G^{ A }_{ \enspace \rho \alpha } G^{ B \alpha \beta } D_{ \sigma \pi } G^{ C }_{ \enspace \beta \mu } G^{ D }_{ \enspace \tau \nu } $} & $-$ &  &  \multirow{2}{*}{$ Tr(T^4) $} &  &  \\

39 & \scriptsize{$ \varepsilon^{ \mu \nu \rho \sigma } \left( D^{ \tau } G^{ A }_{ \enspace \rho \alpha } \overleftrightarrow{D}^{ \pi } D_{\sigma} G^{ C }_{ \enspace \beta \mu } \right) D_{ \pi }G^{ B \alpha \beta }  G^{ D }_{ \enspace \tau \nu } $} & $-$ &  &  &  &  \\

\hline

40 & \scriptsize{$ \left( G^{ A \mu \nu } \overleftrightarrow{ D }^{ \alpha \beta \tau} G^{ B \rho \sigma } \right) D_{ \alpha \beta \tau } G^{C}_{ \enspace \mu \nu } G^{ D }_{ \enspace \rho \sigma} $} & $+$ & \multirow{10}{*}{$ 14 $} & \multirow{1}{*}{$ Tr (T^2) Tr (T^2) $} & \multirow{5}{*}{ \scriptsize{$  D^{6} G^{ \mu \nu } G^{ \mu \nu } G^{ \mu \nu } G^{ \mu \nu } $} } & \multirow{10}{*}{ $ \frac{2 \times 10^{ -5 }}{E_\text{TeV}^{10}} $ } \\

\cline{5-5}

41 & \scriptsize{$ \left( G^{ A \mu \nu } \overleftrightarrow{ D }^{ \alpha \beta \tau } G^{ C \rho \sigma } \right) D_{ \alpha \beta \tau }G^{B}_{ \enspace \mu \nu } G^{ D }_{ \enspace \rho \sigma} $} & $+$ & & \multirow{4}{*}{$ Tr(T^4) $} &  &  \\

42 & \scriptsize{$ \left( D_{ \alpha } G^{ A \mu \nu } \overleftrightarrow{D}^{ \beta \tau} G^{ C \rho }_{ \quad \mu } \right) D_{ \sigma \beta \tau }G^{B}_{ \enspace \nu \rho } G^{ D \sigma \alpha} $} & $+$ & & & &  \\

43 & \scriptsize{$ \left( D_{ \alpha \tau }G^{ A \mu \nu } \overleftrightarrow{D}^{ \beta } D^{ \tau } G^{ C \rho }_{ \quad \mu } \right)D_{ \sigma \beta }G^{B}_{ \enspace \nu \rho }  G^{ D \sigma \alpha} $} & $+$ & & & &  \\

44 & \scriptsize{$ \left( D_{ \alpha }G^{ A \mu \nu } \overleftrightarrow{D}^{ \beta \tau} D_{ \sigma } G^{B}_{ \enspace \nu \rho } \right) D_{ \beta \tau }G^{ C \rho }_{ \quad \mu } G^{ D \sigma \alpha} $} & $+$ & & & &  \\

\cline{5-6}

45 & \scriptsize{$ \left( D^{ \rho}_{ \enspace \alpha \tau }G^{ A \mu \nu } \overleftrightarrow{D}^{ \beta } D^{ \tau } \widetilde{G}^{ C }_{\enspace \mu \sigma } \right) D_{ \beta }G^{B}_{ \enspace \nu \rho } G^{ D \sigma \alpha} $} & $-$ &  & \multirow{1}{*}{$ Tr(T^4) $}  & \multirow{1}{*}{ \scriptsize{$  D^{6} G^{ \mu \nu } G^{ \mu \nu } G^{ \mu \nu } \widetilde{G}^{ \mu \nu } $} } &  \\

\cline{5-6}

46 & \scriptsize{$ \varepsilon^{ \mu \nu \rho \sigma } \left( D^{ \tau } G^{ A }_{ \enspace \rho \alpha } \overleftrightarrow{D}^{ \pi \eta} G^{ B \alpha \beta } \right) D_{ \sigma \pi \eta} G^{ C }_{ \enspace \beta \mu } G^{ D }_{ \enspace \tau \nu } $} & $-$ &  & \multirow{1}{*}{$ Tr (T^2) Tr (T^2) $} & \multirow{3}{*}{ \scriptsize{$  \varepsilon D^{4} G^{ \mu \nu } G^{ \mu \nu } G^{ \mu \nu } G^{ \mu \nu } $}} &  \\
\cline{5-5}

47 & \scriptsize{$ \varepsilon^{ \mu \nu \rho \sigma } \left( D^{ \tau } G^{ A }_{ \enspace \rho \alpha } \overleftrightarrow{D}^{ \pi \eta }D_{ \sigma } G^{ C }_{ \enspace \beta \mu } \right) D_{ \pi \eta }G^{ B \alpha \beta } G^{ D }_{ \enspace \tau \nu } $} & $-$ &  &  \multirow{2}{*}{$ Tr(T^4) $} &  &  \\

48 & \scriptsize{$ \varepsilon^{ \mu \nu \rho \sigma } \left( D^{ \tau \eta} G^{ A }_{ \enspace \rho \alpha } \overleftrightarrow{D}^{ \pi } D_{ \sigma \eta } G^{ C }_{ \enspace \beta \mu } \right) D_{ \pi }G^{ B \alpha \beta }  G^{ D }_{ \enspace \tau \nu } $} & $-$ &  &  &  &  \\

\hline

49 & \scriptsize{$ \left( D_{ \alpha }G^{ A \mu \nu } \overleftrightarrow{D}^{ \beta \tau \pi } G^{ C \rho }_{ \quad \mu } \right)D_{ \sigma \beta \tau \pi }G^{B}_{ \enspace \nu \rho }  G^{ D \sigma \alpha} $} & $+$ & \multirow{2}{*}{$ 16 $} & \multirow{2}{*}{$ Tr(T^4) $} & \multirow{1}{*}{ \scriptsize{$  D^{8} G^{ \mu \nu } G^{ \mu \nu } G^{ \mu \nu } G^{ \mu \nu } $} } & \multirow{2}{*}{ $ \frac{ 10^{ -6 }}{E_\text{TeV}^{12}} $ } \\

\cline{6-6}

50 & \scriptsize{$ \left( D^{ \rho}_{ \enspace \alpha }G^{ A \mu \nu } \overleftrightarrow{D}^{ \beta \tau \pi }  \widetilde{G}^{ C }_{\enspace \mu \sigma } \right) D_{ \beta \tau \pi }G^{B}_{ \enspace \nu \rho } G^{ D \sigma \alpha} $} & $-$ &  &   & \multirow{1}{*}{ \scriptsize{$  D^{8} G^{ \mu \nu } G^{ \mu \nu } G^{ \mu \nu } \widetilde{G}^{ \mu \nu } $} } &  \\

\hline

\end{tabular}
\begin{minipage}{5.7in}
\medskip
\caption{\label{tab:ggggg2} \scriptsize
\footnotesize Primary dimension 12, 14, and 16 operators for the
$gggg$ interaction. There are two
redundancies that appear at dimension 14 and four at dimension 16. To form a set of independent operators, the operators and descendants $x^n y^m \mathcal{ O }^{gggg}_{ 4 }$, $x^n y^m \mathcal{ O }^{gggg}_{ 9  }$, $x^n y^m \mathcal{ O }^{gggg}_{ 12 }$, $x^n y^m \mathcal{ O }^{gggg}_{ 14 }$, $x^n y^m \mathcal{ O }^{gggg}_{ 19 }$, $x^n y^m \mathcal{ O }^{gggg}_{ 20 }$  with $x = s^2 + t^2 + u^2$, $y = s t u$, $n \geq 0$ and $m \geq 1$, should be omitted.}

\end{minipage}
\end{center}
\end{adjustwidth}
\end{table}

\section{Decays of the $Z$ Boson}
\label{sec:Zdecays}
Now that we have the amplitudes and unitarity bounds, we can continue the analysis of modifications to the decays of the $Z$ boson, taking into account the upper bounds on coupling strengths from the Tables.
First, we start with the decays $Z\to \bar{f}f(\gamma, g)$.  Such decays occur in the Standard Model through radiation of the gauge boson off of the fermions.  Such emissions are collinear enhanced, so there could be hope that the contact amplitudes in Ref.~\cite{Bradshaw_2023} could be distinguished in differential distributions.  In this part of phase space, we estimate $\mathrm{BR}_{\mathrm{SM}}(Z\to \gamma \bar{f}f) \approx \frac{\alpha}{4\pi} \mathrm{BR}_{\mathrm{SM}}(Z\to \bar{f}f)$.  Assuming just the irreducible background from the Standard Model in the channel $Z\to \gamma \bar{\mu}\mu$, we find that to get a 1$\sigma$ fluctuation at the HL-LHC with $\sim 6\times 10^9$ $Z$ bosons, would require unitarity violation at about 5 TeV for the dimension 6 and dimension 7 amplitudes, which have two field strengths.  Of course, reducible backgrounds will reduce this estimate, but this suggests that for reasonably high unitarity violating scales, this could be searched for at the LHC.  For $Z\to \bar{f}f g,$ the fermions would need to be a quark-antiquark pair, for which the $b$ quark would be the most promising.  However, these would have substantial QCD backgrounds, so our optimistic analysis would be entirely too unrealistic.    

We now turn to $Z$ decays into three gauge bosons.  The $Z \rightarrow 3 \gamma$ decay is allowed in the SM, but is a higher-order process that is only possible through at least a single $W$ boson or fermion loop \cite{BAILLARGEON1991158, PhysRevD.46.5074, Yang:1994nd, PhysRevD.23.2795, VanderBij:188077, Bernreuther:199833, Altarelli:1989hv, glover_z_1993}, leading to a predicted branching ratio of  $\sim5 \times 10^{ -10 }$ \cite{glover_z_1993}.\footnote{See \cite{PhysRevD.49.3775} for a study of effective Lagrangians for the $Z\gamma\gamma\gamma$ interaction and \cite{PhysRevD.50.602} for a study where the decay is mediated by a scalar loop in a two Higgs doublet model using a minimally supersymmetric Standard Model.} Most recently, an ATLAS search found the bound $\textrm{BR}(Z \rightarrow 3 \gamma) < 2.2 \times 10^{ -6 }$ \cite{ATLAS:2015rsn}. In this analysis, the background consists of a irreducible part with three or more prompt photons and  a reducible component of a combination of photons and electrons or hadronic jets misidentified as photons. With the HL-LHC luminosity of $\sim 3000 \textrm{ fb}^{ -1 }$ \cite{bruning_high-luminosity_2019, arduini_machine_2020}, this bound should naively improve to about $\textrm{BR}(Z \rightarrow 3 \gamma) \lesssim 3 \times 10^{ -7 }.$  This is much larger than the SM prediction, so we need the BSM amplitude to dominate, which we estimate requires unitarity violation at 500 (330) GeV for dimension 8 (10) operators.  Such low unitarity violating scales would require new physics to hide itself from LHC searches, so disfavors the possibility of observing this at the HL-LHC without substantial improvements in reducing reducible backgrounds.  In fact, if we impose our unitarity bounds to be violated at a TeV, this predicts $\mathrm{BR}(Z\to 3 \gamma) \lesssim 2\times 10^{-9}$ for dimension 8 amplitudes, which is extremely challenging to observe at a hadron collider.  However, one could hope that a future lepton collider producing $10^{12}$ $Z$'s might have more controllable backgrounds that could reach these small branching ratios.   

Next, we look at $Z \rightarrow \gamma gg$ and $Z\to ggg$ decays. The branching ratios are predicted to be $\sim 4.9 \times 10^{ -6 }$ \cite{1983PhRvD..27..196L, VanderBij:188077, perez_new_2004} and
$\sim 1.8 \times 10^{ -5 }$ \cite{1983PhRvD..27..196L, VanderBij:188077, Bernreuther:199833, perez_new_2004}, respectively.  Since our unitarity bound estimates are the same for $Z\to (\gamma\gamma\gamma, \gamma gg, ggg)$, even if we optimistically assume that there is only the irreducible background from the Standard Model prediction and that unitarity is violated at a TeV, we find that there is at most a $2 \sigma$ fluctuation for the HL-LHC run for the dimension 8 interactions.  Since these would be likely to have substantial reducible backgrounds, this shows that these are unlikely to be observable.  Such a conclusion was also reached in \cite{VanderBij:188077}, which showed that measuring the associated  $gg \rightarrow Zg$ interaction at a hadron collider was ``rather remote" due to there being too much background, particularly from $q \bar{ q } \rightarrow Z g$, $qg \rightarrow Zq$ and $\bar{ q }g \rightarrow Z\bar{ q }$ processes. They also concluded that attempts at measuring the coupling from observations of the $Z \rightarrow ggg$ process at a lepton collider would suffer too greatly from $Z \rightarrow q \bar{q} g$ background.

\section{Conclusions}
\label{sec:conclusions}
This paper has determined the allowed on-shell amplitudes for 4-point interactions of gauge bosons in the Standard Model.  
Following ~\cite{Chang:2022crb, Bradshaw_2023}, this has completed the analysis of all 3- and 4-point interactions for the Standard Model content.  For certain couplings, this required studying Lagrangian operators up to mass dimension 16, which demonstrates the efficacy of the numerical approach used in these papers.  
  
The characterization of these amplitudes holds the promise of allowing the most general model independent collider searches by studying the interactions of Standard Model particles.  They also serve as a useful intermediary between experimental and theoretical analyses since, in comparison to EFT operators, they are interpretable and do not suffer from basis ambiguities. As an illustration of phenomenological study, we investigated the potential for discovering new physics in $Z$ decays.  In these estimates, we showed that $Z\to \gamma \bar{\ell}\ell$ is of interest at the HL-LHC but other modes, like $Z\to \gamma\gamma\gamma$ would require unitarity violation well below a TeV, which are likely in violation of direct search constraints.

Moving forward, it will be useful to perform realistic phenomenological studies at the HL-LHC and future colliders.  Another direction is to use the Mandelstam descendants as a model for theoretical uncertainties.  Finally, the utilization of on-shell amplitudes in realistic analyses will undoubtedly require solutions to practical challenges along the way.  We and our collaborators are currently exploring such questions and hope that this work has motivated others to do the same.  
\section*{Acknowledgements}
We would like to thank M.~Luty for useful discussions.
The work of C.A. and S.C. was supported in part by the U.S. Department of Energy under Grant Number DE-SC0011640.

\bibliographystyle{utphys}
\bibliography{references}

\end{document}